\documentclass{emulateapj}
\usepackage{graphicx}

\newcommand{\xmm}{{\it XMM-Newton} }
\begin{document}
\title{X-ray properties of an Unbiased Hard X-ray Detected Sample of AGN}

\author{Lisa M. Winter\altaffilmark{1}, Richard F. Mushotzky\altaffilmark{2}, Jack Tueller\altaffilmark{2}, \& Craig Markwardt\altaffilmark{1, 2}}
\email{lwinter@astro.umd.edu}

\altaffiltext{1}{University of Maryland, College Park, MD}
\altaffiltext{2}{ NASA Goddard Space Flight Center, Greenbelt, MD}

\begin{abstract}
The SWIFT gamma ray observatory's Burst Alert Telescope (BAT) has detected a
sample of active galactic nuclei (AGN) based solely on their hard X-ray flux (14-195\,keV).
In this paper, we present for the first time {\it XMM-Newton} X-ray spectra for 22 BAT AGNs
with no previously analyzed X-ray spectra.  
If our sources are a representative sample of the BAT AGN, as we claim, our
results present for the first time global X-ray properties of an unbiased towards absorption (n$_H < 3 \times 10^{25}$\,cm$^{-2}$), local ($<z> = 0.03$), AGN sample.
We find 9/22 low absorption 
(n$_H < 10^{23}$\,cm$^{-2}$), simple power law model
sources, where 4 of these sources have a statistically significant soft component.  Among these sources, we find the presence of a warm absorber statistically significant for only one Seyfert 1 source, contrasting with the ASCA results of \citet{rey97} and \citet{geo98}, who find signatures of warm absorption in half or more of their Seyfert 1 samples at similar redshifts.
Additionally, the remaining sources (14/22) have more complex spectra, well-fit by an absorbed power law at $E > 2.0$\,keV.  Five of the complex sources are classified as Compton-thick candidates.  Further, we find four more sources  with properties consistent with the hidden/buried AGN reported by Ueda {\it et al.} (2007).  Finally, we include a comparison of the {\it XMM-Newton}
EPIC spectra with available SWIFT X-ray Telescope (XRT) observations.  From these
comparisons, we find 6/16 sources with varying column densities, 6/16 sources
with varying power law indices, and 13/16 sources with varying fluxes, over periods of
hours to months.  Flux and power law index are correlated for objects where both parameters vary.

\end{abstract}

\section{Introduction}

Active galactic nuclei (AGN) surveys are
typically dominated by two selection effects: (1) dilution by starlight
from the host galaxy and (2) obscuration by dust and gas in the host
galaxy and/or the AGN itself (see Hewett \& Foltz 1994 and Mushotzky
2004).  These factors previously kept an unbiased AGN sample from
reach.  However, with the capabilities of SWIFT's Burst Alert
Telescope (BAT), this has changed. The BAT surveys the sky in the hard
X-ray range of 14-195\,keV.  As of early 2006, BAT detected $\sim$150
AGN with a median redshift of 0.03 \citep{mar05, tue07}.  These AGN
were selected purely by their hard X-ray flux, and thus, all but the most
heavily absorbed sources (n$_H > 3 \times 10^{25}$\,cm$^{-2}$) are not
affected by obscuration from gas and dust, which prevents them from
being easily detected in UV, optical, or soft band X-ray surveys.  Thus, the BAT survey
will provide the information to derive the true distribution of AGN characteristics across
the entire electromagnetic spectrum.

The BAT detection limit is a few $\times 10^{-11}$\,erg\,s$^{-1}$\,cm$^{-2}$.  Thus, the BAT
AGNs are powerful sources, with X-ray luminosities over the full range of AGN luminosities.
For the closest Seyfert galaxies (z$\approx0.01$), the BAT flux limit corresponds to a
14 -- 195\,keV luminosity of roughly $3\times10^{42}$\,erg\,s$^{-1}$.
Most of the sources ($\approx 97$ \%)
have been optically detected as relatively bright nearby objects with 
$<z> \approx 0.03$, while most of the sources without bright optical counterparts are blazars, of which there are 15/153 \citep{tue07}.  Since the BAT sources are mostly bright and nearby, they are good sources with which to study multi-wavelength properties. 

Despite the X-ray brightness of the BAT AGN, a number of these sources had yet to be observed spectroscopically
in the X-rays.  We present an analysis of {\it XMM-Newton} EPIC spectra of 23 sources, 2 of which correspond to a pair of interacting galaxies where the BAT source is confused, i.e. likely a combination of the AGN, for
which spectra were obtained through discretionary project scientist time.  These sources had
no previous X-ray spectrum, were clearly detected by the BAT, and had clear optical
counterparts from Digital Sky Survey (DSS) images. Many of these sources now also
have SWIFT X-ray Telescope (XRT) spectra available.  Thus, we are able to compare the
XRT spectra with the EPIC spectra.  In Section 2, we detail the observations and data analysis
for the XMM-Newton and SWIFT observations.  In Section 3, we describe spectral fits to 
the X-ray spectra, including a search for variability between the XRT and {\it XMM-Newton} observations.  We summarize our findings in Section 4.  


\section{Observations and Data Analysis}
\subsection{XMM-Newton and SWIFT XRT Spectra}
We obtained proprietary {\it XMM-Newton} observations of 22 out of 26 proposed BAT AGN 
sources through
discretionary project scientist time.  These particular BAT AGNs were selected based on: their
detection in BAT with high significance ($\sigma > 5$), a clear optical counterpart to
the BAT source in Digital Sky Survey images, and no previous X-ray spectrum.  
A list of these sources, whose X-ray spectra 
were observed for the
first time, is in Table~\ref{tbl-1}.  Here we list the source name, co-ordinates, redshift,
Galactic column density towards the source, AGN type, and host galaxy
type.  For one of the BAT sources, the ``BAT source'' is an interacting galaxy system with two
AGNs, MCG +04-48-002 and NGC 6921.
In addition to the {\it XMM-Newton} observations, we downloaded archived SWIFT XRT
observations (48) for the 23 AGNs from the NASA HEASARC archives.  
In Table~\ref{tbl-2} we include
details on the observations examined.

We reduced the {\it XMM-Newton} data using the Science
Analysis System (SAS) version 7.0.  We created calibrated photon
event files for the EPIC-MOS and PN cameras using the
observation data files (ODF) with the commands {\tt emchain}
and {\tt epchain}.  Following this, the events tables were filtered
using the standard criteria outlined in the {\it XMM-Newton ABC Guide}\footnote{\url{http://heasarc.gsfc.nasa.gov/docs/xmm/abc/}}.  For
the MOS data (both MOS1 and MOS2 cameras), good events constitute
those with a pulse height in the range of 0.2 to 12 keV and event
patterns that are characterized as 0-12 (single, double, triple, and
quadruple pixel events).  For the PN camera, only patterns of 0-4
(single and double pixel events) are kept, with the energy range for
the pulse height set between 0.2 and 15 keV.  Bad pixels and events
too close to the edges of the CCD chips were rejected using the
stringent selection expression ``FLAG == 0''. 

Light curves of the observations were produced with {\tt xmmselect} and
examined for flaring events (distinguished by high count rates).  Time
filtering was required only for the sources SWIFT J0641.3+3257, SWIFT J0911.2+4533,
MCG +04-22-042, MRK 417, WKK 1263, and NGC 6921/MCG +04-48-002 (the interacting system).  The light curves were filtered using the task
{\tt tabgtigen}, as outlined in the SAS ABC guide, with the parameters `RATE$<50$' counts/s
for the PN and `RATE$<10$' counts/s for MOS observations.
 
Spectra of the sources were extracted using the SAS task {\tt especget}.  This routine extracts
source and background spectra from a defined region as well as response and ancillary
response matrices.  We extracted the source spectra from circular regions 
with radii between $\approx 30-125$\,\arcsec.  The extraction radius depended upon the 
location of the source with respect to other sources and the edge of a CCD chip.  Background
regions were extracted from annular regions surrounding the source, where possible.  In cases
were the source was near the edge of a chip or another source, circular regions near the
source were used, of the same size as the source region and on the same chip.

Of the 22 target observations, there were a few cases where the position of the source fell
within a gap in the CCD chip (for either the PN or a MOS detector).  This occurred for
the MOS1 observation of SWIFT J0641.3+3257 and the PN observation of SWIFT J0911.2+4533.
Thus, there are no corresponding spectra from these CCDs.  Also, the position of source 
SWIFT J0904.3+5538 was located such that a large fraction of the light (66\%) was located in a gap in the PN
CCD.  Due to a problem with the ODF files for  NGC 6860, the PN
observation of this source could not be extracted. 

For the XRT data, we extracted spectra of target sources from the cleaned photon counting 
event files downloaded from the public High Energy Astrophysics Science Archive Research Center (HEASARC) archive.  We extracted spectra for observations that had XRT 
exposure times of at least 1000\,s, following the instructions outlined in {\it The
SWIFT XRT Data Reduction Guide}\footnote{\url{http://heasarc.nasa.gov/docs/swift/analysis/xrt\_swguide\_v1\_2.pdf}}.  Spectra were extracted for the sources using the FTOOL
XSELECT.  With this tool, a source region was created in DS9, with a radius of $\approx$ 50 -- 70\,\arcsec.  A
background region was created in a source free region close to the source, with a radius of
95\,\arcsec.  We used the standard response and ancillary response files available for the photon
counting event files with grades 0 to 12.  These are publicly available in the
SWIFT XRT calibration database (CALDB).

For all of the spectra, we binned the source and background spectrum and response files for each observation
with the FTOOL {\tt grppha}.  All spectra were binned with 20\,counts/bin.  We then fit
the spectra in the 0.3-10\,keV range with XSPEC version 11.  The details of these fits are
described in the following section.  

\subsection{SWIFT BAT Spectra}

BAT is a wide field ($\approx 2$ steradians) coded aperture hard X-ray instrument. During normal operations, it usually covers $\approx 60$\% of the sky each day at $< 20$ milliCrab sensitivity.  The BAT spectra were derived from an independent all sky mosaic map in each energy bin, averaged over 22 months of data, beginning on Dec 5 2004 (Tueller in preparation). The survey was processed using the BAT Ftools and additional software to make mosaic maps that will be released soon. The energy bin edges are 14, 20, 24, 35, 50, 75, 100, 150, 195 keV. The energies are calibrated in-flight for each detector using an on-board electronic pulser and the 59.5 keV gamma and  La L and M lines from a tagged 241Am source. The average count rate in the map bin that corresponds to the known position of the counterpart was used. Due to the the strong correlation of the signal in adjacent map bins of the oversampled coded aperture image, it is not necessary to perform a fit to the PSF.  Each rate was normalized to the Crab nebula rate using an assumed spectra of $10.4 \times E^{-2.15}$\,ph\,cm$^{-2}$\,s$^{-1}$\,keV$^{-1}$ for the BAT energy range. Due to the large number of different pointings that contribute to any position in the map, this is a good approximation of the response. This has been verified by fitting sources with known spectra (Cas-A compared to BeppoSax and
Integral, PSR1509-589 compared to Integral, NGC2110 and Cen-A compared to
Suzaku) and generally produces a good connection to X-ray spectra in sources that do not vary much with time.  Error estimates were derived directly from the mosaic images using the RMS image noise in a region around the source of roughly 3 degrees in radius.  This is the optimum procedure due to the residual systematic errors of 1.2 to 1.8 times statistical values in the current BAT mosaics.  Analysis of the noise in the images suggests that the variations in noise are small on this scale.  Analysis of negative fluctuations shows that the noise is very will fit by a Gaussian distribution and that this normalization is very accurate on average.  All fitting of the BAT data was performed on this normalized data using a diagonal instrument response matrix. This procedure correctly accounts for instrumental systematics in sources with spectral indices similar to the Crab. While there may be significant systematic errors for sources with spectra that are much flatter than the Crab, this is not a significant problem for any of the sources presented in this paper.

\section{Spectral Fitting}

\begin{figure}
\epsscale{1.0}\plotone{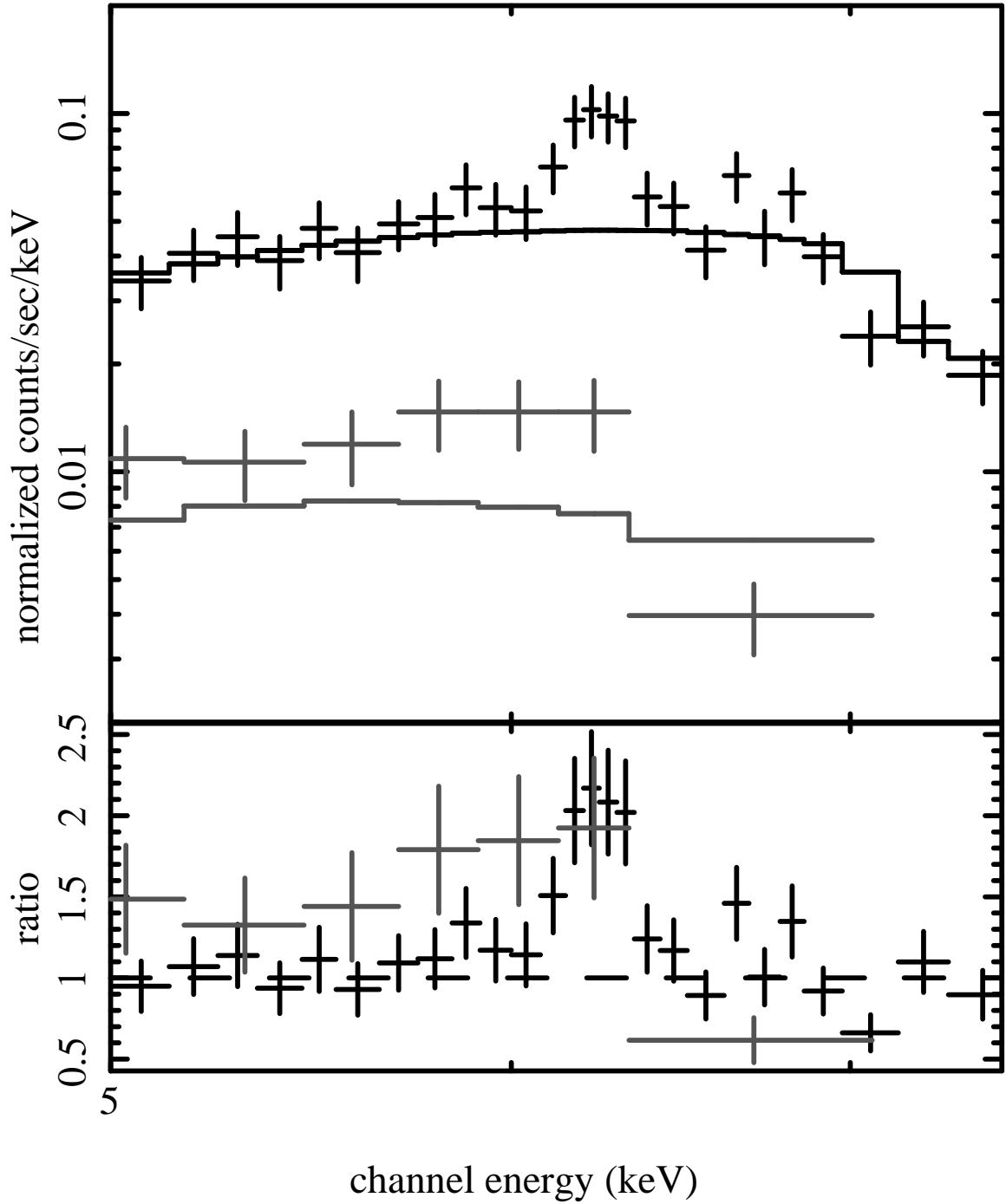}
\caption{{\it XMM-Newton} PN and the highest quality XRT spectrum for NGC 1142 centered on the 6.4\,keV fluorescent Fe K line.  The lines represent a simple absorbed power law model.  In the $\approx 10$\,ks PN, the Fe K line is clearly distinguishable requiring the addition of a Gaussian component.
However, the $\approx 7$\,ks XRT spectrum, binned as the PN spectrum with 20 photons/bin, does
not have the spectral resolution required to distinguish this feature.  We found this to be the case with
all of the XRT spectra examined for these 22 BAT AGNs.
\label{fig1}}
\end{figure}

In examining the X-ray spectra of these BAT AGNs, there are two main goals: to determine
how the sources vary between observations and to determine the spectral properties
of the source (the hydrogen column density, spectral shape, and properties of the Fe K line and/or other lines if
present).  Since the extracted \xmm spectra have, on average, ten times the number of
counts from the XRT spectra, we will focus on the \xmm spectra for a more detailed analysis.  The effective spectral resolution for the XRT is lower, since there are fewer counts and at 20\,cts per bin even fewer counts remain, such that Fe K
lines which are clearly visible in 10\,ks \xmm observations are not resolved in comparable
XRT observations (see Figure~\ref{fig1}).  Thus, in order to examine the variability 
between the XRT and \xmm observations we need to rely on simple models for the AGN spectra.

As a first fit to the spectra and as a means to compare variability between XRT and \xmm
observations, we
fit each observation separately with a simple absorbed power law (absorption model * {\tt pow}).  
In XSPEC, we used
the absorption model {\tt tbabs} to account for cold absorption in the Milky Way.  Thus, we fixed 
the {\tt tbabs} model hydrogen absorption to the \citet{dic90} value (see Table~\ref{tbl-1}).  
We allowed
the parameter of a second {\tt tbabs} model to float in order to account for the combination of absorption from
the AGN host galaxy and local environment.  

To apply this model to the \xmm EPIC observations, we simultaneously
fit the PN and MOS observations, using a constant value to account for differences in flux calibration.  
The best-fit spectral parameters for these observations are recorded
in Table~\ref{tbl-4}, including the observed flux values at
soft (0.3-2\,keV) and hard (2-10\,keV) energies.  We include the best-fit spectral
parameters for the XRT observations in Table~\ref{tbl-3}.  All quoted errors represent the
90\% confidence level.  We excluded from our spectral fits any observation with $\leq 50$\,counts.
Also, we note that this simple power law model is a very poor fit to the \xmm spectra of 
NGC 612, NGC 1142, ESO 362-G018, MRK 417, ESO 506-G027, and MCG +04-48-002.
For each of these five sources the reduced $\chi^2$ value indicate that the simple power
law fit was not a good description of the data.  Thus, no errors
were calculated for the spectral parameters.  We will discuss the spectra of these sources
and more acceptable models below.

\subsection{Detailed Spectral Properties}
From our initial examination of the spectra, we found that simple power law models (see Table~\ref{tbl-4}) were not sufficient in describing the spectra of all of our sources.  In particular, there are
three main features that were not accounted for by a simple absorbed power law: a soft excess,
line emission, and additional complexity from emission with varying column densities.  
A ``soft excess''
in an AGN may be the result of thermal emission (for instance, from star formation), 
low temperature Comptonization, blurred reflection \citep{cze03, ros05}, or 
blurred absorption \citep{gie04}.
Line emission can be produced by a number of mechanisms, in particular from fluorescent, photo-ionized or
collisionally ionized gas.  Most importantly for AGN, is fluorescence from lowly ionized iron,
 with the strongest
feature being the Fe K line at 6.4\,keV.  The final type of ``feature'', a complex spectrum with 
different column densities, is difficult to interpret.  Such spectra, which appear to have a ``double
power law'' shape, could be the result of contamination of the AGN light by a less absorbed X-ray binary spectrum, a dusty environment where the AGN emission is partially covered by absorbing material, or scattering
of some of the AGN light.

In order to better characterize our sources, we first grouped our sources in basic categories: (1)
pure power law sources (with or without absorption beyond the Galactic Milky Way values), 
(2) power law sources with a soft excess
(with or without absorption), and (3) more complex ``double power law'' shaped spectra.
Based on the F-test, any additional components added to our spectral models improved the
fit by a $\chi^2$ of at least 8 (this is the 99\% level for 2 additional degrees of freedom).  The
only exception is the addition of a gaussian line ({\tt zgauss} in XSPEC) to characterize
the Fe K-$\alpha$ line at 6.4\,keV.  Where the addition of the line was not significant, we derived upper limits on the strength and intensity (indicated by the normalization which is the integrated
photon flux of the line) of a possible emission line.  Therefore,
for all of our sources, we fixed the energy of a gaussian line to 6.4\,keV in the source's rest frame
with a fixed width (FWHM) of 0.01\,keV, corresponding to an unresolved line. 

\subsubsection{Simple Power Law Sources}
A total of five of our sources (5/22) had \xmm spectra best-fit by simple power law models.  None of
these sources showed evidence of a strong Fe K line.  We record the equivalent width and normalization value for these lines in Table~\ref{tbl-5}.  
Since the redshift for the source SWIFT J0216.3+5128 is unknown
and there is no evidence for a strong line, we exclude this source from the analysis.  We attempted
to obtain an optical spectrum for this source as part of our BAT AGN follow-up program with
the 2.1-m telescope at Kitt Peak National Observatory (Winter {\it et al.} in preparation).  We found,
however, that the probable Digital Sky Survey optical counterpart within the BAT and \xmm error
circles was a star.  Likely, the true optical counterpart of this source is faint and below the
DSS detection limit.  Based on the lack of a bright optical counterpart and the featureless X-ray
power law spectrum, this source is probably a blazar.

\subsubsection{Soft Excesses}
Four sources (4/22) had \xmm spectra best represented by a power law with a soft excess.  Since the
spectra do not have the counts necessary to distinguish between a reflection or absorption model,
we used a simple blackbody ({\tt bbody} in XSPEC) model to characterize this component.  The
best-fit spectral parameters are shown in Table~\ref{tbl-6}.  The soft excess is significant for all
of these sources with $\Delta\chi^2$ values, between the simple power law and the power law with
a blackbody model, from $\approx 220 - 1400$.  As with the pure power law sources, we fit an
Fe K line at 6.4\,keV with a Gaussian.  The results are recorded in Table~\ref{tbl-6}.  

\subsubsection{Warm Absorbers}
In our models for AGN spectra, we assume that the absorption along the line of sight is cold, neutral hydrogen (with the {\tt tbabs} model).  However, signatures of warm, optically-thin absorption from photo-ionized gas, have been detected in half of an ASCA Seyfert 1 AGN sample studied by \citet{rey97} and 13/18 of the ASCA Seyfert 1 AGN sample of \citet{geo98}.  The main signatures of a warm absorber are 
the \ion{O}{7} and \ion{O}{8} K edges at 0.74\,keV and 0.87\,keV.  Since warm absorbers are seen in such a large fraction of ASCA observations of Seyfert 1 sources and our data are of high enough quality to distinguish the warm absorber signatures, or at least constrain upper limits,  we looked for these signatures among the sources with a high number of counts below 2\,keV (with the exception of possible blazar SWIFT J0216.3+5128).  

Following \citet{rey97}, we added two edge models ({\tt zedge} in XSPEC) to account for the \ion{O}{7} and \ion{O}{8} K edges.  We fixed the edge at the energies of these warm absorber signatures (0.74\,keV and 0.87\,keV), allowing the optical depth to vary.  In Table~\ref{tbl-11}, we record the errors on optical depth and the change in $\chi^2$.  In the \citet{rey97} sample, optical depths for clearly detected edges ranged from $\approx 0.10 - 1.5$.  From our spectral fitting, half of the sources have upper limits of $\tau < 0.10$ for \ion{O}{7} and 7/8 have upper limits  of $\tau < 0.10$ for \ion{O}{8}.  In fact, the only source with a clear detection of both edges is ESO 490-G026.  Clearly, much less than half of our low absorption/simple X-ray model sources show evidence of warm absorbing material in their spectra.  We will discuss our results further in the Discussion section.

\subsubsection{Complex X-ray spectra sources}

The remaining 13 BAT sources ($\approx 1/2$) had spectra we classified as ``double power law'' spectra, similar to \citet{tur97}.  Earlier,
we mentioned three scenarios that could create this type of a spectrum.  Based on these three
possibilities (viewing another source(s) such as X-ray binaries/diffuse galactic emission
along with the absorbed AGN, cold gas partially covering
the AGN light, and scattering of the AGN light from regions of different column densities)
 we decided to fit the spectra of these sources with the partial covering model, {\tt pcfabs} in
 XSPEC.    The partial
 covering model (see: \citet{hol80}) has two parameters:
 the hydrogen column density and the covering fraction.  In addition to describing a partial
 absorber spectrum, this model is also useful for describing a spectrum where AGN light has been
 scattered, where the power law index of the direct, heavily absorbed spectrum is the same as
 that of the scattered component (which is not heavily absorbed). This model, to summarize, 
 allows for flexibility in the soft spectrum which can fit all of the three physical origins mentioned for a 
 complex spectrum (assuming the X-ray binary/diffuse emission has the same power law slope as the
 AGN emission, which is not expected).
 Thus, we fit the spectra of the
 remaining 13 sources with the model {\tt tbabs}*{\tt pcfabs}*({\tt pow} + {\tt zgauss}) (Table~\ref{tbl-7}).
 
 As we noted, the partial covering model gives an equivalent result to a model with a power
 law modified by different amounts of absorption and two flux normalizations.  We also wanted to 
 test whether the power law spectral indices for these two components are the same or vary.  
 To this end, we
 fit the sources with the model {\tt tbabs}*({\tt tbabs}*{\tt pow} + {\tt tbabs}*({\tt pow} + {\tt zgauss})).
 This model allows the power law indices, normalizations, and column densities to vary for two 
 separate power laws.  The results do not differ significantly 
 for half of the sources between the partial covering model (Table~\ref{tbl-7}) and the separate
 power laws model (Table~\ref{tbl-8}), with $\Delta\chi^2 < 8$.   Each of these seven sources (excluding MCG +04-48-002) show a
 ratio of the low absorption power law to the more highly absorbed power law flux (N$_{\Gamma_1}$/N$_{\Gamma_2}$) less than 0.14 with an average value of 0.03.  These results could be consistent
 with any of the three possible physical models, where the portion of scattered light or additional 
 non-AGN emission or unabsorbed AGN light is very small.  This is true of all of the sources, with the exception of ESO 362-G018 and NGC 6860, whose spectra are more complicated.
 
 For the BAT source corresponding to the interacting system (NGC 6921 and MCG +04-48-002), both sources are clearly absorbed.  For MCG +04-48-002, the absorbed power law component in the double power law model was not well constrained due to the weak contribution from the low absorption power law component.  This source was much weaker than NGC 6921 in the \xmm observation, by an order of magnitude.  Thus, throughout the remainder of the paper we will distinguish NGC 6921 as ``the BAT source''.  This will not change any results, since both sources are absorbed sources with similar spectral results.  We added MCG +04-48-002 to the spectral fits because, though it is clearly the weaker source in the XMM observations, it is brighter than NGC 6921 in the XRT observations.  Further, in recently obtained Suzaku observations, which we are currently analyzing, we found that MCG +04-48-002 was the brighter source.  These findings and a detailed analysis will be discussed in a future paper.
 
\begin{figure}
\epsscale{1.0}\plotone{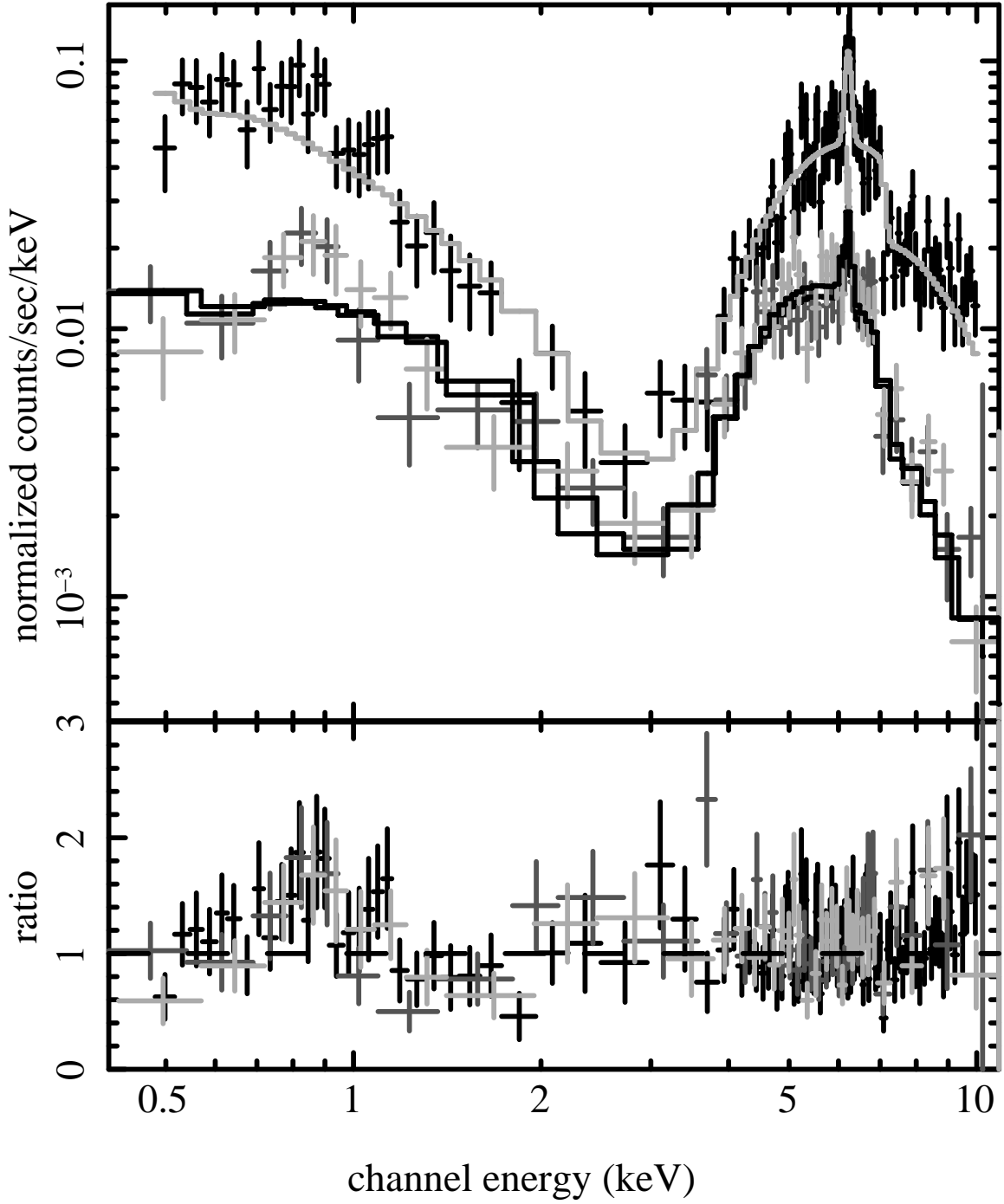}
\caption{\xmm spectrum (PN, MOS1, and MOS2) of NGC 1142 fit with the model
{\tt tbabs}*{\tt pcfabs}*({\tt pow} + {\tt zgauss})*{\tt const}.  There is a clear soft excess with
possible unresolved lines (in PN).  We find that a better fit to this source is obtained with the addition of a blackbody model.
\label{ngc1142}}
\end{figure}

Based on reduced $\chi^2$, three of the ``double power law'' sources (NGC 1142, ESO 362-G018, and ESO 506-G027) require additional/alternative models.  For NGC 1142, a soft excess is clearly
present (see Figure~\ref{ngc1142}).  The addition of a blackbody component with a temperature
kT$= 0.123$\,keV improved the separate power laws fit (Table~\ref{tbl-8}), yielding an acceptable fit with $\chi^2$/dof of 217.13/191.  We note that this blackbody temperature, kT$ = 0.123$\,keV, is similar to the values
seen in Table~\ref{tbl-6} for the sources fit with simple blackbody and power law models.
The spectrum of ESO 362-G018, however, is still even more complicated.  This spectrum
appears to have well defined lines, particularly a strong line measured with an energy of
0.56\,keV, which is consistent with \ion{O}{7} and improves the separate power laws fit by $\Delta\chi^2 = 50$.  However, the power
law spectral index for this source is still extremely flat ($\Gamma = 0.67$) where the
typical value for $\Gamma$ is $\approx 1.80$ \citep{mus82}.  This is also true of
NGC 612, MRK 417, ESO 506-G027, and NGC 6860.
 
Flat power law indices have been noted, in addition to high column densities ($n_H \geq 10^{24}$\,cm$^{-2}$) and strong Fe K lines (EW greater than a few hundred eV), as indicators of Compton-thick
AGN \citep{mat96}.  For NGC 612, MRK 417, and ESO 506-G027, all of these factors are met.  However,
even though ESO 362-G018 has a strong Fe K line and a flat spectrum, the fitted column density is
only $6.3\times10^{22}$\,cm$^{-2}$.  The spectrum of NGC 6860 is even odder, with both power
law components (see Table~\ref{tbl-8}) having a flat slope with very low hydrogen column densities (n$_H < 10^{22}$\,cm$^{-2}$) and no strong Fe K line.  

Though high column densities are possible indicators of Compton-thick sources, it is also possible
for the X-ray spectrum of a Compton-thick source to have a lower measured column density.  Such a model,
as was proposed for the Seyfert 1 source MRK 231 by \citet{mah00}, could be applicable
for ESO 362-G018 and NGC 6860.  In this model, the central power law source is blocked by Compton-thick material.  The resulting reflection component is then scattered and absorbed elsewhere, outside the Compton-thick region.  As a result, the measured X-ray column density is from this second absorbing region.  Therefore, we include ESO 362-G018 and NGC 6860 in our list of Compton-thick candidates, despite their low column densities. 

For our Compton-thick candidates, the sources with flat power law indices (ESO 362-G018, NGC 6860, 
NGC 612, MRK 417, and ESO 506-G027), we simultaneously fit the 8-channel BAT spectrum along with the XMM-Newton spectra, allowing a constant factor to vary for the BAT data, as we did for the MOS spectra.  Increasing the energy range to 195\,keV allows for better constraints on
the power law component at high energies.  Also, since a Compton-thick source spectrum should
be a heavily reflected spectrum, it is extremely important to have higher energy data to determine
the cutoff energy of the power law.  This is evident considering that the reflection spectrum ({\tt pexrav}
in XSPEC \citep{mag95}) depends upon the cutoff energy of a power law, in addition to iron abundance,
reflection factor, and geometry of the system.  When we use this model, we fixed the iron abundance to
the solar value and the inclination angle of the system to the default ($60 \degr$).

We note that in using the BAT spectra we are
assuming that the individual 14 -- 195\,keV spectra do not vary over the period of 22 months used to create
the BAT spectrum.  Future AGN observations with Suzaku, which can obtain simultaneous spectra from 0.3--200\,keV, will allow us to test the accuracy of this assumption.  To this end, we have obtained and are processing the Suzaku spectrum for one of our Compton-thick candidates, MRK 417.  A paper is in preparation.

\subsubsection*{i. Low column density Compton-thick candidates}
\paragraph{ESO 362-G018} 
The X-ray spectrum of ESO 362-G018 was not well-fit by either the partial covering or double power law model.  
While the double power law model provided the best fit, with a reduced $\chi^2$ value of 1.35,
this is not satisfactory.  In the residuals from the fit,  at least two emission line features were present.
Adding gaussians for these lines, the fit improved by $\Delta\chi^2 = 72$.  The energies of these lines (0.56\,keV and 0.90\,keV),
which have fluxes on the order of the Fe K line flux, correspond to helium-like oxygen and possibly helium-like neon lines.  

The flat power law and strong Fe K line suggest a reflection spectrum.  Simultaneously fitting the \xmm spectra
with the BAT spectrum, we replaced the heavily absorbed power law model with a reflection model
({\tt pexrav}).  This model is an acceptable fit to the data with a reduced $\chi^2$ of 1.14.  However,
though the reflected power law component is more typical of AGN ($\Gamma = 1.99^{+0.15}_{-0.27}$), the column
density is extremely low for a Compton-thick source (n$_H = 6.2^{+9.0}_{-2.6} \times 10^{21}$\,cm$^{-2}$).  As mentioned earlier, the observed low column density could be the result of scattering of the reflection spectrum through a second absorbing region of lower column density.  The details of this fit are listed in Table~\ref{tbl-10} along with the other Compton-thick candidates.     

From the HST observation of this source, \citet{deo06} describe the image as showing dusty 
lanes that are interspersed with star-forming regions.  This complex environment could partially
cover some of the X-ray emission as well as contribute line-emission from young stars.  
Similarly, ESO 362-G018 could be a very Compton-thick source whose flat, reflection-dominated
spectrum is scattered and viewed through an absorber \citep{mah00}.
From
the $\approx 10$\,ks \xmm exposure, it is clear that the spectrum is complex.  For a more
accurate description of the source spectrum, a longer observation with higher signal-to-noise
is necessary.

\paragraph{NGC 6860}
From a literature search, we found that NGC 6860 is well studied
in the optical and IR.  \citet{ben06} find optical emission line diagnostics of this source indicative of an intermediate state between an AGN and starburst galaxy.  Indeed, they state that the AGN dominates only in the
inner 10$\arcsec$ in the IR/optical.  While it is unclear whether this is also true in the X-ray band, we necessarily extracted a spectrum from a much larger, 85$\arcsec$ region.  
The Optical Monitor pipeline processed images (U, UVW1, UVM2) also confirm the presence of star
formation, where the nucleus is seen surrounded by a ring of star forming regions.  From
the evidence of the optical and IR observations, it is likely that the spectral form of
NGC 6860 is composite, with both star burst/star formation and AGN contributions.

From the double power law model, we found that residuals to the fit indicated a soft excess.  Given
the optical and IR evidence of star formation, we added an {\tt apec} model to fit this excess.
The {\tt apec} model in XSPEC is a model for collisionally-ionized diffuse gas.  Since the
quality of the \xmm spectra is too low to distinguish between collisionally and photo-ionized gas,
we used the simpler, collisionally ionized model.  As input parameters, the {\tt apec} model requires a plasma temperature,
metal abundance, redshift, and normalization.  We fixed the redshift to the source's value and set
the abundance to the solar value.  Adding this model, with a best-fit plasma temperature of
kT$ = 0.14$\,keV and normalization of $2 \times 10^{-4}$, improved the double power law fit by 
$\Delta\chi^2 = 6$.  Therefore, this is not significant at the 90\% level.  One thing to note, however,
is that the addition of the {\tt apec} model causes $\Gamma_1$, the low absorption power law
component, to steepen from  0.47 to 3.60.  The higher absorption power law model components, column density
and power law index, do not change.

Adding a reflection model to this fit (double power law with an {\tt apec} model), improves the
fit by $\Delta\chi^2 = 10$.  With this model, we could not constrain the reflection factor
or folding energy.  We then added the BAT data to the \xmm spectra.  Still, the reflection factor
and folding energy were not constrained (they continued to increase to unphysical values).
We fixed these values to a folding energy of 100\,keV and complete reflection.  
Though this model does not make physical sense, since a strong Fe K line is expected in a
reflection dominated spectrum, it was clear from the fitting that a high reflection factor is preferred
statistically.  This reflection dominated model obtains
a good statistical fit to the data with $\Gamma_1 = 1.14$ and $\Gamma_2 = 2.31$ and $\chi^2/dof = 444.5/408$ (see Table~\ref{tbl-10}).
However, we stress that since there is no strong Fe K line and the column density is low, we do not believe that this model is a good physical
description of the data.

\begin{figure}
\epsscale{1.0}\plotone{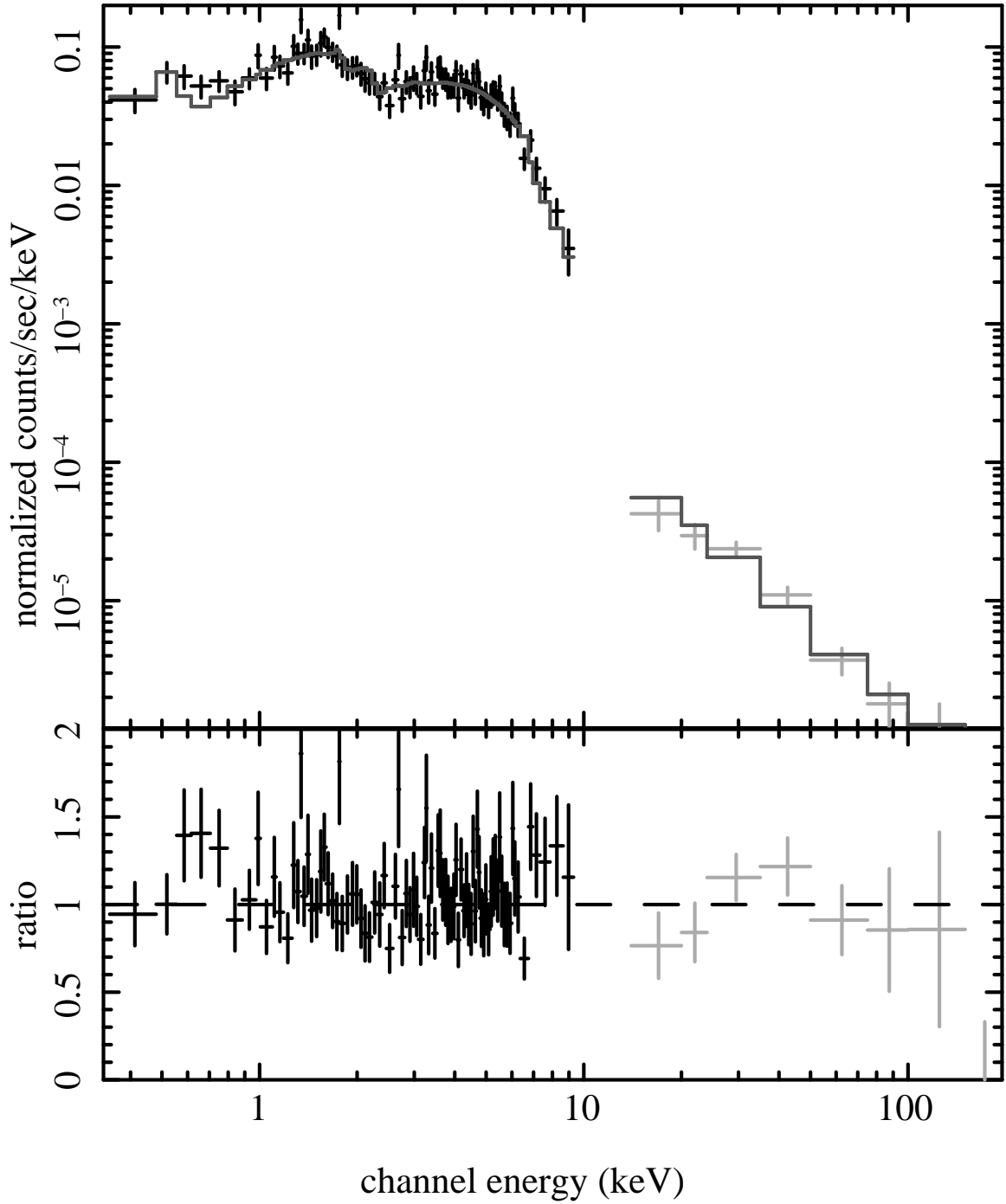}
\caption{\xmm MOS1 (no PN data was available for this source) and SWIFT BAT spectra of NGC 6860 fit with a reflection model.  The
spectrum of this source is very complex and could be adequately fit by a few different models (such
as a reflection model, ionized absorption in place of neutral absorption, and a double partial
covering model).  However, residuals in the model point to complexity that can not be explained
without higher signal-to-noise
observations.
\label{fig-6860}}
\end{figure}

For this source, we also fit the data with a double partial covering model (which could possibly be
justified in a clumpy, dusty environment) and a model where we replaced the neutral absorption
model, {\tt tbabs}, with an ionized absorber, {\tt absori}.  Both models fit the data with similar
$\chi^2$ values as the reflection model (reduced $\chi^2$ of 1.02 and 1.08, respectively).  Both
models also cause the fitted power law indices to steepen to values typical of AGN sources.
We conclude that the spectrum of this source is too complicated (see Figure~\ref{fig-6860}) to quantify with the data available.
A better signal-to-noise spectrum is required to understand this source's X-ray spectrum.

\subsubsection*{ii. High column density Compton-thick candidates}
As mentioned, we simultaneously fit the BAT spectra with the \xmm spectra for sources with
flat power law indices.  Our three additional Compton-thick candidate sources are NGC 612, MRK 417, and ESO 506-G027, all
with n$_H > 5 \times 10^{24}$\,cm$^{-2}$.
For each of these sources, we replaced the heavily absorbed power law component in the
double power law model (Table~\ref{tbl-8}) with the reflection model ({\tt pexrav}).  We record
the absorbed column density, power law index for the reflection component, cutoff energy (which was not
constrained for ESO 506-G027), normalization factor for the BAT spectrum, and goodness of fit in Table~\ref{tbl-10}.

\begin{figure}
\epsscale{1.0}\plottwo{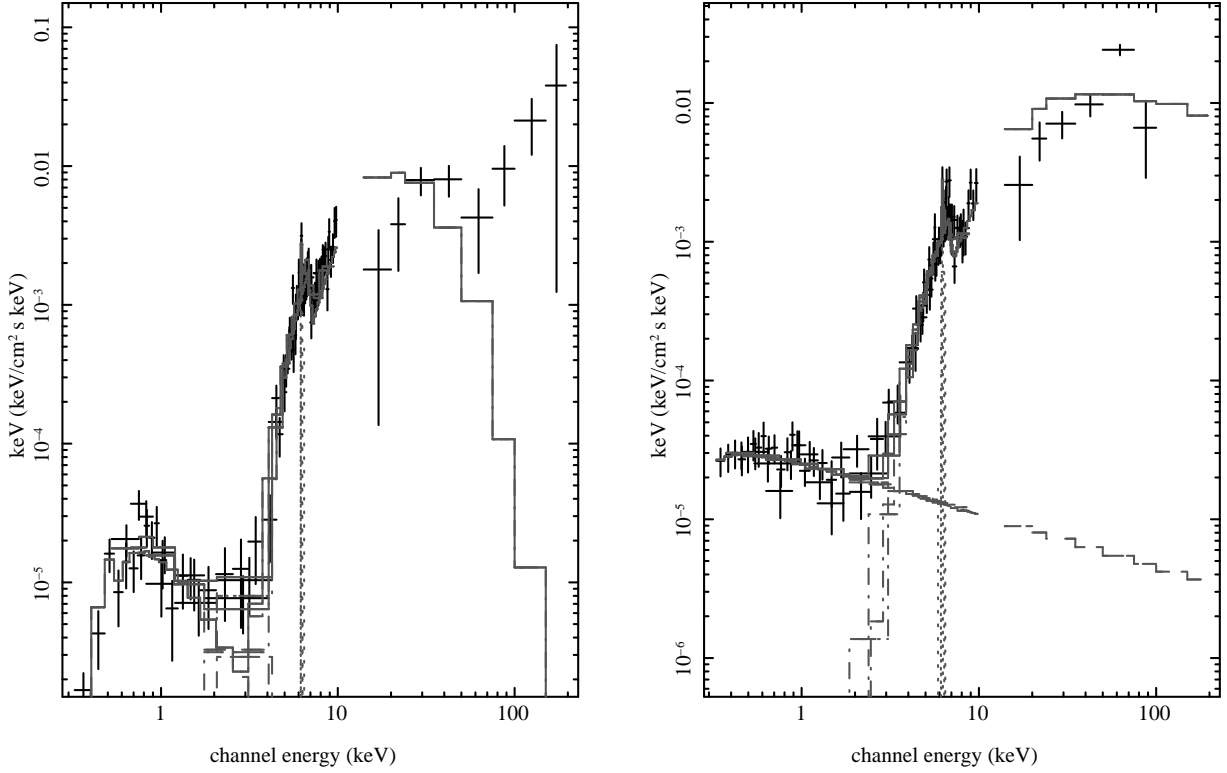}{figure4b.ps}
\caption{\xmm and BAT spectra of the Compton-thick candidate sources NGC 612 (left) and
MRK 417 (right).  The model used is {\tt tbabs}*({\tt tbabs}*{\tt pow} + {\tt tbabs}*({\tt pexrav} + {\tt zgauss}))*{\tt const}.  The unfolded spectrum is plotted (E$^2$\,f(E) vs. E, where f(E) is the model).  The fits to
the sources are described in the text.  These fits were obtained with the constant factor set to 1.0 (normalized to the PN spectrum).  The BAT spectra show some curvature and are not well fit by this model.  This was not true of the remaining 3 Compton-thick candidates.
\label{fig-compton}}
\end{figure}

Allowing the BAT flux normalization to vary by a constant multiplicative factor, we found that the
 values of the factors for NGC 612 and MRK 417 were very low ($<< 0.50$).  Examining the
BAT spectra, there is clear curvature in the BAT energy spectrum of these two sources, which is not well fit by the {\tt pexrav} model (see Figure~\ref{fig-compton}).  This curvature is not seen in the other three Compton-thick candidate spectra (see Figure~\ref{fig-6860} for the spectrum of NGC 6860).  Of particular note, the BAT spectrum
of NGC 612 appears flat (well-modeled by a power law index $<< 1.0$).  For MRK 417, fixing
the BAT multiplicative factor to 1.0 (the same as the PN spectrum) leads to a worse fit to the 
data with $\chi^2$/dof$ = 144.7/85$.  With this fit, the cutoff energy for the {\tt pexrav} model
becomes unconstrained while the column density and power law index increase (n$_H = 3.4^{+0.9}_{-0.8} \times 10^{23}$\,cm$^{-2}$ and $\Gamma = 1.85^{+0.12}_{-0.12}$).  The same effect happens with the spectrum
of NGC 612, where the best fit gives $\chi^2$/dof$ = 126.4/86$ with n$_H = 8.2^{+0.9}_{-2.6} \times 10^{23}$\,cm$^{-2}$ and $\Gamma = 1.09^{+0.34}_{-0.45}$.   It is possible that the curvature seen in the BAT
spectrum is a real feature of the spectrum
above 10\,keV.  However, simultaneous observations for the 2 -- 10\,keV and
15 -- 200\,keV bands are needed to determine whether the BAT spectra correctly represent the
very hard X-ray spectrum.  We have already obtained Suzaku spectra for MRK 417 and are in the process of analyzing the data, which will be presented in an upcoming paper. 

%


 \subsection{Variability}

 The main focus of our variability study is determining how the sources vary between the \xmm and XRT observations, on a timeframe of hours to months.  However, with 10\,ks \xmm observations, we also looked for shorter variability by examining the light curves of our sources.  To this end, we extracted light curves from the filtered PN (or MOS1 where there was no PN data available) event files with the FTOOL XSELECT.  We extracted light curves from the same regions used to extract spectra, binned by 100\,s.
 We also extracted a background light curve from a region of the same size on the same chip as the source.  We excluded SWIFT J0911.2+4533 from our analysis due to the low average count rate in the MOS1 observation (the source is located in a gap in a chip for the PN observation), $< 0.1$ct\,s$^{-1}$, which is on the order of the count rate in the background spectrum.
 
 For the remaining 21 sources, following the analysis of \citet{nan97}, we computed the normalized excess variance and $\chi^2$ values, for the assumption that the flux was constant, to quantify variability.  We list these values as well as the average count rate in Table~\ref{tbl-13}.   
 Within our sample, 8 sources were flagged as possibly variable during the \xmm observation, with reduced $\chi^2 > 1.5$, corresponding to a probability of $< 1$\% of the count rates corresponding to constant count rates.  For each of these 8 sources, we examined both the source and background light curves.  We found that for 5 of the sources the source and background light curves showed identical variability.  For each of these sources, the ratio of average background count rates to average source count rates was relatively large, between 0.2 and 0.8\,ct\,s$^{-1}$.  Thus, the background rates were significant compared to the source rates.  The variability, also seen in the background light curve, was not intrinsic to the source for these sources.  This was not the case for MRK 352, ESO 548-G081, and UGC 6728.  The variability for these bright, Seyfert 1 sources is measured source variability.
 
\begin{figure}
\epsscale{1.1}\plotone{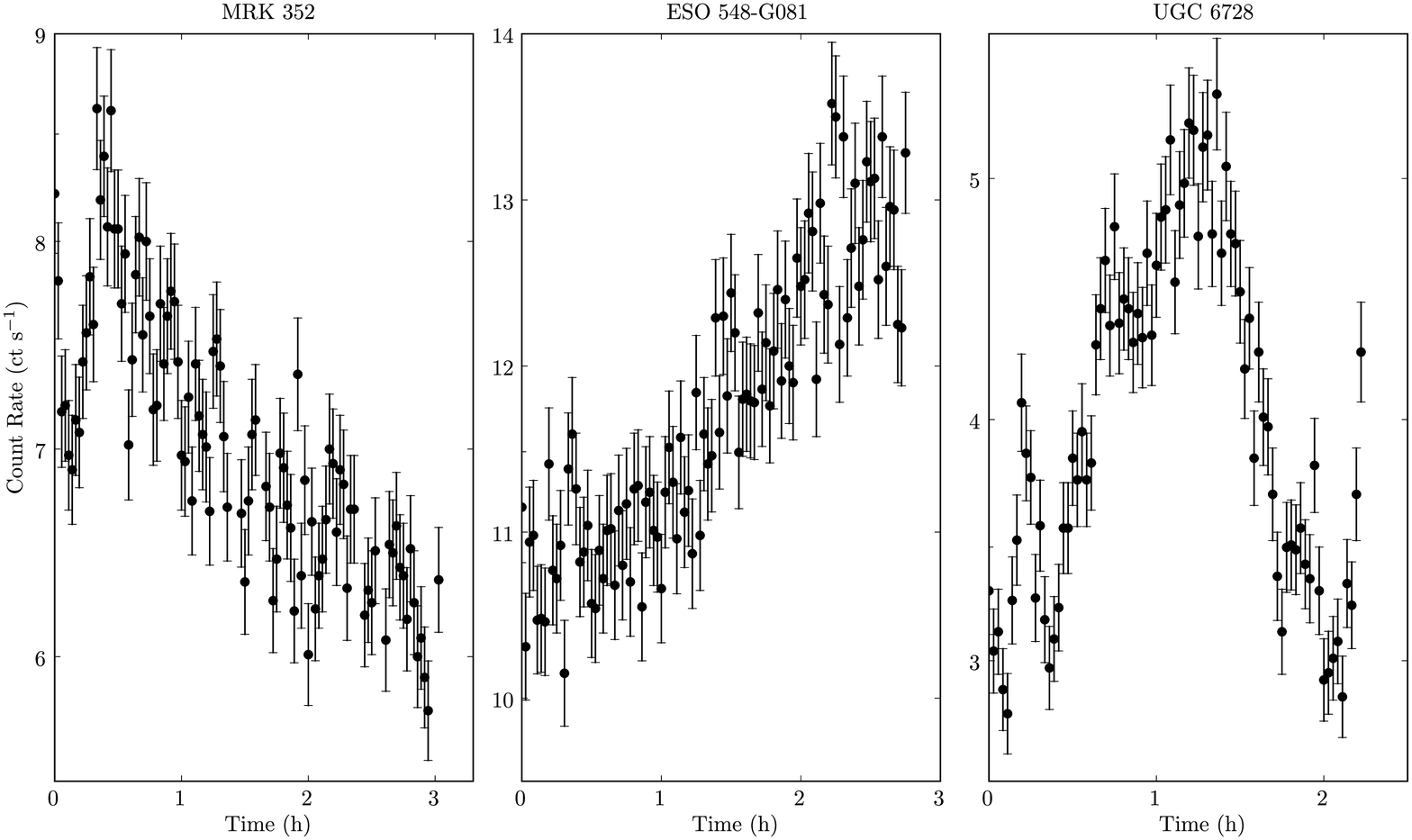}
\caption{Light curves binned by 100\,s for the \xmm PN observations of MRK 352, ESO 548-G081, and UGC 6728.  These two were the only sources to show significant variability above the background level($\chi^2/dof > 1.5$ compared to a constant flux model).  Each of these sources are among the brightest flux sources in our sample (though not the three brightest), all optical Seyfert 1 sources.
\label{fig-lc}}
\end{figure}

In Figure~\ref{fig-lc}, we include the light curves for each of the 3 sources with variability during the \xmm observation.  From the light curves, we estimated an average change in count rate/time, or $\Delta$ R/$\Delta$T, where $\Delta$R$ = $R$_{max} - $R$_{min}$ and $\Delta$T is the corresponding change in time.  These values are 2.5\,ct\,s$^{-1}$/2.4\,hr (MRK 352), 2.8\,ct\,s$^{-1}$/2.7\,hr (ESO 548-G081), and 2.4\,ct\,s$^{-1}$/0.8\,hr (UGC 6728).  For both MRK 352 and ESO 548-G081, these rates are minimums since the light curves are decreasing/increasing monotonically.  UGC 6728, however, shows a definite maximum and is thus the most rapidly variable source, with count rate changing appreciably over less than an hour.

\begin{figure}
\epsscale{1.8}\plottwo{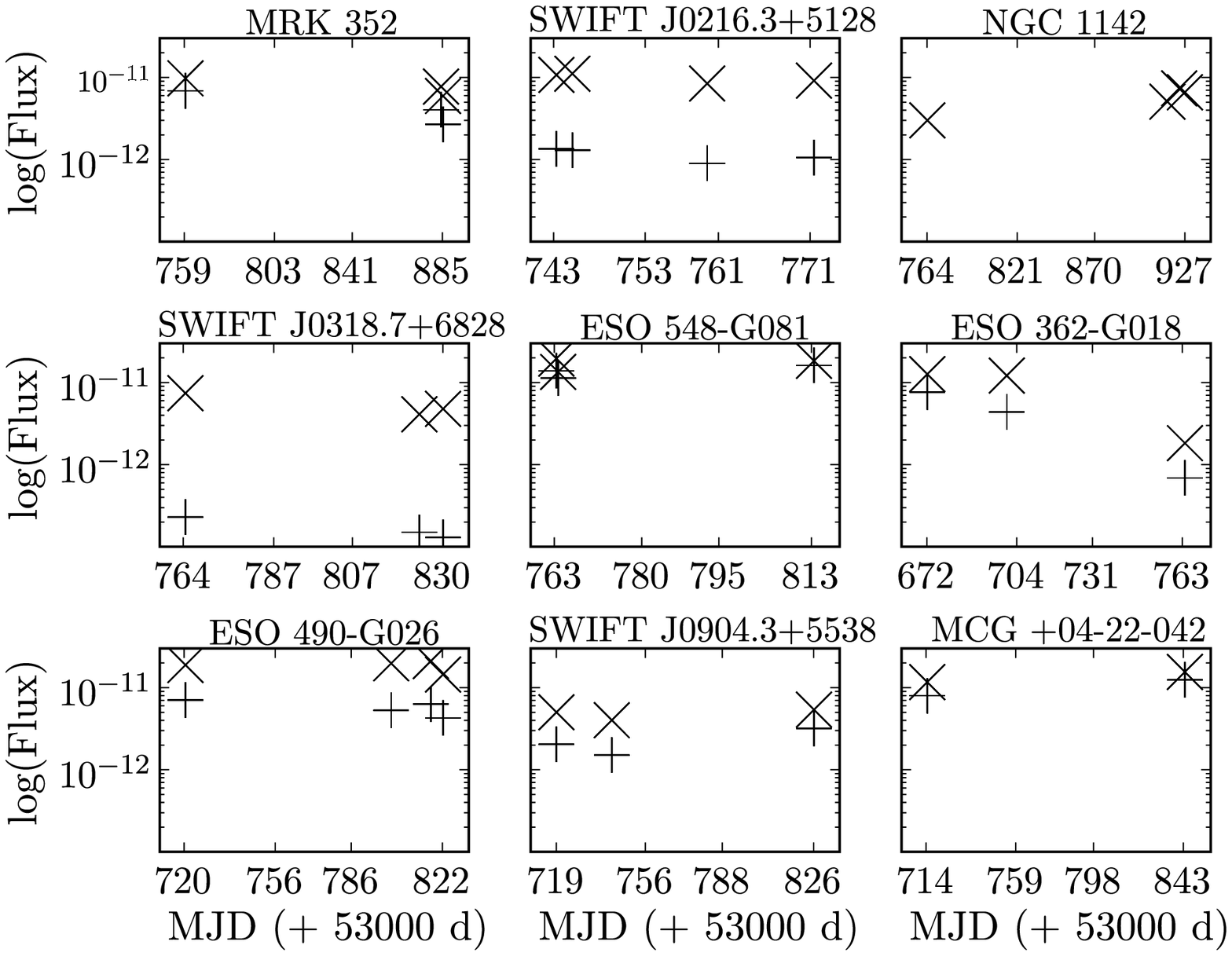}{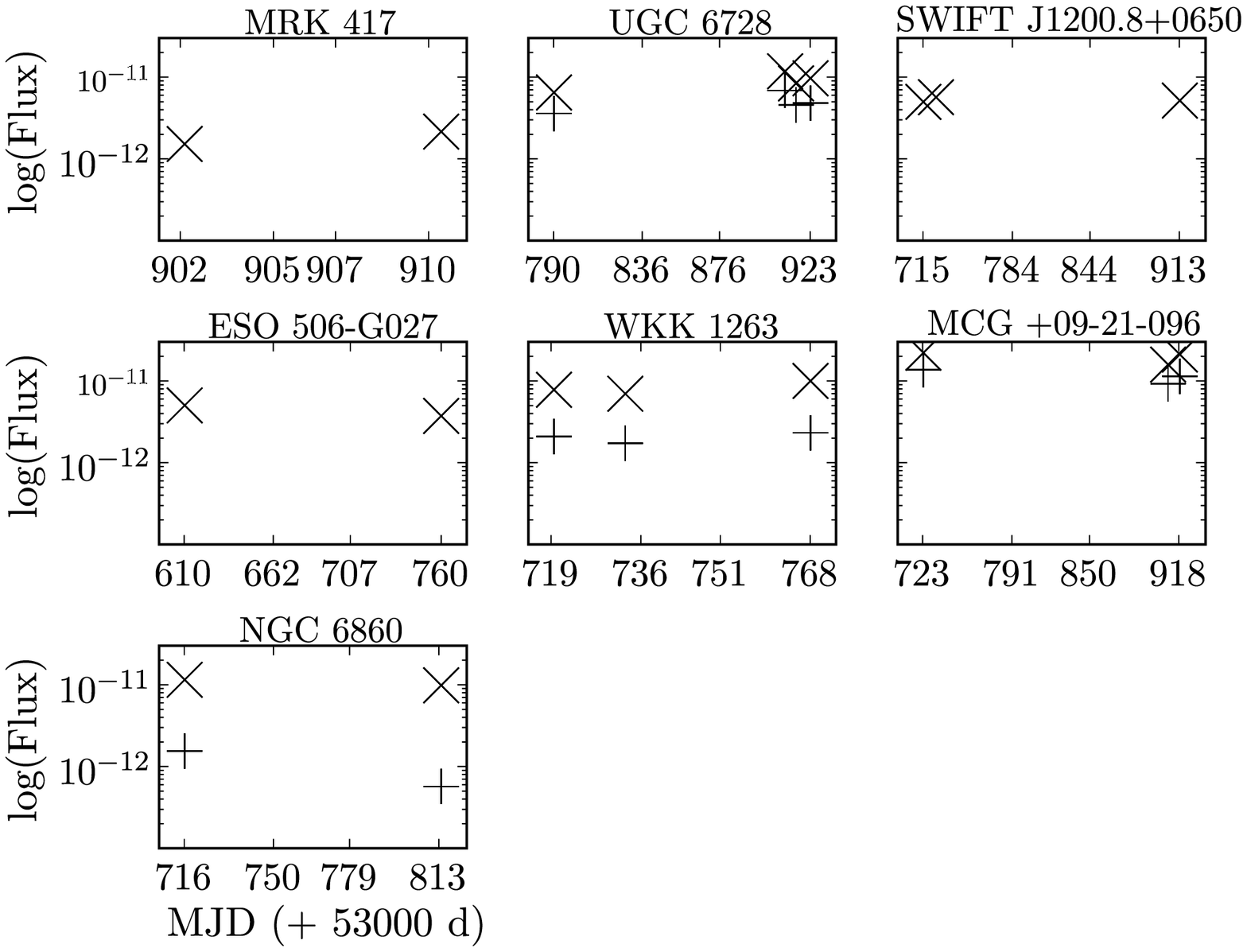}
\caption{Plots of the flux variation in the soft 0.5 -- 2\,keV (+) and hard 2 -- 10\,keV (x) bands for
sources with \xmm and XRT observations ($> 100$\,counts).  These flux values were obtained
from simple absorbed power law fits (see Tables~\ref{tbl-3} and~\ref{tbl-4}).  For the sources with
few/no counts in the soft band, the soft flux was unmeasurable.
\label{fig-var1}}
\end{figure}

In order to compare variability between observations, on time scales of days to months, we compared the
 XRT and \xmm spectral
fits listed in Tables~\ref{tbl-3} and~\ref{tbl-4}.   In Figure~\ref{fig-var1}, we plotted the hard, 2 -- 10\,keV, (x) and soft, 0.5 -- 2\,keV, (+) 
flux for multiple observations of our sources.  We made the initial assumption that any intrinsic
differences in flux between the instruments is less than 10\%.  From the figure, it is clear that
variations greater than this level occurred for all of the 16 sources with $> 100$\,counts in the XRT spectra.  Of these, the most extreme changes are seen for ESO 362-G018,
where both the hard and soft flux drop by an order of magnitude between the last two observations.
However, without error bars on the flux and with a simplified model that is not satisfactory for all
the sources, particularly for the high column density/complex spectra, a simple comparison of the fluxes is only a starting point for our variability study.

\begin{figure}
\epsscale{1.5}\plottwo{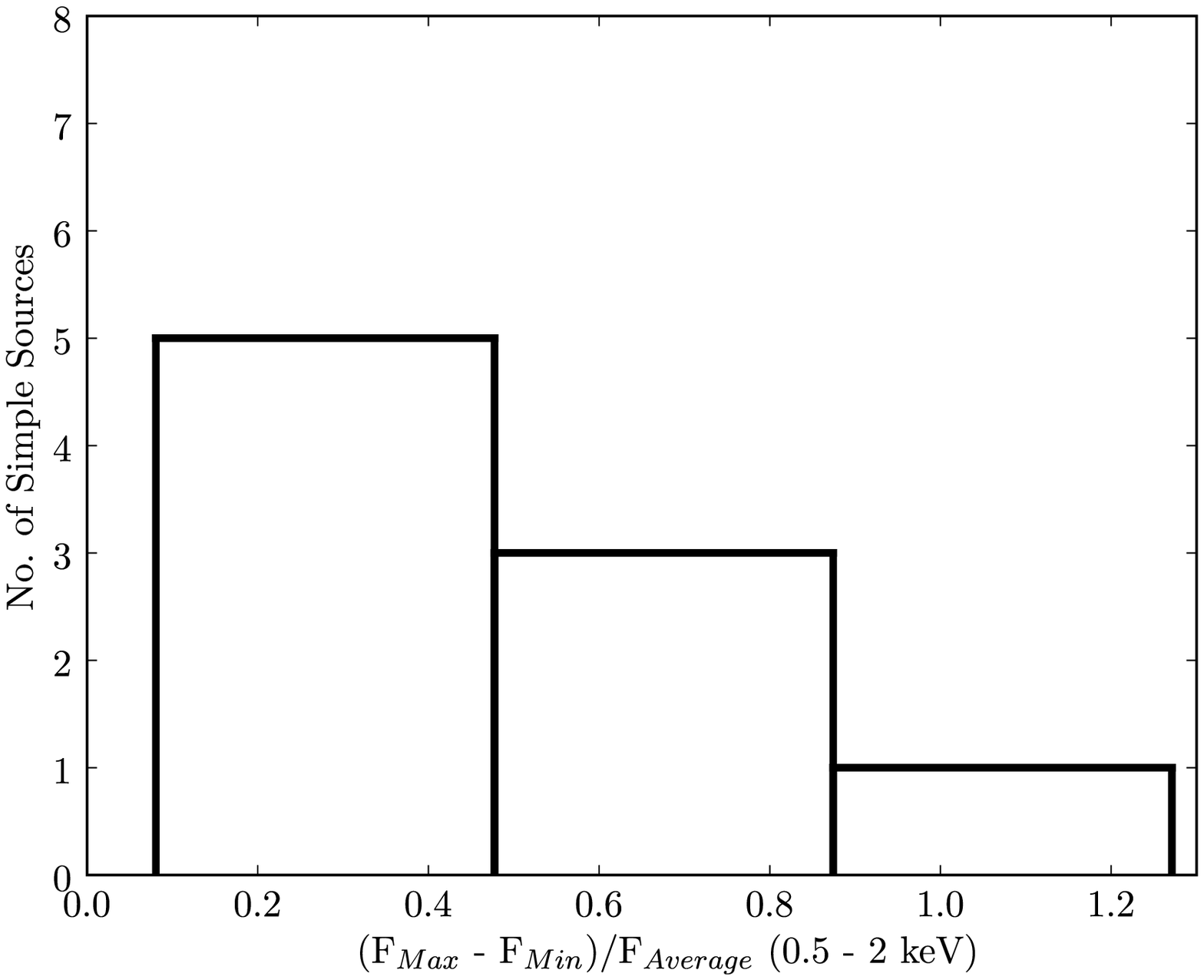}{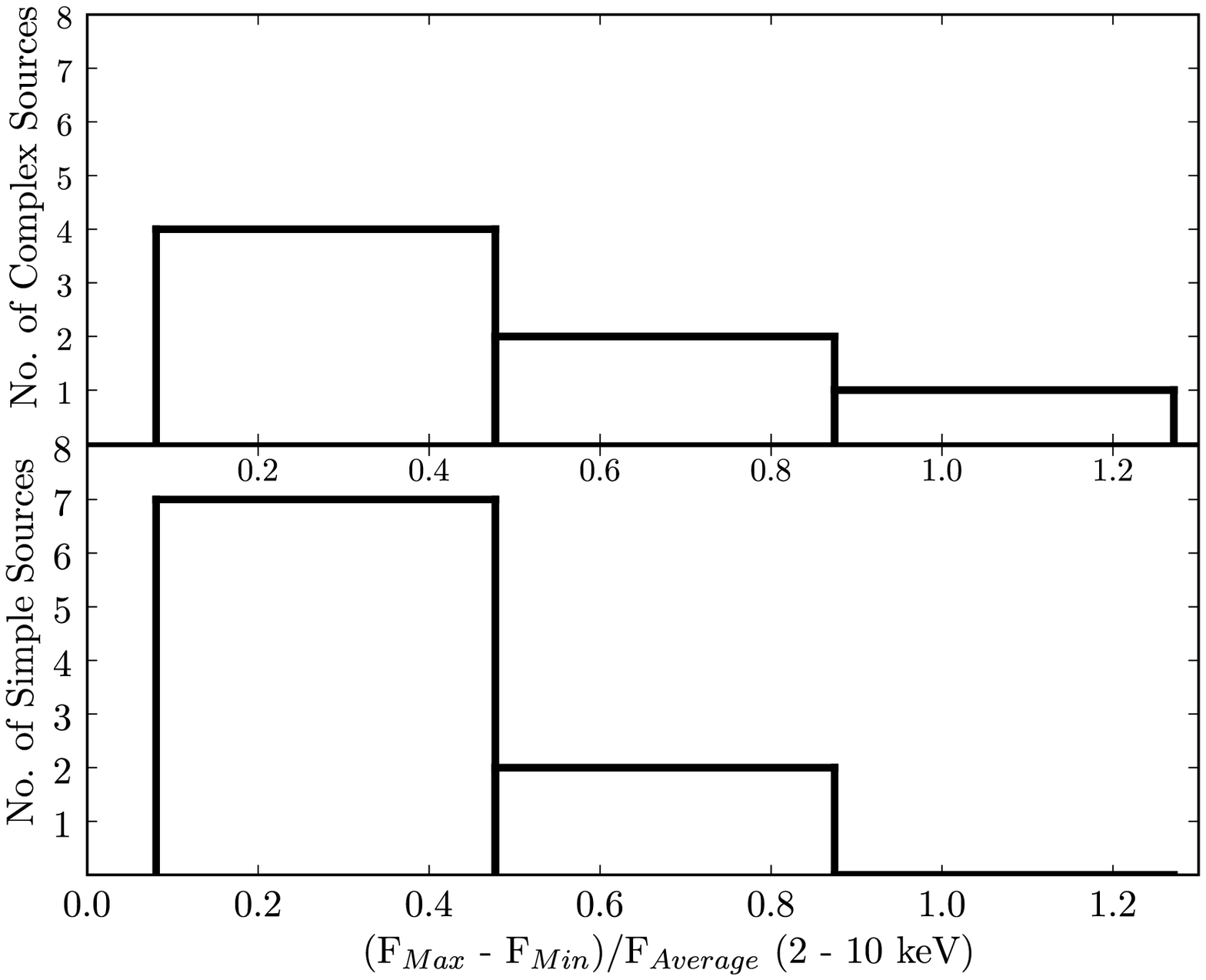}
\caption{Distribution of variability measurements, (F$_{max} -$F$_{min}$)/F$_{avg}$, for all of the sources listed
in Table~\ref{tbl-9}.  The first histogram shows the values for the soft (0.5 -- 2.0\,keV) band while
the second set of histograms show the distribution in the hard (2.0 -- 10.0\,keV) band.  For the soft band, we show the distribution only for the low column density/simple model sources.  The more absorbed/complex sources had fewer counts in the soft band, making the (F$_{max} -$F$_{min}$)/F$_{avg}$ unreliable for most of these sources.  Further, only half of the absorbed sources had XRT observations with $> 100$\,counts over all bands, so the sample is not complete even in the hard band.  For the simple model sources, we find that the variability, estimated from (F$_{max} -$F$_{min}$)/F$_{avg}$, is higher for many of the sources in the soft band than the hard band.  
\label{fig-fmax}}
\end{figure}

In addition to the flux, both the power law index and hydrogen column density introduce other sources of variation measured by the simple power law model.  Changes in these parameters also affect the measured
flux.  Thus,  we computed a statistic to quantify the flux variations between observations.
To this end, we determined the value
(F$_{max} - $F$_{min}$)/F$_{avg}$ and the corresponding $\Delta$t$_{max}$ ($|$t$_{max} -$t$_{min}|$ in days or the change in time for the greatest difference in observed flux between two observations)  for each source in both the hard and soft bands.  These values are listed in Table~\ref{tbl-9} and the distributions of the values are
plotted in Figure~\ref{fig-fmax}.  From the histograms, there is no measured difference between
the low column density sources (simple sources) and the high column density (complex sources).
The values of (F$_{max} - $F$_{min}$)/F$_{avg}$, however, are much smaller in the hard
band than the soft.  We note that the heavily absorbed sources, with much lower count rates in the
soft band, have much less accurate soft flux measurements as well as fewer sources with
observations $> 100$\,counts.  For the low column density sources, we find
an average  (F$_{max} - $F$_{min}$)/F$_{avg}$ value of  0.52 in the soft band and 0.37 in the hard
band.  

Based on the (F$_{max} - $F$_{min}$)/F$_{avg}$ values for our sources,
our results indicate that the AGN spectra vary more in the soft band than the hard band.  This claim was also made based on ASCA observations of Seyfert 1 sources by \citet{nan97}.  
In addition to this result, we find that the hard flux variability for the low absorption and more
complex sources is similar.  Unfortunately, due to the lower number of counts in the XRT observations,
we can only compare half of the complex sources to the complete sample of low column density
sources.  

\begin{figure}
\epsscale{1.5}\plottwo{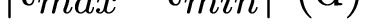}{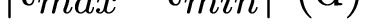}
\caption{Plots of the flux variation in the soft 0.5 -- 2\,keV and hard 2 -- 10\,keV bands, measured by  (F$_{max} -$F$_{min}$)/F$_{avg}$, versus number of days between observations for the maximum and minimum flux.  In this figure, sources with X-ray spectra best described by a simple model (triangles), complex model (circles), and the two complex sources with n$_H < 10^{23}$\,cm$^{-2}$ (squares) are plotted.  The (F$_{max} -$F$_{min}$)/F$_{avg}$ values are lower in the hard band.  In the soft band, three complex sources are not plotted due to uncertainty in measuring their soft fluxes.  The average value of $\Delta$t$_{max}$ or $|$t$_{max} -$t$_{min}|$ is about 100 days for both hard and soft flux.
\label{fig-deltat}}
\end{figure}

In Figure~\ref{fig-deltat}, we plot the variability measure (F$_{max} - $F$_{min}$)/F$_{avg}$ versus the change in time between the observations of F$_{max}$ and F$_{min}$, $\Delta$t$_{max}$.  From this plot, it is clear that, as already stated, the variability measurement (F$_{max} - $F$_{min}$)/F$_{avg}$ is smaller in the hard band than the soft band.
However, there is no significant difference in $\Delta$t$_{max}$ for the total sample, with average values of 
100.1 days for the soft band and 81 days for the hard band.

As a next step in our analysis,  we simultaneously fit the \xmm PN spectrum
with all of the corresponding XRT observations for the sources listed in Table~\ref{tbl-9}.  We began by fixing all XRT fit parameters to
the best-fit \xmm PN power law (+ gaussian where there is a strong line and blackbody where it was required) model.  We accounted for absorption using the {\tt tbabs} model
for low column sources and the {\tt pcfabs} model for the heavily absorbed/complex sources.  We then allowed the flux to vary between these observations by adding a {\tt const} model.
Where the addition of this model significantly changed $\chi^2$ ($\Delta\chi^2 > 10$) we flagged
the source as having a varying flux.  We then tested variability in column density and power law
photon index by allowing each of these parameters, along with their normalizations, to vary.
Again, we noted significant changes in $\chi^2$.

In order to measure the amount each model parameter varied between observations, we obtained 
error measurements for $n_H$, $\Gamma$, and the 0.3 -- 10\,keV flux.  We used the XSPEC
model {\tt pegpwrlw} in place of the {\tt pow} model.  The {\tt pegpwrlw} model is similar to the simple
power law model, however, the parameters $E_{min}$ and $E_{max}$ are used to indicate the 
energy range for the power law component.  The normalization is then the flux from the pegged power law
in units of $10^{-12}$\,erg\,s$^{-1}$\,cm$^{-2}$.  Since the normalization is a parameter in the model,
errors are easily computed for the flux with the XSPEC command {\tt err}.  For all of the sources,
we fixed $E_{min} = 0.3$\,keV and $E_{max} = 10.0$\,keV.  We indicate variability based on our
model fits in n$_H$, $\Gamma$, and flux (from the {\tt pegpwrlw} model) in Table~\ref{tbl-9}.  
Details on the model fitting for the individual variable sources are discussed in the appendix.   

\begin{figure}
\epsscale{1.0}\plottwo{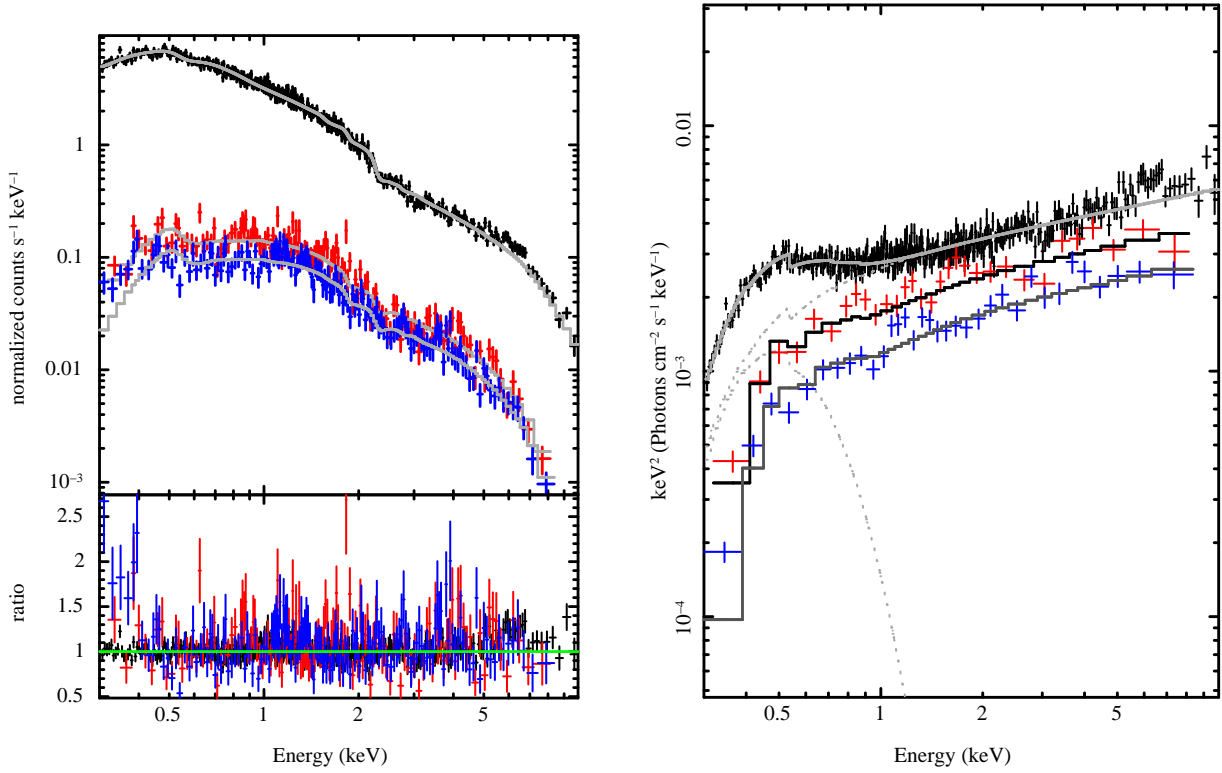}{figure9b.ps}
\caption{{\bf (left)} \xmm PN spectrum (black) with two XRT (red and blue) observations for MRK 352 fit with the model
{\tt tbabs}*{\tt tbabs}*({\tt pegpwrlw} + {\tt bbody}).  The best-fit model required
both the hydrogen column density and flux to vary between the three spectra.
{\bf (right)} The unfolded spectrum (E$^2$\,f(E) vs. E, where f(E) is the model) is plotted for the same source, MRK 352.
\label{fig-vary}}
\end{figure}

For the sources with low column densities, n$_H < 10^{23}$\,cm$^{-2}$, and simple
spectral shapes (Tables~\ref{tbl-5} and~\ref{tbl-6}),
all had at least one XRT observation to compare with the \xmm spectra.  All of these sources
showed some form of variability.  As an example, Figure~\ref{fig-vary} shows the XRT and PN 
normalized observed and unfolded spectra with the best-fit model for MRK 352.  This source showed
more variability than any other low column density source.
For MRK 352, the \xmm spectrum, taken five months prior to the XRT observations, shows no absorption and is nearly double
the flux level of the second XRT observation.  In the XRT observations, which are taken only
a day apart,  the flux changes by 40\%.  The column densities also change between these two observations,
by approximately 30\%, where the XRT columns are an order of magnitude higher than the
\xmm observation's measured column density.
The observations for the low column sources indicate variability in: flux for all of these sources (8), column density 
for half, and power law index for 3 sources.
  
For the sources with hydrogen column densities higher than $10^{23}$\,cm$^{-2}$ (Table~\ref{tbl-7}), five of the AGN had XRT observations, all with less than 100\,counts, while NGC 4992 had no XRT
observations.  Of the remaining six sources, MRK 417, SWIFT J1200.8+0650, and ESO 506-G027
did not vary, in that, allowing n$_H$, the power law, and flux to vary yielded $\Delta\chi^2 < 10$.  Using
the {\tt pegpwrlw} model, the errors on the flux for each observation (\xmm and XRT) were within
range of the other observations.  For example, SWIFT J1200.8+0650 showed an unabsorbed flux from the power law component ranging from: 1.01 -- 1.25 $\times 10^{-11}$\,erg\,s$^{-1}$\,cm$^{-2}$ (XMM), 1.01 -- 1.29 $\times 10^{-11}$\,erg\,s$^{-1}$\,cm$^{-2}$ (XRT-1), and 1.04 -- 1.50 $\times 10^{-11}$\,erg\,s$^{-1}$\,cm$^{-2}$ (XRT-2) with $\chi^2$/dof$ = 235.7/241$ for the {\tt tbabs}*{\tt pcfabs}*({\tt pegpwrlw}) model.     

The sources NGC 1142, SWIFT J0318.7+6828, ESO 362-G018, and NGC 6860 did show significant variability
between the \xmm and XRT observations (as detailed in the appendix).  To summarize the variability,
3 showed no variability, 4 had variable fluxes, 2 had varying column densities, and 3 showed
varying spectral indices.  Given the complex shape of the spectra of ESO 362-G018 and NGC 6860,
we are uncertain of how to interpret the variability.  We simply noted the sources as varying under all
of our criteria, but again note their complexity.  

ESO 362-G018 showed the most variability of the objects in our study.  We conclude this section
with a discussion of this source's spectrum.  In Figure~\ref{fig-ufspec}a, we plotted the observed 
spectra of this source.  The shape of the spectra varied considerably below 2\,keV for all observations.
In the hard band, the XRT observations show no evidence of the Fe K line which is so prominent
in the PN spectrum.  In Figure~\ref{fig-ufspec}b, we plot the unfolded spectra.  Here
the y-axis corresponds to E$^2$\,f(E).  In this plot, we can see that the XRT spectra (taken about
2 months before the \xmm observation) are much brighter.  If the Fe K line remained at the same 
flux level, it would be completely dominated by the power law component.  This is one possible explanation for the disappearance of the Fe K line.

To summarize our findings, few sources (3/21) varied appreciably on the $\approx 3$\,hr time scale of the \xmm observations.  Those sources that did vary were bright, X-ray sources with spectra well-fit by simple power law models.
From a comparison of the XRT and \xmm spectra, taken a day through months apart, it is clear
that most of the sources vary on longer time scales.  In the extreme case of ESO 362-G018, the source varies drastically in flux,
column density, and overall shape in two months time.  All of the low column density sources varied
in flux, while half showed evidence for varying column densities.  Unfortunately, given the lower
count rates in comparable exposure times, we have less data on the high column density sources.
From a comparison of the  (F$_{max} - $F$_{min}$)/F$_{avg}$ distributions, they appear to vary
similarly to low column sources.  However, higher quality data is necessary to draw firm conclusions.

\section{Discussion}
In this study, we examined the X-ray properties of a sub-sample of BAT detected AGN from
the 9-month BAT catalog.  These sources, selected based on their 14-195\,keV flux, probable
optical identifications with the Digital Sky Survey and 2MASS, and their lack of an archival X-ray spectrum, are probably representative of the whole 9-month sample.  Having a bright
DSS or 2MASS counterpart does not impart a significant selection effect, considering that all but 20 of the BAT AGN 9-month sample fit this criteria.  The 20 sources without a bright optical/IR counterpart are mostly blazars, of which there are only 15 in the 9-month catalog.  Since blazars are less than 10\% of the sample, we decided to focus on the majority, non-blazar sources.  For the BAT AGN sample, \citet{tue07} list 2MASS K$_s$ magnitudes for the sources with available values from NED.  The average value for the BAT sample is 11.47 with a variance of 4.04.  Our sub-sample is slightly dimmer, with an average magnitude of 12.10 and a variance of 1.80, but well within the distribution of BAT AGN magnitudes.

The column density distribution of our sub-sample, shown in Figure~\ref{fig-bat}, is representative of the larger sample.  In the
plot of hydrogen column density versus 14-195\,keV flux (values from \citet{tue07}), our
22 {\it XMM-Newton} follow-up sources span the range of hydrogen column densities.  
Roughly half of the sources have low column densities (n$_H < 10^{23}$\,cm$^{-2}$), while
half are more heavily absorbed.  We find the same ratio of absorbed to non-absorbed sources
in our \xmm follow-up sample.  While the low absorption sources in our sample span the lower range
of hard X-ray fluxes (from BAT), this is expected, since the sources were not previously
studied in the X-ray regime.  For the absorbed sources, however, we find that our sample spans
the full range of BAT X-ray fluxes.

Given that our sample of sources is a representative sample of the 9-month BAT catalog, it
is worthwhile to discuss the general properties of our sources.  To begin, the optical
host galaxy classifications of our sources are listed in Table~\ref{tbl-1}.  Examining
the host galaxy classifications, 17/22 of the hosts are classified as spirals or peculiar spirals.  
This result
is interesting, considering that \citet{gro05} found the hosts of X-ray selected, z $\approx 0.4-1.3$, Chandra Deep Field sources to be dominated by ellipticals.  If the BAT AGN hosts are predominately spirals, as our sample suggests, this could imply an evolutionary effect in AGN host galaxies between the z$ \approx 0.03$ and 
z $\approx 0.4-1.3$ universe.  

From our detailed X-ray spectral fits, we found that 9 of the 22 sources had column densities below
$10^{23}$\,cm$^{-2}$ and spectra well-fit by simple power law or power law with a soft excess
models.  Nearly half of these sources showed evidence of having a soft excess.  Optically,
all of these sources except for SWIFT J0216.3+5128 and WKK 1263 are Seyfert 1 -- 1.2 sources.
As discussed in the detailed spectral fitting section, SWIFT J0216.3+5128 is most likely a
blazar.  Though, WKK 1263 is identified as a Seyfert 2 in NED, no optical spectrum is available in the
literature to confirm this.  It is possible that this source was simply misclassified.

The remaining
sources in our study had more complex X-ray spectra.  For most of these sources, half of the sample, an absorbed power law component model fit is unacceptable.  This was particularly true for ESO 362-G018 and NGC 6860.  The column densities and power law indices computed from the simple model, for 
these sources, are drastically different than values from more complex models.  Both of these sources, optically Sy 1.5, had optical images indicating dust clouds
interspersed with star formation.  Likely, the complex environment contributed to the complexity
seen in the X-ray observation.  However, without higher signal-to-noise observations we were
unable to resolve the complex spectral components.  These results illustrate the danger of using
low quality data/simple models to determine the properties of complex sources.  

The remaining 11 sources, half of the sample, had column densities 
clearly above $10^{23}$\,cm$^{-2}$.  We 
classified the observed spectra of these sources as having a ``double power law'' shape, similar to \citet{tur97}.
Optically, these sources are Seyfert 2s.  The exceptions are NGC 612, a weak-lined, giant radio galaxy, and NGC 4992, which
have ``galaxy'' spectra or optical spectra showing no emission lines indicative of AGN emission.  NGC 612 was specifically classified as a ``non-LINER'', e.g. having non-AGN line ratios, by \citet{lew03}, while NGC 4992 is an INTEGRAL source whose optical spectrum led \citet{mas06} to classify the source as an X-ray bright, optically normal galaxy.
Considering the X-ray column densities for these
sources, the optical AGN emission could be hidden or obscured
by the high column of gas, n$_H > 5 \times 10^{23}$\,cm$^{-2}$, in the line of sight.

For all of the complex X-ray spectra sources, 
we fit the spectra with (1) a partial covering absorption model and (2) a
double power law model, where each power law component had a separate absorption model.
Most of the sources showed no significant difference in $\chi^2$ between these two models.
This makes it impossible to determine whether the soft flux is the result of scattering of the
AGN light, partial covering of the AGN light, or other X-ray sources (such as X-ray binaries
or diffuse galaxy emission) contaminating the AGN spectrum.  Particularly for sources, such as NGC 612, with low 0.5 -- 2\,keV fluxes, e.g., F$_{0.5 - 2 keV} = 2 \times 10^{-14}$\,erg\,s$^{-1}$\,cm$^{-2}$ corresponding to a luminosity, L$_{0.5 - 2 keV} \approx 4 \times 10^{36}$\,erg\,s$^{-1}$, within the observed range of Galactic X-ray binaries, since the X-ray luminosity function extends to $\approx 3 \times 10^{38}$\,erg\,s$^{-1}$ \citep{gri02}.  

For sources with flat spectra (low photon index, $\Gamma$),  we 
fit the spectra with a Compton thick model (a reflection dominated model, {\tt pexrav}).  We added the BAT
spectra to the \xmm data in order to extend the energy range to 200\,keV.  In addition to providing
an adequate fit to the data, replacing the heavily absorbed power law component with a reflection
model resulted in higher spectral slopes more consistent with average AGN photon indices (where the BAT spectra were not curved, see Table~\ref{tbl-10}).  Based on the detailed model fits,
NGC 612, ESO 362-G018, MRK 417, ESO 506-G027, and NGC 6860, are classified as Compton-thick candidates.  Two of these sources, ESO 362-G018 and NGC 6860, have measured column
densities n$_H << 10^{23}$\,cm$^{-2}$.  While a column density this low is not expected from
reflection in a Compton-thick region, an alternate model where the reflection component is scattered and then absorbed outside of the Compton-thick region, such as employed for MRK 231 \citep{mah00}, could explain the spectra.

In addition to the Compton-thick candidates, four of the complex spectra sources had high partial 
covering fractions ($> 0.99$) with the partial covering model and very low ratios of the unabsorbed
power law to the absorbed power law component (N$_{\Gamma_1}$/N$_{\Gamma_2} < 0.02$) with the double power law model.  Thus, emission from the sources SWIFT J0641.3+3257, SWIFT J0911.2+4533, SWIFT J1200.8+0650, and NGC 4992, was extremely low in the soft band (0.5 -- 2\,keV) compared to the hard band (2 -- 10\,keV).  These sources are consistent with the new class proposed by \citet{ued07} of hidden or buried AGN.  \citet{ued07} predict that these sources should have lower [\ion{O}{3}] luminosities than typical Seyfert 2 sources.  Archival optical spectra of NGC 4992 show very weak [\ion{O}{3}], in fact, so much so that, as discussed above, the spectrum of this source appears as a typical galaxy.  In our optical study (in preparation), we will explore this issue further.  For now, it is important to note that more of these hidden sources exist.  If our sample, showing 4/22 hidden AGN, is representative of the larger BAT sample, we expect that about 1/5th of local AGN have these same properties, making them nearly undetectable in optical samples.

Having classified the sources into categories, we now describe the general properties
of our sample as a whole.  To begin, in Figure~\ref{fig-fluxratio} we plotted the column
densities versus two different flux ratios (similar to a diagnostic in \citet{mal07}).  The column densities we used are listed in Tables~\ref{tbl-5},~\ref{tbl-6}, and~\ref{tbl-8}, where we used the column density of the more heavily absorbed power law component for the complex spectra.  The flux ratios plotted are
the ratios of F$_{2 - 10\,keV}$/F$_{14 - 195\,keV}$ (medium/hard) and F$_{0.5 - 2\,keV}$/F$_{2 - 10\,keV}$ (soft/medium).  In
the plots, we represent the three classes of objects (simple power law or power law and blackbody
fit sources with low columns (blue), complex heavily absorbed spectra (black), and complex spectra
that require more complicated models (red)).  We find that the low absorption sources have
average values of medium/hard and soft/medium flux of 0.38 and 0.48, respectively.  There is little change between the medium/hard and soft/medium flux ($\approx 20$\%)
for the low absorption sources.   The heavily absorbed sources, however, have average
values of medium/hard and soft/medium flux of 0.08 and 0.02.  This is a 75\% change in the values.
Obviously, there is much less soft flux for the absorbed sources.  The complex sources with
poorly defined spectral models, have intermediate values of the medium/hard and soft/medium colors
of 0.17 and 0.11.

In terms of use as a diagnostic, we find that the plot of column density versus ratio of F$_{2 - 10\,keV}$/F$_{14 - 195\,keV}$ (medium/hard) is a good diagnostic of column density, for n$_H < 10^{24}$\,cm$^{-2}$.  Sources with similar column densities occupy areas close to the regions of constant power law index plotted (for $\Gamma = 1.5$ and 1.9).  This is not true for the plot of column density versus ratio of F$_{0.5 - 2\,keV}$/F$_{2 - 10\,keV}$ (soft/medium).  This appears to be a poor diagnostic, despite its wide spread use in deep X-ray surveys, with a large spread in the soft/medium color particularly seen in the sources with n$_H > 10^{23}$\,cm$^{-2}$.  These results are not surprising, since the 2--10\,keV flux is also affected by absorption.

From these color diagrams, we decided to construct a color-color diagram of the soft/medium
flux ratio versus the hard/medium flux ratio in attempts to construct a better diagnostic diagram for sources with too few counts to measure column density.  In Figure~\ref{fig-color}, we plot this diagram using the same
symbols as in the previous diagram to indicate low absorption (blue), complex (red), and
more heavily absorbed (black) sources.  In this figure, it is clear that the different types of sources
are clearly separated.  The low absorption sources occupy the left hand upper corner, where
the soft/medium and hard/medium colors are nearly equal.  The heavily absorbed sources are
closer to the right bottom corner, where the hard/medium flux ratio is much higher than the soft/medium
flux ratio.  Between these values, the complex sources as well as a source from each of the
other two categories, reside.  All of these sources have measured column densities from $10^{22} -
10^{23}$\,cm$^{-2}$, intermediary between the two classes.  This result is very nice in that it provides
a good diagnostic tool for observations with few counts, but requires data above 15\,keV.  

In addition to the flux and column density measurements, we have measured power law
indices, blackbody components (where present), and Fe K equivalent widths (with the physical width of the line fixed to 0.01\,keV at 6.4\,keV).  From our sample,
we found no correlation between the hard band (2 -- 10\,keV) luminosity and power law
indices.  An important point to note is that the measured power law index for the complex spectra depends very much on the 
model used.  Comparing the results of the partial covering model with the double power law model (Tables~\ref{tbl-7} and~\ref{tbl-8}),  the average power law index for the partial covering model is significantly
higher ($<\Gamma> = 1.74$ compared to $<\Gamma_2> = 1.36$) with smaller associated error bars.
For the remaining 9 sources, the sources with spectra modeled by absorbed simple power law or
power law and blackbody models, $<\Gamma> = 1.75$, similar to the results from the partial covering
model.  The values for the simple model/low absorption sources and those from the partial
covering model are consistent with average photon indices for AGN ($\approx 1.8$ from \citet{mus82}).

Soft excesses, modeled with a blackbody component, were statistically significant in half of the spectra modeled
by a simple power law model.  We find a significantly smaller fraction compared to the ROSAT sample of \citet{gal07}, who find soft excesses in all of their sources, and a significantly larger fraction compared to the Lockman Hole \xmm survey \citep{mat05}, where only 11\% of type 1 and 25\% of type 2 sources show a soft excess.  Due to the low number of counts for our heavily absorbed sources,
we can not quantify with certainty how many complex/heavily absorbed sources have this component, but at least one source (NGC 1142) has a statistically significant soft excess.  For the low absorption
sources, $<kT> = 0.08$\,keV, which is similar to but slightly lower than that seen for PG selected
QSOs ($<kT_{BB}> = 0.14 \pm 0.02$\,keV) \citep{por04, pic05}.  If the soft excess is the result of
a thermal process, the lower kT values in our sample could be related to the lower luminosities of
our sample, compared to the PG QSOs.  In fact, our average black body temperature is directly in the range of those found for type 1 AGN in the Lockman Hole sample, $<kT> = 0.09\pm0.01$\,keV \citep{mat05}.

The final spectral component measured for our entire sample is the Fe K equivalent width (EW) at 6.4\,keV.
In Figure~\ref{fig-iron}, we plot the Fe K EW versus the hard band (2 -- 10\,keV) luminosity.
We fit a line with the ordinary least squares bisector method to the upper limits of the EW measurements (see plot),
yielding a fit of $\log EW = (-0.697 \pm 0.144) \times \log L_{2 - 10 kev} + (32.045 \pm 6.164)$.
The significance of this fit, indicated by R$^2 = 0.22$, where R$^2$ is the coefficient of determination, is very low.  Thus, our results show no
indication of the X-ray Baldwin or IT effect \citep{iwa93}, an anti-correlation of Fe K EW and
hard band luminosity.

For the low absorption/simple model sources, we measured the significance of the \ion{O}{7} and \ion{O}{8} K edges to search for evidence of a warm absorber (see Table~\ref{tbl-11}).  All of the simple model sources, with the possible exception of WKK 1263, are classified optically as Seyfert 1s and thus can be directly compared with the \citet{rey97} sample.  In Figure~\ref{fig-rey}, we plot the values of optical depth for each of the edges versus L$_{2-10 keV}$ for our sources as well as the \citet{rey97} sources with luminosities in the same range.  As the figure shows, the optical depths we found for our sources are much lower than those from the Reynolds sample.  Only one source, ESO 490-G026, had a clear detection ($\Delta\chi^2 = 25$) with the optical depths of both edges having upper limits above 0.10.  Thus, where
half of the Reynolds sample and 13/18 of the \citet{geo98} sample showed evidence of a warm absorber in the line of sight, we find only 1/8 of our Seyfert 1 sources to show significant evidence of a warm absorber.  This result could be due to an incomplete or biased sample of Seyfert 1 sources, since these were among the low absorption sources with the lowest BAT flux in the 9-month sample.  Alternatively, our result could be representative of the entire BAT sample.  In this case, it is possible that the previous AGN samples showed more detections because they were from an optically selected/soft X-ray selected sample.  Thus, a possibility is that the emission that ionizes the gas, creates a region of warm, ionized gas, that also destroys dust.  An optical or softer X-ray survey could preferentially select these sources, missing more obscured sources.  Analysis of the remaining BAT sources will allow us to verify whether our result of few warm absorbers is consistent with  the properties of the entire 9-month catalog.

In addition to the spectral properties, we examined our sources for two types of variability: (1) during the \xmm observations and (2) variability in spectral form and brightness between the \xmm and XRT observations.  For the first type of variability, we created binned light curves for each object in our sample.  We found that 3/21 sources showed significant variability, with rates varying by 2.4 -- 2.8\,ct\,s$^{-1}$ over time scales of 0.8 -- 2.7\,hr, in the PN observations.  The sources that varied the most were all Seyfert 1 sources with low absorption and X-ray spectra well-fit by simple power law or power law and a blackbody models.  These sources were among the brightest in our sample.  While only three sources showed short term variability, during the \xmm observation which lasted $\approx 10$\,ks, nearly all of the sources (13/16) exhibited variability on longer time scales, of hours to months, from comparisons of the \xmm and XRT observations.  This result agrees with earlier studies which found AGN more variable on long time scales than short time scales \citep{bar86, nan97}. 

From our comparison of the \xmm and XRT spectra, we found 13/16 sources
had varying fluxes, 6/16 had varying column densities, and 6/16 sources had varying power law
indices.  Sources tended to vary more in the soft band than the hard band (see the Variability section).  Unfortunately, due to lower count rates, measuring variability for the heavily absorbed sources
was more uncertain, particularly in the soft band (0.5 -- 2\,keV).  In addition, we had less XRT observations with $> 100$\,counts for these sources.  Based on the result that our sources varied more in the soft than hard band, it is likely that if we had more counts in the soft band for the heavily absorbed sources, as well as more observations for comparison, our results would agree with other AGN variability studies which found $ > 90$\% of their sources to vary over the time scale of months to years (such as the AGN from the Lockman Hole \citep{mat07} and Chandra Deep Fields \citep{bau03, pao04} studies).

In Figure~\ref{fig-varavg}, we plot column density versus flux and photon index versus flux
for the sources that vary.  The plots show the observed parameter for each observation/average parameter for the 
source, where the observed values for each source are plotted with a different symbol.  In
the column density figure, there is clearly no correlation seen between the column density and
flux.  Similar results were found by \citet{ris02} for a sample of Seyfert 2 galaxies, where they
conclude that the variations in column density can not be caused by varying ionization states
but by a clumpy absorber.  In the plot of spectral indices, however, we do find a correlation
between the spectral index and the flux.  Therefore, we find that higher fluxes correspond
to higher spectral indices.  This result has been seen for individual sources \citep{mus93}.  Based on a variability study of the AGN sources in the Lockman Hole, \citet{mat07} find no correlation between spectral variability and flux variability.  Further, they find flux variability much more prevalent in their sample than spectral variability, finding spectral variability in only $14 \pm 8$\% of Seyfert 1s and $34 \pm 14$\% of Seyfert 2s.   However, \citet{mat07} note that the detection of spectral variability is related to the quality of the spectrum.  When they consider this factor, they predict a higher fraction of $\approx 40$\% to exhibit spectral variability.  We do not have a complete sample for Seyfert 2 sources, due to the low number of counts in the XRT observations for heavily absorbed sources.  However, we find 7/8 classified Seyfert 1 sources to exhibit a variation in either column density or power law index, much higher than the \citet{mat07} estimated value.  Additionally, as we stated earlier, there is a clear correlation between changing flux and power law index for individual sources.

For the Compton-thick sources, variability or a lack thereof, gives clues to the size and location of the Compton-thick gas.  For our heavily obscured Compton-thick candidates, only MRK 417 had enough counts in an XRT observation ($> 100$\,counts) to test for long term variability.  We found no statistically significant evidence of variability for this source between the \xmm and XRT observations, taken 6 months apart.  This lack of variability in two observations does not give us much information.  For the low absorption Compton-thick candidates, however, we find a great deal of variability between the \xmm and XRT observations.  In particular, for NGC 6860, the flux and spectral index are higher in the XRT observations, while the column density is lower.  Since the XRT observation was taken 4 months earlier, this puts a limit on the suggested change from Compton-thin to a reflection-dominated spectrum.  Similarly, significant changes are seen between the XRT and \xmm observations of ESO 362-G018.  In this source, the most significant change is the disappearance of the strong Fe K line seen in the \xmm observation.
A smaller time constraint is placed on this source, 2 months between the last XRT and the \xmm observation, for a change from a Compton-thin to a reflection-dominated spectrum.

Changes from Compton-thin to Compton-thick spectra have been noted before, particularly by \citet{mat03}.  They discuss two possible scenarios to explain the changes in spectra, a change in column density of the absorber \citep{ris02} and a ``switched-off'' source, that is, a state where the emission from the central source drastically decreases below our detection threshold.  For both ESO 362-G018 and NGC 6860, Seyfert 1.5 sources embedded in dusty host galaxies, a changing absorber is a more appealing explanation.
 

 \section{Summary}
From our analysis of the \xmm and XRT spectra of 22 BAT-selected AGN, the complexity of the spectra of a large fraction of
nearby ($<z> \approx 0.03$) AGN is clear.  Based on the range of X-ray column densities and
BAT (14 -- 195\,keV) fluxes \citep{tue07}, our sources are a representative sample of the 9-month BAT
catalog.  In analyzing their properties, we are presenting for the first time the global X-ray properties of an unbiased, local AGN sample.  

Within our sample, we find half of the sources
to have low absorption ($n_H < 10^{23}$\,cm$^{-2}$) and spectra well-described by simple
power law models.  Half of these sources statistically show evidence of a soft excess.  We tested
these sources for the presence of a warm absorber, finding only one statistically significant detection out of 8 low absorption sources.  This is at odds with the studies of \citet{rey97, geo98} who found half or more of their samples consistent with warm absorbers.  If our result of few warm absorbers is found in the entire Seyfert 1 BAT AGN sample, the detection of a large number of warm absorbers is likely a selection effect of optical/soft X-ray AGN samples.

The remaining
13 sources, which had too few soft counts to test for the presence of a warm absorber, have more complex spectra.  Within the class of complex sources, we find five
Compton-thick candidates (based on a flat spectrum above 2\,keV), two of these sources with spectra too complex to model successfully with the available signal-to-noise.  Additionally, four other sources are consistent with the hidden/buried AGN described in \citet{ued07}.  Since $\approx 1/5$ of our sample fits in this category, we agree with \citet{ued07} that these types of sources are a significant fraction of local AGN.  If these sources have weak [\ion{O}{3}] emission, as \citet{ued07} predicts, they would be easily missed in optical surveys and require very hard X-ray surveys, such as the BAT and Integral surveys, for detection.

On short time scales, during the $\approx 3$\,hr \xmm observations, we found that only 3/21 sources varied significantly, all of which were bright, low absorption X-ray sources.
Comparing the XRT and \xmm observations of 16/22 sources, which were separated by hours to months, we were able to compare the spectra for longer time scale variability.  Most of the sources varied in flux (13/16), such that our results agree with previous studies which found AGN to vary more on longer time scales than short time scales \citep{bar86, nan97}.  In terms of spectral variability, nearly half of the sources varied in both column density (6/16) and power law index (6/16).  We found no correlation between column density and flux between observations for the individual sources.  However, there was a strong correlation between power law index and 0.3 -- 10\,keV flux, where steeper slopes correspond to higher fluxes.  Contrary to the variability study by \citet{mat07} who estimate $\approx 40$\%
 of their sample to vary with respect to spectral shapes, we find 7/8 identified Seyfert 1s to vary in either column density or power law index.  We note, as \citet{mat07} point out, that the detection of this variability depends on the quality of the data.  Thus, similar comparisons with Seyfert 2s were not plausible since the data quality was much lower.

Optically, the Seyfert type of the sources match the X-ray column densities.  Thus, the Seyfert 1
sources have n$_H < 10^{22}$\,cm$^{-2}$ and the Seyfert 2 sources have n$_H > 10^{22}$\,cm$^{-2}$.
The two sources with no optical AGN signatures are heavily absorbed sources with
n$_H > 5 \times 10^{23}$\,cm$^{-2}$.  The host galaxies of our sample are mostly spirals,
contrasting with the results of \citet{gro05}, who find elliptical hosts dominating the $z \approx 0.4 - 1.3$
universe.

We are continuing to analyze the XRT spectra, as well as search the literature, for the X-ray
properties of the complete sample of 9-month BAT AGN.  With the X-ray properties of our complete sample, we will compare the BAT AGN properties with those of other AGN samples, such as the ASCA sample of Seyfert 1 and Seyfert 2 sources presented in \citet{tur97} and \citet{tur97b}. 
In addition, we are analyzing
the optical spectra of a sub-sample of the sources to determine the optical properties
of our sample.  With these data, we will present the optical and X-ray properties of a
local AGN sample.  

\acknowledgments
We thank Christopher Reynolds for useful discussions. \\
Also, we acknowledge the work that the Swift BAT team has done, which has made this work possible.
{\it Facilities:} \facility{Swift (BAT)}

\clearpage
\appendix
\section{Details on Variability of Individual Sources}
Below we describe the variability between observations for the AGN listed in Table~\ref{tbl-9}.
The low column density sources are those whose best-fit spectra were fit by a simple absorbed power law or a power law and blackbody model.  The high column density/complex sources are those
that we fit with the partial covering/double power law models.  In the following discussion, XMM is used
to denote the PN spectrum while XRT-1 denotes, for example, the first XRT observation for the
source, as listed in Table~\ref{tbl-2}.

\subsection{Low Column Density Sources}
\paragraph{MRK 352 --}
The PN and two XRT spectra for MRK 352 were not well fit until the flux was allowed to vary.  A
varying flux improved the fit by  $\Delta\chi^2 \approx 6500$.  The fit was then greater improved
by allowing the column density to change ($\Delta\chi^2 = 320$).  Changing the power law
index and normalizations improved the fit by $\Delta\chi^2 = 30$, however, the power law photon
indices were the same within the errorbars.  The best-fit {\tt tbabs}*{\tt tbabs}*({\tt pegpwrlw} + {\tt bbody})
model is shown in Figure~\ref{fig-vary} where $\chi^2/dof = 1289.01/1189$.  The black body and
power law components were those seen in Table~\ref{tbl-6}.  Hydrogen column density changes between
the observations as: 0.00 -- 0.02 $\times 10^{21}$\,cm$^{-2}$ (XMM),  0.83 -- 1.04 $\times 10^{21}$\,cm$^{-2}$ (XRT-1), 1.28 -- 1.51 $\times 10^{21}$\,cm$^{-2}$ (XRT-2).  The flux errors from
the pegged power law component were:  1.96 -- 2.00 $\times 10^{-11}$\,erg\,s$^{-1}$\,cm$^{-2}$ (XMM), 1.37 -- 1.47 $\times 10^{-11}$\,erg\,s$^{-1}$\,cm$^{-2}$ (XRT-1), and 0.98 -- 1.05 $\times 10^{-11}$\,erg\,s$^{-1}$\,cm$^{-2}$ (XRT-2).  Considering that the XRT observations were taken only a day apart,
it is clear that this source varies considerably.  Five months earlier, the XMM spectra show the
source nearly twice as bright with no absorption.

\paragraph{SWIFT J0216.3+5128 --}
All of the observations for SWIFT J0216.3+5128 took place within the span of a month.  While
no variations were seen between the column densities and power law indices ($\Delta\chi^2 < 3$),
the flux did vary.  The flux errors from
the pegged power law component were:  1.63 -- 1.75 $\times 10^{-11}$\,erg\,s$^{-1}$\,cm$^{-2}$ (XMM), 2.25 -- 2.52 $\times 10^{-11}$\,erg\,s$^{-1}$\,cm$^{-2}$ (XRT-1), 2.15 -- 2.43 $\times 10^{-11}$\,erg\,s$^{-1}$\,cm$^{-2}$ (XRT-2), and 1.72 -- 2.12 $\times 10^{-11}$\,erg\,s$^{-1}$\,cm$^{-2}$ (XRT-3).  From
the first two observations, the flux drops about 30\% over two weeks and then remains at about the
same level through the last two observations.

\paragraph{ESO 548-G081 --}
The spectra of ESO 548-G081 were found not to vary in column density.  However, they did vary
in both flux ($\Delta\chi^2 = 625$) and $\Gamma$ ($\Delta\chi^2 = 60$).  We fit this source
with a {\tt tbabs}*{\tt tbabs}*({\tt pegpwrlw} + {\tt bbody} + {\tt zgauss}) model with $\chi^2$/dof = 1428.1/1295.
We note that the blackbody parameters were not fixed for observation XRT-2.  This spectrum showed
a soft excess that was fit with a blackbody much lower than the value for the XMM observation (see Table~\ref{tbl-6} with kT in the range of 0.034 -- 0.049\,keV.  There were remaining residuals in this fit for XRT-2, leaving us unsure of the nature of this feature which could be the result of a hot pixel or the instrument.  The higher energy spectrum (above 0.5\,keV) appeared well fit by the model used.  We found the error on the photon index as: 1.851 -- 1.877 (XMM), 1.716 -- 1.802 (XRT-1), and 1.930 -- 2.002 (XRT-2).  The flux errors from
the pegged power law component were:  2.91 -- 2.96 $\times 10^{-11}$\,erg\,s$^{-1}$\,cm$^{-2}$ (XMM), 3.76 -- 4.04 $\times 10^{-11}$\,erg\,s$^{-1}$\,cm$^{-2}$ (XRT-1), and 4.13 -- 4.33 $\times 10^{-11}$\,erg\,s$^{-1}$\,cm$^{-2}$ (XRT-2).  It is unclear whether the changes between the XMM observation
and XRT-1 reflect differences in the instruments or the source.  Both observations were taken on the
same day, nine hours apart.  Two months later, the XRT-2 observation shows the source to have a steeper power law index and a higher flux.

\paragraph{ESO 490-G026 --}
Allowing flux, column density, and power law indices to vary between the four observations
of ESO 490-G026 greatly improved the fit with $\Delta\chi^2$ values of 171, 21, and 62.  
We used an absorbed power law + blackbody model (kT set at the value in Table~\ref{tbl-6}) with
an Fe K line.  Using the {\tt pegpwrl} for a power law component, the best fit gave $\chi^2$/dof =
1281.8/1313.
The errors on n$_H$ were: 2.96 -- 3.59 $\times 10^{21}$\,cm$^{-2}$ (XMM), 3.10 -- 3.81 $\times 10^{21}$\,cm$^{-2}$ (XRT-1), 2.54 -- 3.84 $\times 10^{21}$\,cm$^{-2}$ (XRT-2), and 4.12 -- 6.74 $\times 10^{21}$\,cm$^{-2}$ (XRT-3).  The errors on the photon index were: 1.66 -- 1.71 (XMM), 1.88 -- 2.00 (XRT-1),
1.63 -- 1.86 (XRT-2), and 1.74 -- 2.09 (XRT-3).  Finally, the errors on the flux from the pegged power
law component were: 3.23 -- 3.32 $\times 10^{-11}$\,erg\,s$^{-1}$\,cm$^{-2}$ (XMM), 3.79 -- 4.09 $\times 10^{-11}$\,erg\,s$^{-1}$\,cm$^{-2}$ (XRT-1), 3.53 -- 3.97 $\times 10^{-11}$\,erg\,s$^{-1}$\,cm$^{-2}$ (XRT-2), and 2.71 -- 3.37 $\times 10^{-11}$\,erg\,s$^{-1}$\,cm$^{-2}$ (XRT-3).  Observation XRT-1
occurred about 3 months before the XMM observation.  In this time, the power law index flattened while
the flux decreased.  The column density of the sources does not vary between the first three observations.  However, between XRT-2 and XRT-3, 5 days apart, the column increased by nearly twice the previous amount with the flux decreasing.

\paragraph{SWIFT J0904.3+5538 --}
As with ESO 490-G026, all of the parameters (column density, flux, and power law photon index)
varied for SWIFT J0904.3+5538.  For this source, we fixed the parameters of the blackbody
component to the best-fit values of the XMM PN observation.  No Fe K line was required in this
spectrum.  The $\Delta\chi^2$ values allowing flux, n$_H$, and $\Gamma$ to vary were
523, 12, and 20.  Errors for column density were:  0.61 -- 1.29 $\times 10^{21}$\,cm$^{-2}$ (XMM), 1.01 -- 1.93 $\times 10^{21}$\,cm$^{-2}$ (XRT-1), and 1.37 -- 2.41 $\times 10^{21}$\,cm$^{-2}$ (XRT-2).  The errors on the photon index were: 1.80 -- 1.97 (XMM), 1.45 -- 1.69 (XRT-1), and
1.49 -- 1.75 (XRT-2).  Finally, the errors on the flux from the pegged power
law component were: 0.94 -- 1.02 $\times 10^{-11}$\,erg\,s$^{-1}$\,cm$^{-2}$ (XMM), 0.76 -- 0.88\,$\times 10^{-11}$\,erg\,s$^{-1}$\,cm$^{-2}$ (XRT-1), and 0.60 -- 0.69 $\times 10^{-11}$\,erg\,s$^{-1}$\,cm$^{-2}$ (XRT-2).  Between XRT-1 and XRT-2, approximately a month apart, the source dimmed
by $\approx 20$\% and then brightened more than twice that amount 3 months later in the XMM
observation.  With the higher flux, the XMM observation showed less absorption and a steeper slope.

\paragraph{MCG +04-22-042 --}
MCG +04-22-042 was well-fit with a simple absorbed power law ({\tt pegpwrlw}) with 
$\chi^2$/dof = 1534.8/1190.  This fit required flux and column density to vary between the
XMM and XRT observations with $\Delta\chi^2$ of 938 and 298, respectively.  Errors for 
column density were:  0.00 -- 0.02 $\times 10^{20}$\,cm$^{-2}$ (XMM) and 1.77 -- 3.36 $\times 10^{20}$\,cm$^{-2}$ (XRT-1).  Though these values are small, there was clearly a change in column
density, evidenced by the very significant change in $\chi^2$.  The errors on the flux for the
pegged power law component were: 3.38 -- 3.42 $\times 10^{-11}$\,erg\,s$^{-1}$\,cm$^{-2}$ (XMM) and 2.29 -- 2.44\,$\times 10^{-11}$\,erg\,s$^{-1}$\,cm$^{-2}$ (XRT-1).  The two observations were
approximately 5 months apart, showing that the flux and column changed while the photon index
remained roughly the same.  The flux increased (in the XMM obs.) while the column density 
decreased.

\paragraph{UGC 6728 --}
In the four UGC 6728 spectra, the only statistically significant variation is in flux ($\Delta\chi^2 = 680$).
The absorbed pegged power law model yields a best-fit $\chi^2$/dof of 855.3/863.  This source
varies by a high amount with errors in flux of: 1.15 -- 1.19 $\times 10^{-11}$\,erg\,s$^{-1}$\,cm$^{-2}$ (XMM), 2.12 -- 2.28 $\times 10^{-11}$\,erg\,s$^{-1}$\,cm$^{-2}$ (XRT-1), 1.40 -- 1.60 $\times 10^{-11}$\,erg\,s$^{-1}$\,cm$^{-2}$ (XRT-2), and 1.53 -- 1.89 $\times 10^{-11}$\,erg\,s$^{-1}$\,cm$^{-2}$ (XRT-3).
Thus, the flux doubled between the four months from the XMM observation and XRT-1.  It then 
decreased by about 25\% over 5 days, remaining at about the same level in the XRT-3 observation a week later.

\paragraph{WKK 1263 --}
The only significant change in $\chi^2$ for the combined spectral fits to WKK 1263 was in flux
($\Delta\chi^2 = 150$).  The best-fit pegged power law fit had $\chi^2$/dof = 826.8/782.
  The errors on the flux from the pegged power
law component were: 1.52 -- 1.57 $\times 10^{-11}$\,erg\,s$^{-1}$\,cm$^{-2}$ (XMM), 1.18 -- 1.39\,$\times 10^{-11}$\,erg\,s$^{-1}$\,cm$^{-2}$ (XRT-1), and 1.07 -- 1.21 $\times 10^{-11}$\,erg\,s$^{-1}$\,cm$^{-2}$ (XRT-2).  The source was brighter (by as much as 50\%) in the XMM observation taken about a month after the XRT 
observations.

\paragraph{MCG +09-21-096 --}
MCG +09-21-096 showed significant variation in $\chi^2$ when allowing variations in flux
and power law photon index ($\Delta\chi^2 = 797$ and $12$).  The model ({\tt tbabs}*{\tt pegpwrlw})
was a good fit with $\chi^2$/dof = 1289.2/1334.  Errors on $\Gamma$ were: 1.78 -- 1.79 (XMM),
1.77 -- 1.82 (XRT-1), and 1.67 -- 1.74 (XRT-2).  The errors on the flux from the pegged power
law component were: 2.77 -- 2.81 $\times 10^{-11}$\,erg\,s$^{-1}$\,cm$^{-2}$ (XMM), 3.97 -- 4.16\,$\times 10^{-11}$\,erg\,s$^{-1}$\,cm$^{-2}$ (XRT-1), and 3.50 -- 3.74 $\times 10^{-11}$\,erg\,s$^{-1}$\,cm$^{-2}$ (XRT-2).   Between the XRT-1 and XMM observations, 7 months apart, the photon index
is the same while the flux decreased by about 40\%.  Then, in the week between the XMM and
XRT-2 observation, the source photon index flattened slightly while the flux increased to nearly the level
in the XRT-1 observation.
\\
\subsection{High Column Density/Complex Sources}
\paragraph{NGC 1142 --}
Since NGC 1142 has a strong Fe K line (see Table~\ref{tbl-7} for EW and normalization), we fixed the Fe K parameters for the three XRT observations to the best fit values for the PN spectrum.  Allowing the
flux to vary between these observations improved the fit by $\Delta\chi^2 = 130$.  There is no
evidence of variability in column density, however, varying power law components improved the
fit by $\Delta\chi^2 \approx 30$.  Fitting the spectra with a pegged power law model, the best-fit
model has $\chi^2$/dof $= 132.6/105$.  Errors on the power law photon index, $\Gamma$, were:
1.54 -- 2.29 (XMM), 2.34 -- 3.85 (XRT-1), 2.15 -- 3.98 (XRT-2), and 1.08 -- 2.94 (XRT-3).  Though
the error bars for the photon index are large, due to the few counts for this heavily absorbed source, it is clear that observation XRT-1 has a steeper power law component than the XMM observation.
 The errors on the {\tt pegpwrlw} flux were:  1.97 -- 6.01 $\times 10^{-11}$\,erg\,s$^{-1}$\,cm$^{-2}$ (XMM), 10.87 -- 322.3 $\times 10^{-11}$\,erg\,s$^{-1}$\,cm$^{-2}$ (XRT-1), 10.10 -- 478 $\times 10^{-11}$\,erg\,s$^{-1}$\,cm$^{-2}$ (XRT-2), and 3.81 -- 43.13 $\times 10^{-11}$\,erg\,s$^{-1}$\,cm$^{-2}$ (XRT-3).  Once again, the XRT error bars are large due to the few counts.  Noting that the XMM observation occurred 6 months before XRT-1, clearly the flux is higher in the XRT-1 observation
 while the photon index is steeper.  No conclusions can be drawn from the final two XRT observations.

\paragraph{SWIFT J0318.7+6828 --}
SWIFT J0318.7+6828 showed no variability in column density or power law component between
the XMM and two XRT observations ($\Delta\chi^2 < 3$ allowing each to vary).  The variability
was significant in flux with $\Delta\chi^2 \approx 40$ when a constant model was added.  The
errors on the {\tt pegpwrlw} flux were computed as:  1.01 -- 1.25 $\times 10^{-11}$\,erg\,s$^{-1}$\,cm$^{-2}$ (XMM), 0.76 -- 1.00 $\times 10^{-11}$\,erg\,s$^{-1}$\,cm$^{-2}$ (XRT-1), and 0.79 -- 1.04 $\times 10^{-11}$\,erg\,s$^{-1}$\,cm$^{-2}$ (XRT-2) with $\chi^2$/dof$ = 412.9/452$.  Thus, the XRT observations, taken a week apart, did not vary.  However, three months earlier the XMM observations
show the source to be brighter by $\approx 30\%$.

\paragraph{ESO 362-G018 --}
As mentioned, the spectra of ESO 362-G018 showed more variability than any other source in this 
sample.  The value  $F_{max} - F_{min}/F_{avg}$ for this source was 1.64 in the soft band
and 1.22 in the hard band.  This source had a complex spectrum, described in the Detailed Spectral Fitting section.
We fit the XMM and XRT spectra of this source with a partial covering and power law model.
However, we added gaussian components to fit the strong Fe K line (in the XMM observation) and
the helium-like oxygen edge.  The best-fit model required flux, n$_H$, and $\Gamma$ to vary
($\Delta\chi^2 =$ 1890, 254, and 40) with $\chi^2$/dof = 538.4/445.  The errors on column 
density were 24.54 -- 31.23 $\times 10^{22}$\,cm$^{-2}$ (XMM), 6.47 -- 269.3 $\times 10^{22}$\,cm$^{-2}$ (XRT-1), and 2.12 -- 4.71 $\times 10^{22}$\,cm$^{-2}$ (XRT-2), with partial covering fractions
of 0.91 -- 0.93 (XMM), 0.10 -- 0.91 (XRT-1), and 0.45 -- 0.68 (XRT-2).  Errors on the photon index
were: 2.13 -- 2.23 (XMM), 1.76 -- 2.05 (XRT-1), and 1.69 -- 1.99 (XRT-2).  Finally, errors on flux
for the pegged power law component were:  1.46 -- 1.83 $\times 10^{-11}$\,erg\,s$^{-1}$\,cm$^{-2}$ (XMM), 2.17 -- 7.69 $\times 10^{-11}$\,erg\,s$^{-1}$\,cm$^{-2}$ (XRT-1), and 2.25 -- 2.95 $\times 10^{-11}$\,erg\,s$^{-1}$\,cm$^{-2}$ (XRT-2).

\begin{figure}
\epsscale{1.0}\plottwo{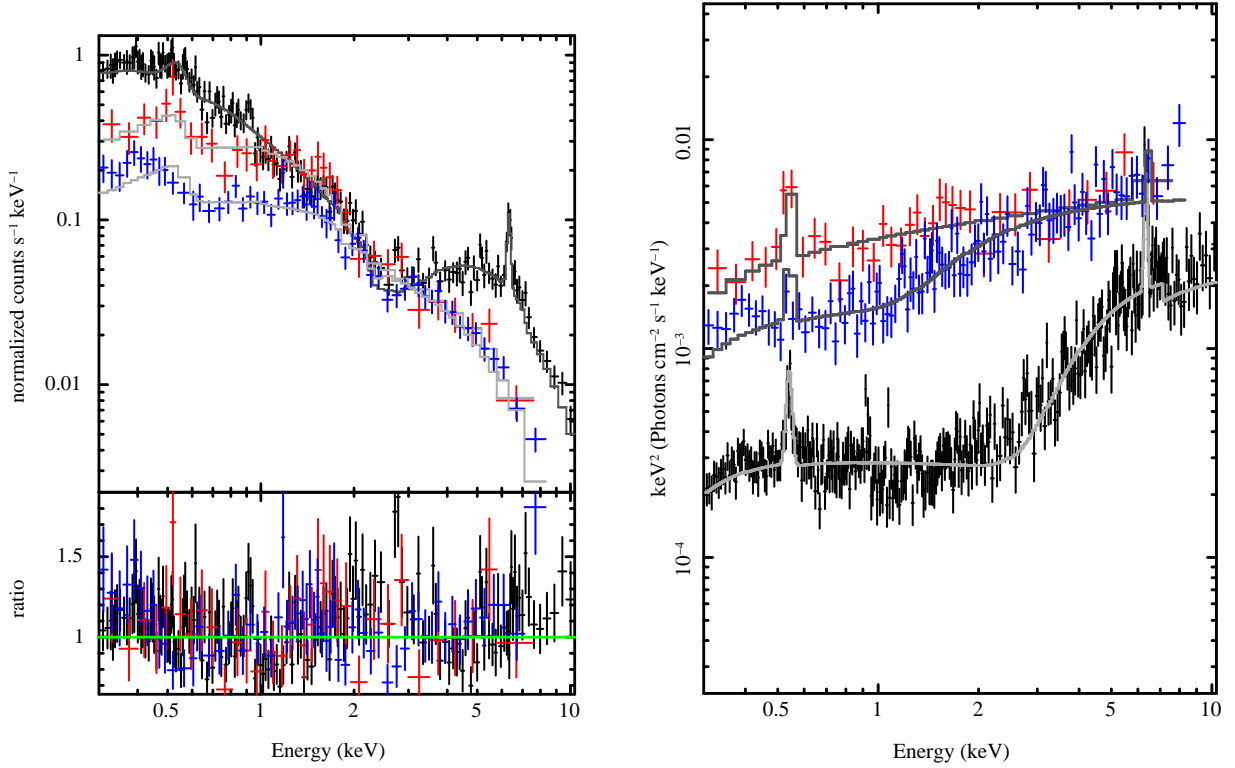}{figure10b.ps}
\caption{{\bf (left)} \xmm PN spectrum (black) with two XRT (red and blue) observations for ESO 362-G018 fit with the model
{\tt tbabs}*{\tt pcfabs}*({\tt pegpwrlw} + {\tt zgauss} + {\tt zgauss}).  This source, with a complex spectrum,
varied considerably between the XRT observations, taken approximately 2 months later, and \xmm
observation.  The \xmm observation shows a strong Fe K line and a column density 10 times that seen
by the XRT observations.  The flux is also lower by $\approx 50$\% in the \xmm observation (the spectra shown are normalized and exhibit the {\bf observed} spectrum). {\bf (right)} The unfolded spectrum for ESO 362-G018 is plotted (E$^2$\,f(E) vs. E).  If the Fe K line
remained at the same level as in the \xmm observation, the increased flux from the power law
component would dominate the line emission in the XRT observations.  This is a possible 
explanation for the appearance of the Fe K line.  The observed spectrum is shown in Figure~\ref{fig-vary}.
\label{fig-ufspec}}
\end{figure}

\begin{figure}
\epsscale{1.0}\plotone{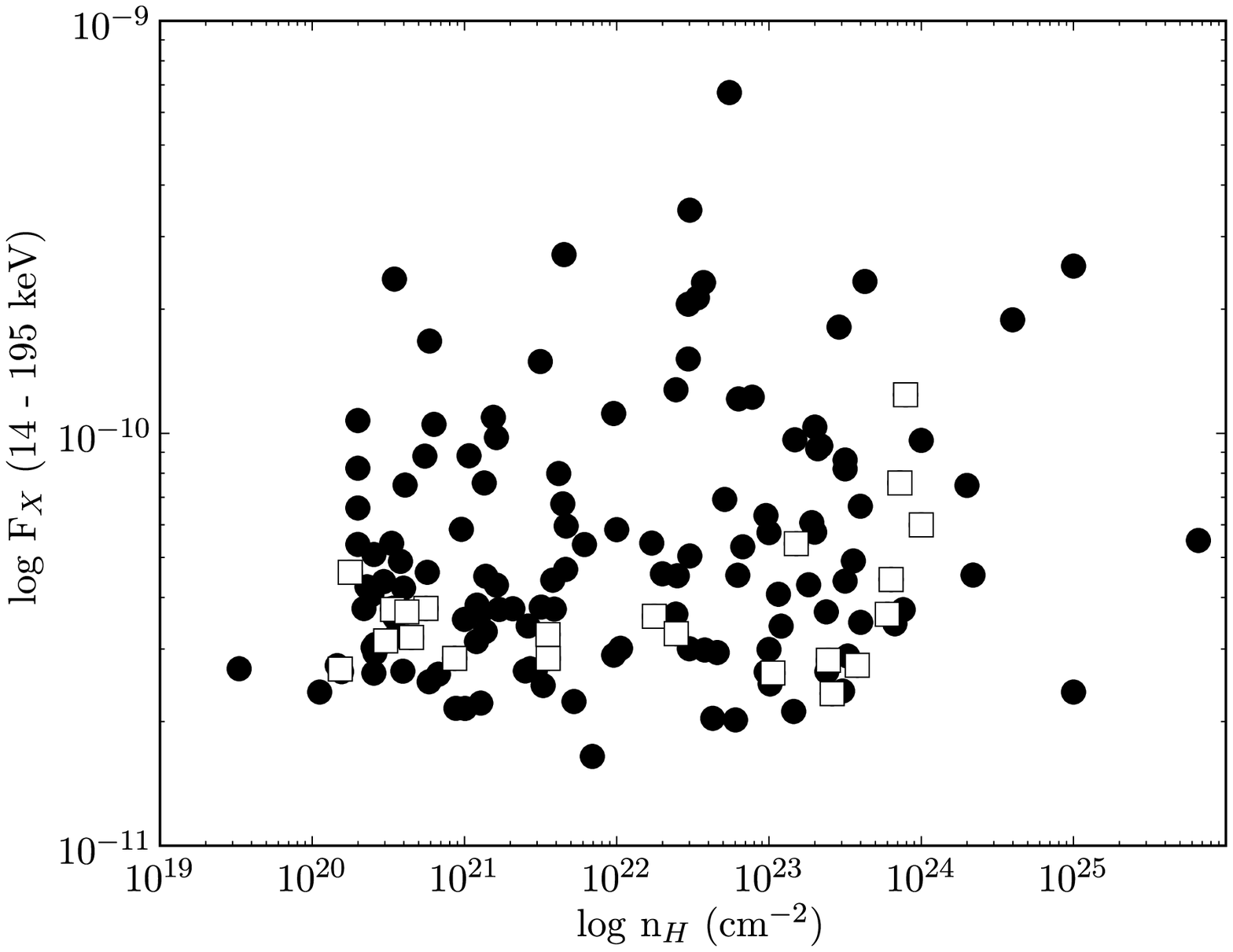}
\caption{Hydrogen column density (cm$^{-2}$) versus the 14-195\,keV flux (erg\,s$^{-1}$\,cm$^{-2}$) measured by SWIFT's BAT instrument.
These values are listed in \citet{tue07}, with the circles representing 9-month catalog
sources and the squares sources from the 9-month catalog with {\it XMM-Newton} follow-ups detailed
in this paper.  We note that for column densities higher than $\approx 10^{24}$\,cm$^{-2}$, the
spectrum is likely optically thick to Compton emission and thus there is a greater uncertainty in
the measured hydrogen column density.
\label{fig-bat}}
\end{figure}

\begin{figure}
\epsscale{1.5}\plottwo{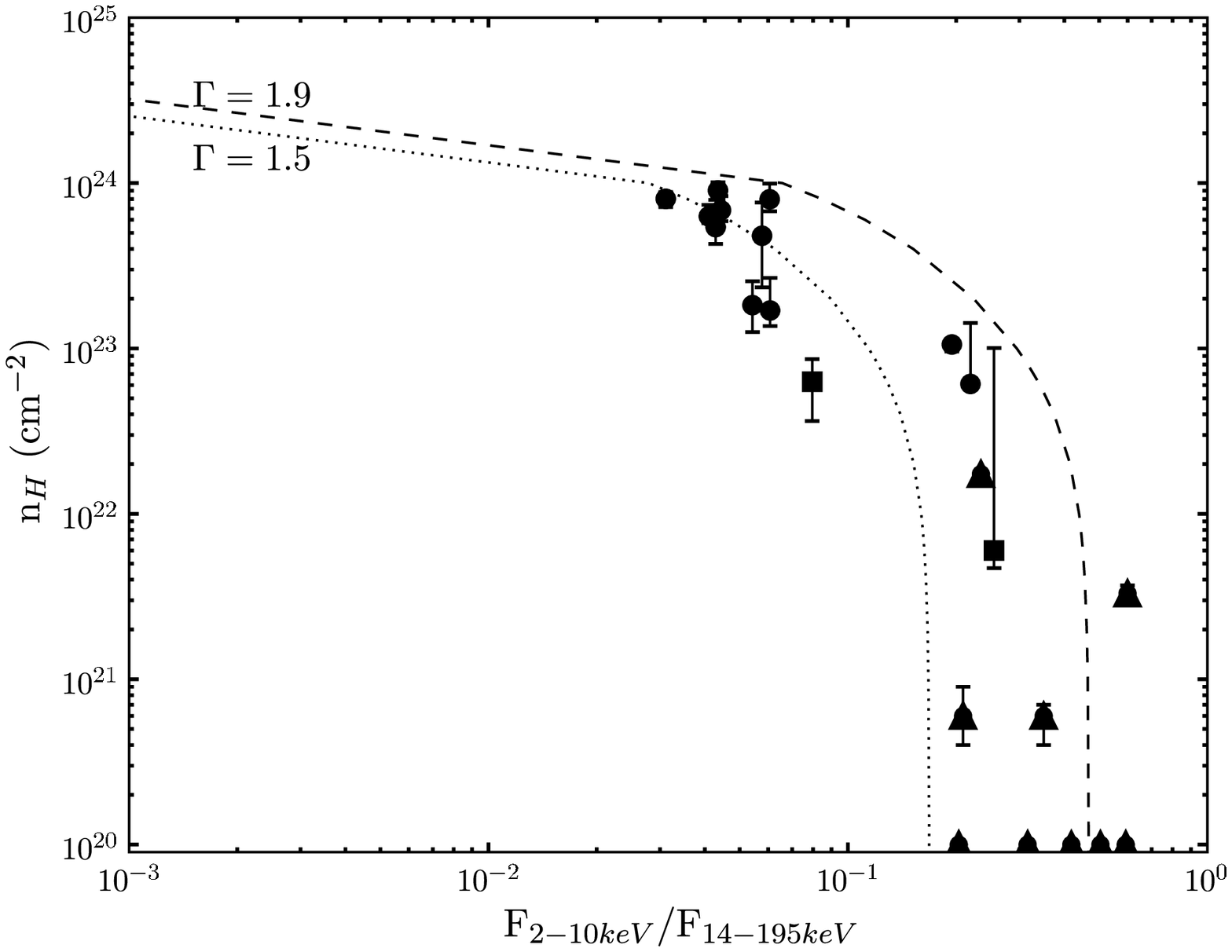}{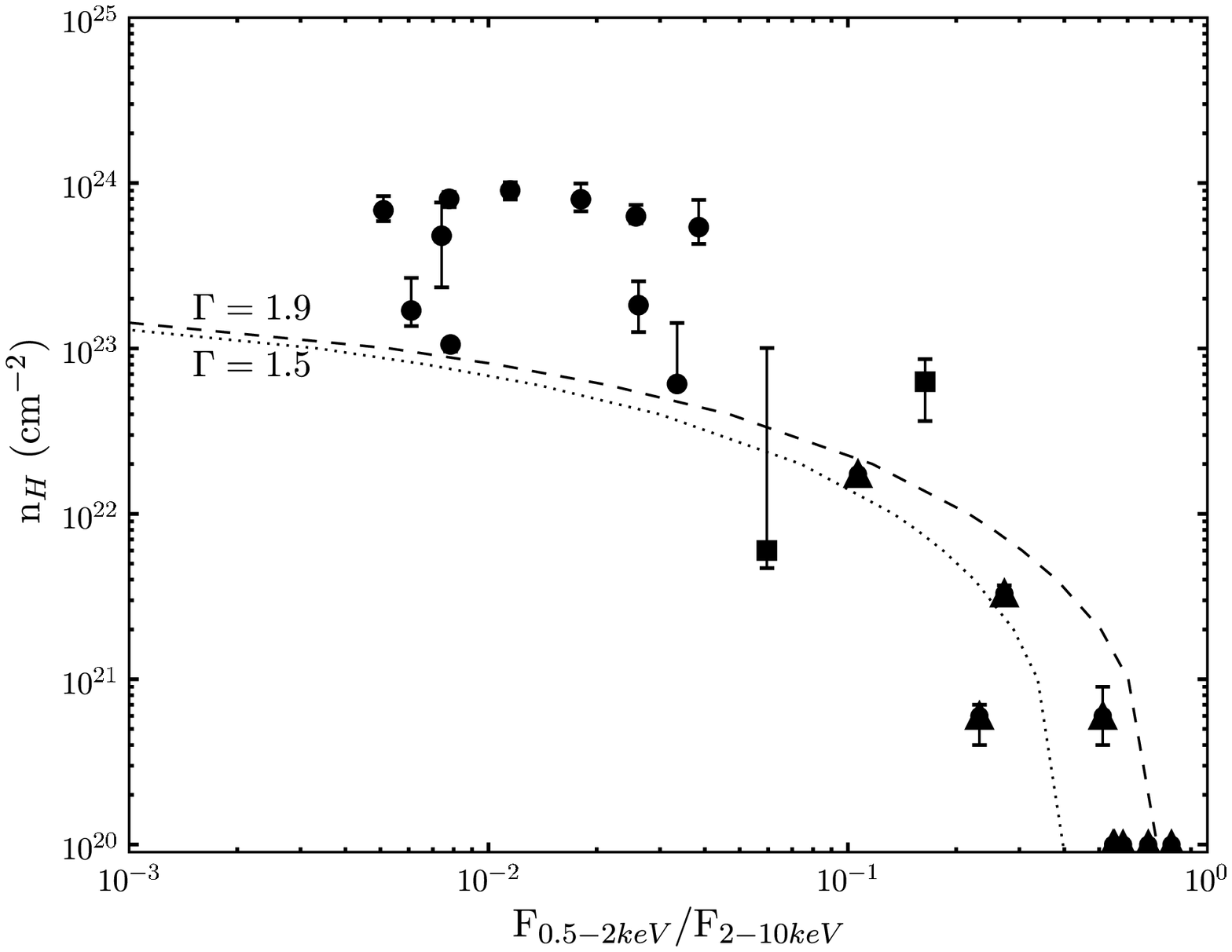}
\caption{Plots of column density vs. flux ratios for the \xmm follow-up sources.  At left,
we show the ratio of the 2 -- 10\,keV flux to the 14 -- 195\,kev (BAT) flux.  The triangles
are the sources best-fit by a simple power law or power law and blackbody model.  The squares
represent the two sources with complex spectra we could not interpret (ESO 362-G018 and NGC 6860).  Finally, the circles represent the absorbed sources with complex spectra.  At right,
the ratio of the 0.5 -- 2\,keV flux to the 2 -- 10\,keV flux is shown.  For the unabsorbed
sources, we plotted the sources as having n$_H = 10^{20}$\,cm$^{-2}$.  These sources
had approximately the same ratio of hard flux/BAT flux as soft flux/hard flux.  For the 
more heavily absorbed sources, the ratio of hard flux/BAT flux is clearly larger than
the soft flux/hard flux ratio.  The lines represent column density vs. flux ratio for constant power law indices ($\Gamma = 1.9$ and
$1.5$, as labeled).
\label{fig-fluxratio}}
\end{figure}

\begin{figure}
\epsscale{1.0}\plotone{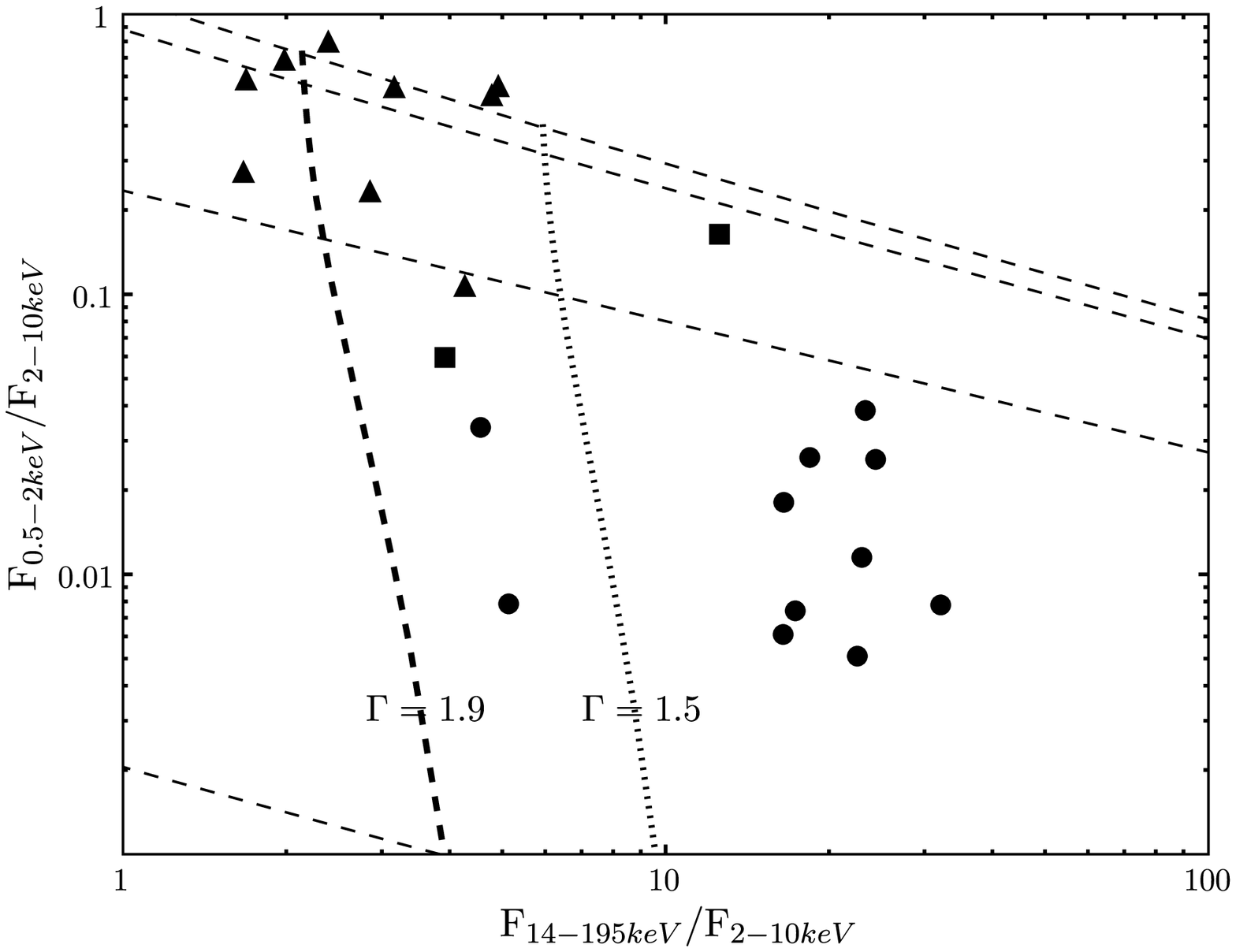}
\caption{Color-color diagram of soft/medium (0.5 -- 2\,keV flux / 2 -- 10\,keV flux) and hard/medium (14 -- 195\,kev (BAT) flux / 2 -- 10\,keV flux) colors.  The vertical lines represent values for constant
power law indices (of $\Gamma = 1.9$ and $1.5$, as labeled) with different absorbing columns.
The other lines on the diagram represent constant column densities for different power law indices
(from top to bottom: $10^{20}$, $10^{21}$, $10^{22}$, and $10^{23}$\,cm$^{-2}$).  The triangles
are the sources best-fit by a simple power law or power law and blackbody model.  The squares
represent the two sources with complex spectra we could not interpret (ESO 362-G018 and NGC 6860).  Finally, the circles represent the absorbed sources with complex spectra.  The unabsorbed sources clearly occupy a region to the top left in the diagram while the more heavily absorbed
sources lie towards the bottom right.  Sources with column densities in the middle (between
$10^{22}$ and $10^{23}$\,cm$^{-2}$) lie between our unabsorbed and heavily absorbed points.
From this result, we present a new diagnostic to describe spectra with low counts.
\label{fig-color}}
\end{figure}

\begin{figure}
\plotone{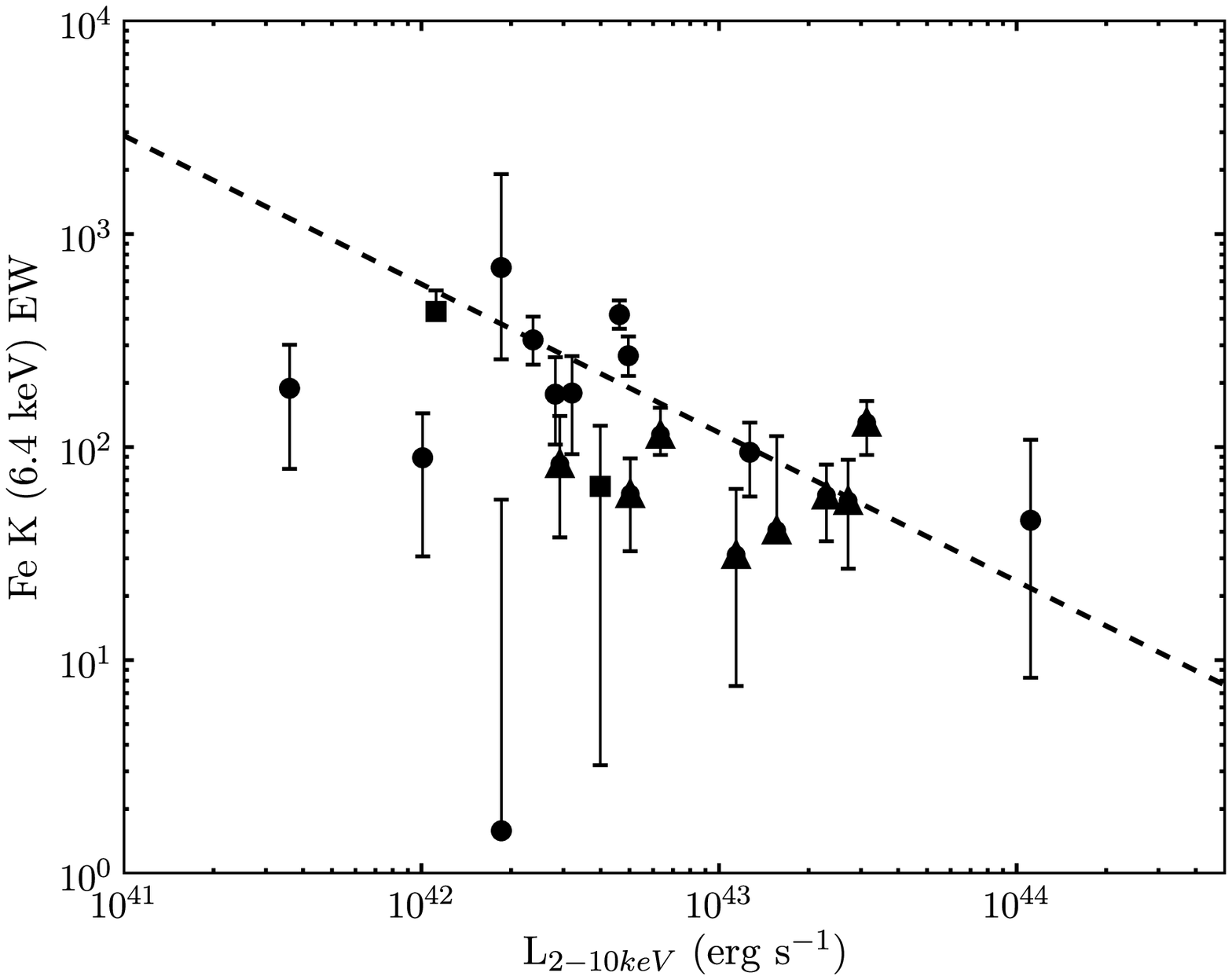}
\caption{Plot of the Fe K equivalent width measurements (eV) versus the luminosity in the
hard band (2 -- 10\,keV).  The equivalent width measurements are from the best-fit models for
the sources shown in Tables~\ref{tbl-5},~\ref{tbl-6}, and~\ref{tbl-8}.  The simple power law/ power law and blackbody, unabsorbed/low absorption sources are plotted as triangles.  The squares
represent the two sources with complex spectra we could not interpret (ESO 362-G018 and NGC 6860).  Finally, the circles represent the absorbed sources with complex spectra.  The source
SWIFT J0216.3+5128 is not plotted since there was no evidence of a line and the redshift is unknown.
The line is the ordinary least squares bisector fit to the data using the upper limits on the Fe K 
equivalent widths. 
\label{fig-iron}}
\end{figure}

\begin{figure}
\epsscale{1.6}\plottwo{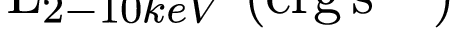}{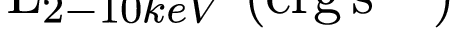}
\caption{Plot of the optical depth of an added \ion{O}{7}, 0.74\,keV, and \ion{O}{8}, 0.87\,keV, K edge vs. 2 -- 10\,keV luminosity.  This model was added for the low absorption/simple model sources (circles, solid lines for upper limits), which are optical Seyfert 1 sources (with the possible exception of WKK 1263 for which we found no archived optical spectrum).  We compare our values to those from \citet{rey97} (squares, dashed lines for upper limits).  Clearly, we find much weaker optical depths among our sample, most noticeably for \ion{O}{8} where the upper limits on optical depth are well below $\tau = 0.1$ for all but one source.  Further, in Table~\ref{tbl-11}, the addition of the two edge models gives a statistically better fit ($\Delta\chi^2 > 10$) for only two sources (ESO 490-G026 and MCG +04-22-042).  Only ESO 490-G026, has upper limits for both  \ion{O}{7} and \ion{O}{8} with $\tau > 0.1$. 
\label{fig-rey}}
\end{figure}

\begin{figure}
\epsscale{1.3}\plottwo{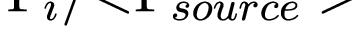}{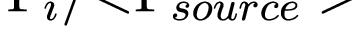}
\caption{Plots of varying column density (left) and photon index (right) with flux. 
For sources
that showed variations in column density or photon index, we plotted the
observation's value divided by the average for the source (i.e. F$_i / <$F$_{source}$), for each individual observation.  The flux is the 0.3 -- 10\,keV flux
from the pegged power law component.  The lines mark the area where each parameter is
1.0 (where the observation value is the average value).  A different symbol is used for
each source.  No correlation is seen between column
density and flux, however, there is a strong correlation between photon index and flux. 
\label{fig-varavg}}
\end{figure}

In Figure~\ref{fig-ufspec}a, we show the normalized XMM and XRT spectra with best-fit model.  It
is clear from this figure that this source varied a great deal in these observations.  Both
XRT-1 and XRT-2, despite the large error bars on XRT-2, have similar spectra.  These observations
were taken approximately a month apart.  Taken two months later, the XMM observation looks
like a different source altogether.  In this time, the column density increased by a factor of 10.
Additionally, the photon index became steeper and the flux dropped by about 50\%.  Along
with these changes, the Fe K line (not distinguishable in the XRT observations) became extremely prominent.  One likely explanation for the appearance of the Fe K line is that it was simply
too dim to be distinguishable at the higher flux levels exhibited in the XRT observations.
This is illustrated in the unfolded spectrum shown in Figure~\ref{fig-ufspec}b.  In this plot, where
the y-axis shows E$^2$\,f(E), it is clear that if the Fe K line remained at the same flux level
as in the XMM observation it would be completely dominated by the power law component.     

\paragraph{NGC 6860 --}
As mentioned in the Detailed Spectral Fitting section, the spectrum of NGC 6860 is quite complex.  Due to this
complexity and a lack of signal to noise (especially considering that the PN data was corrupted),
we are unsure of the true nature of this spectrum.  Therefore, we decided to compare the
XMM and XRT spectra with the {\tt pcfabs}*{\tt pow} model.  This may not be the most
valid description of the data, but it gives a basis to compare the spectra.  Using this model,
we fit both of the MOS observations (with the parameters n$_H$ and $\Gamma$ tied together
while the flux was allowed to vary) simultaneously with the XRT observation.  Variations were
statistically significant for flux, n$_H$, and $\Gamma$ with $\Delta\chi^2 =$ 83, 56, and 15,
respectively.  The best-fit for the partial covering, pegged power law model yielded 
$\chi^2$/dof of 513.3/430.  Errors on the column density were 3.33 -- 6.09 $\times 10^{22}$\,cm$^{-2}$ (XMM)
and 0.59 -- 1.73 $\times 10^{22}$\,cm$^{-2}$ (XRT) with covering fraction errors of 0.0 -- 0.66 (XMM) and
0.73 -- 0.92 (XRT).  The photon index errors were: 0.64 -- 0.88 (XMM) and 1.16 -- 1.71 (XRT).
Finally, errors on the flux from the pegged power law component were: 1.06 -- 1.19 $\times 10^{-11}$\,erg\,s$^{-1}$\,cm$^{-2}$ (XMM MOS-1), 0.99 -- 1.12 $\times 10^{-11}$\,erg\,s$^{-1}$\,cm$^{-2}$ (XMM MOS-2), and 1.44 -- 1.88 $\times 10^{-11}$\,erg\,s$^{-1}$\,cm$^{-2}$ (XRT).  These observations
were taken 4 months apart.  Interestingly, the XRT observation is well-fit by the partial covering model,
giving photon index and covering fractions similar to the typical values seen in Table~\ref{tbl-7}.
However, the column density is much lower (by a factor of 10 from the other sources).  Between the
XRT and XMM observations, the column density seems to have doubled while the photon index
flattened and the flux decreased by nearly half.

\clearpage

\clearpage

\clearpage   

\clearpage 

\clearpage

\clearpage

\clearpage

\clearpage

\clearpage

\clearpage 

\clearpage

\clearpage

\clearpage

\clearpage

\clearpage

\clearpage
 
\clearpage

\begin{deluxetable}{lllllll}
\tabletypesize{\scriptsize}
\tablecaption{{\it XMM-Newton} EPIC and SWIFT XRT Observation Information\label{tbl-1}}
\tablewidth{0pt}
\tablehead{
\colhead{Source} & \colhead{RA (h m s)}  & \colhead{Dec ($\circ\ \prime\ \prime\prime$)} &
\colhead{Redshift} & \colhead{n$_{H (GAL)}$\tablenotemark{1}} &
\colhead{Type\tablenotemark{2}} & \colhead{Host Galaxy\tablenotemark{2}} 
}

\startdata
\object{MRK 352} 			   &	00 59 53.28 &	31 49 36.87   &	0.014864     &  5.59 & Sy1 & SA0\\
\object{NGC 612}			   &  	01 33 57.74 &  	-36 29 35.70  & 	0.029771     & 1.85  & Gal & SA0+ pec\\
\object{SWIFT J0216.3+5128} &	02 16 29.84 &  	51 26 24.70   &  	UNKNOWN & 1.57 & -- & ? \\
\object{NGC 1142}  		   &	02 55 12.196 & -00 11 00.81  &  	0.028847     &  6.00 & Sy2 & S pec\\
\object[NVSS J031818+682927]{SWIFT J0318.7+6828} &	03 18 18.98 &	68 29 31.42   &	0.090100     & 35.1 & Sy2 & S?\\
\object{ESO 548-G081}             &	03 42 03.72 &	-21 14 39.70  &	0.014480     & 3.04 & Sy1 & Sba\\
\object{ESO 362-G018}  	   &	05 19 35.82 &	-32 39 27.90  &	0.01264       &  1.78 & Sy1.5 & S0/a\\
\object{ESO 490-G026}             &	06 40 11.69 &	-25 53 43.30  &	0.024800     &  11.7 & Sy1.2 & Pec \\
\object[CGCG 145-003]{SWIFT J0641.3+3257} &	06 41 23.04 &	32 55 38.60   &	0.017195     & 11.6  & Sy2 & E? \\
\object{MRK 18}   			   &	09 01 58.39 &	60 09 06.20   &	0.011088     & 4.49 & Sy2 & S\\
\object[SDSS J090432.19+553830.1]{SWIFT J0904.3+5538} &	09 04 32.94 &	55 38 30.63   &	0.037142     & 2.78 &  Sy1 & ?\\
\object{SWIFT J0911.2+4533} &	09 11 29.97 &	45 28 05.00   &	0.026782     & 1.64 & Sy2 & S? \\
\object{MCG +04-22-042}         &	09 23 43.00 & 	22 54 32.50   &	0.032349     & 3.37 & Sy1.2 & E \\
\object{MRK 417}			   &	10 49 30.93 &  	22 57 51.90   &	 0.032756    & 2.06 & Sy2 & Sa \\
\object{UGC 6728}		   &	11 45 16.02 &	79 40 53.42   &	0.015300     & 4.49 & Sy1.2 & SB0/a \\
\object{SWIFT J1200.8+0650} & 	12 00 57.92 & 	06 48 23.11   &	 0.036045    & 1.43 & Sy2 & S?\\
\object{ESO 506-027}                &	12 38 54.59 &	-27 18 28.20  &	0.025024     &  6.60 & Sy2 & S pec s\\
\object{WKK 1263}  		   &	12 41 25.74 &	-57 50 03.50  &	0.024430     & 35.5 & Sy2\tablenotemark{3} & Sc\\
\object{MCG +09-21-096}         &	13 03 59.47 & 	53 47 30.10   &	 0.02988 & 1.53 & Sy1 & SABb\\
\object{NGC 4992}		   &	13 09 12.95 &  	11 38 45.32   &	 0.025137    & 2.09 & Gal & Sa \\
\object{NGC 6860}		   &	20 08 46.89 &	-61 06 00.70  &	0.014884     & 4.19 & Sy1.5\tablenotemark{4} & SB ab\\
\object{NGC 6921}		   &	20 28 28.86 & 	25 43 24.30 	 &	0.014287     & 26.0 & Sy2 & SA 0/a\\
\object{MCG+04-48-002}    &    20 28 35.0   &   25 44 00.0          &     0.013900     & 26.0  & Sy2 & S \\
\enddata

\tablenotetext{1}{Galactic column density towards the source, 
in units of $10^{20}$\,cm$^{-2}$, as obtained from the web version of the
n$_H$ FTOOL.  These are the values from \citet{dic90}.\\}
\tablenotetext{2}{AGN type and host galaxy type from \citet{tue07}.  For AGN types, optical identifications
are listed, where available.  Where ``Gal'' is indicated, there are no optical emission lines indicative
of the presence of an AGN.  The optical spectrum looks like a galaxy spectrum.  Additional host
galaxy classifications were obtained from the LEDA database.  Where ``?'' is indicated, there is no available
classification.\\}
\tablenotetext{3}{While WKK 1263 is classified as a Sy2 in NED, we could find no optical spectrum
to confirm this.\\}
\tablenotetext{4}{NGC 6860 is classified as a Sy1.5 by \citet{lip93} contrary to NED's classification
as a Sy1.}
\end{deluxetable}

\clearpage

\begin{deluxetable}{llllll}
\tabletypesize{\scriptsize}
\tablecaption{{\it XMM-Newton} EPIC and SWIFT XRT Observation Information\label{tbl-2}}
\tablewidth{0pt}
\tablehead{
\colhead{Source} & \colhead{Telescope}  & \colhead{Observation ID} & \colhead{Start Date} &
\colhead{Exposure Time (s)\tablenotemark{1}} & \colhead{Total Counts\tablenotemark{1}} 
}

\startdata
MRK 352		& XMM	& 0312190101	& 2006-01-24 & 9773, 12382, 12360 & 71296, 23572, 23175 \\
MRK 352		& XRT	& 00035243001	& 2006-05-29 & 11527 & 3456 \\
MRK 352		& XRT	& 00035243002	&  2006-05-30 &19385 & 3983  \\
NGC 612		& XMM	& 0312190201 & 2006-06-26 & 9744, 12456, 12457 & 1164, 285, 352 \\
NGC 612		& XRT	& 00035627001	& 2006-06-02 & 6743 & 20 \\
NGC 612		& XRT	& 00035627002	& 2006-06-12 & 4875 & 20 \\
SWIFT J0216.3+5128& XMM & 0312190301	& 2006-01-24& 8921, 11485 ,11495 & 14412, 6454, 6053 \\
SWIFT J0216.3+5128& XRT & 00035247001	& 2006-01-08 & 8291 & 1516 \\
SWIFT J0216.3+5128& XRT & 00035247002	& 2006-01-10 & 5559 & 982 \\
SWIFT J0216.3+5128& XRT & 00035247003	& 2006-02-05 & 2113 & 288 \\
NGC 1142	& XMM	& 0312190401	& 2006-01-28 & 8921, 11485, 11496 & 2481, 847, 907 \\
NGC 1142	& XRT	& 00035248001 & 2006-06-29 & 6808 & 245\\
NGC 1142	& XRT	& 00035248002 & 2006-07-07 & 5434 & 201 \\
NGC 1142	& XRT	& 00035248003 & 2006-07-08 & 4670 & 145 \\
NGC 1142	& XRT	& 00035248004 & 2006-07-11 & 1888 & 40 \\
SWIFT J0318.7+6828& XMM & 0312190501	& 2006-01-29 & 6578, 11163, 11170  & 9654, 4512, 4491 \\
SWIFT J0318.7+6828& XRT	& 00035249001	& 2006-03-29& 9247 & 413 \\
SWIFT J0318.7+6828& XRT	& 00035249002	& 2006-04-05 & 8061 & 385 \\
ESO 548-G081& XMM & 0312190601	& 2006-01-28 & 8924,11485, 11499 & 106660, 32699, 32404 \\
ESO 548-G081& XRT & 00035250001	& 2006-01-28 & 3561 & 3632 \\
ESO 548-G081& XRT & 00035250002	& 2006-03-19 & 6605 & 7821 \\
ESO 362-G018	& XMM & 0312190701	& 2006-01-28 & 8921, 11483, 11498 & 8497, 2905, 2884 \\
ESO 362-G018	& XRT & 00035234001	& 2005-10-29 & 1379 & 772 \\
ESO 362-G018	& XRT & 00035234002	& 2005-11-26 & 6979 & 2587 \\
ESO 490-G026	& XMM & 0312190801	& 2006-03-07 & 9192, 11812, 11823 & 52246, 21057, 19713 \\
ESO 490-G026	& XRT & 00035256001	& 2005-12-16 & 8246 & 4448 \\
ESO 490-G026	& XRT & 00035256002	& 2006-03-23 & 2809 & 1434 \\
ESO 490-G026	& XRT & 00035256003	& 2006-03-28 & 2747 & 954 \\
SWIFT J0641.3+3257& XMM & 0312190901	& 2006-03-11 & 10696, -, 13507 & 2686, -, 786 \\
SWIFT J0641.3+3257& XRT & 00035257002	& 2005-12-18 & 7784 & 20 \\
SWIFT J0641.3+3257& XRT & 00035257003	& 2005-12-26 & 14864 & 82 \\
SWIFT J0641.3+3257& XRT & 00035257006	& 2006-01-07 & 20008 & 60 \\
MRK 18		& XMM	& 0312191001	& 2006-03-23 & 9910, 13387, 13402 & 4990, 1089, 1029 \\
MRK 18		& XRT	& 00035259001	& 2005-12-18 & 5353 & 50 \\
MRK 18		& XRT	& 00035259002	& 2005-12-26 & 3088 & 20 \\
SWIFT J0904.3+5538& XMM	& 0312191101	& 2006-03-31 & 7142, 12072 , 12089 & 10830, 10980, 11221 \\
SWIFT J0904.3+5538& XRT & 00035260001	& 2005-12-15 & 5706 & 916 \\
SWIFT J0904.3+5538& XRT & 00035260002	& 2006-01-06 & 6211 & 736 \\
SWIFT J0911.2+4533& XMM & 0312191201	&2006-04-10 & -, 11530, 11531 & -, 615, 515 \\
SWIFT J0911.2+4533& XRT & 00035261001	& 2006-01-04 & 5487 & 21 \\
SWIFT J0911.2+4533& XRT & 00035261002	& 2006-01-22 & 8875 & 80 \\
MCG +04-22-042	& XMM	& 0312191401	& 2006-04-18 &9012, 11809, 11824 & 126070, 39777 39220 \\
MCG +04-22-042	& XRT	& 00035263001 & 2005-12-10 & 8564 & 4986 \\
MRK 417 & XMM & 0312191501	& 2006-06-15 & 7437, 348, 351 & 1075, 348, 351\\
MRK 417& XRT & 00035264001	& 2005-12-12 & 6306 & 41 \\
MRK 417& XRT & 00035264002	& 2006-03-03 & 3534 & 20 \\
MRK 417& XRT & 00035264003	& 2006-06-24 & 16130 & 140 \\
UGC 6728	& XMM	& 0312191601	& 2006-02-23 & 7220, 11404, 11415 & 30705, 11420, 11109 \\
UGC 6728 	& XRT	& 00035266001 & 2006-06-24 & 6331 & 3090 \\
UGC 6728 	& XRT	& 00035266002 & 2006-06-29 & 3017 & 994 \\
UGC 6728 	& XRT	& 00035266003 & 2006-07-07 & 1433 & 463\\
SWIFT J1200.8+0650 & XMM& 0312191701 &  2006-06-26 & 9777, 12495, 12507 & 4543, 1590, 1751  \\
SWIFT J1200.8+0650 & XRT & 00035267001 & 2005-12-11 & 14961& 638 \\
SWIFT J1200.8+0650 & XRT & 00035267002 & 2005-12-21 & 3156 & 120 \\
ESO 506-G027	& XMM & 0312191801	& 2006-01-24 & 8162, 11139, 11150 & 2114, 630, 645 \\
ESO 506-G027	& XRT & 00035273002	& 2005-06-15 & 3209 & 60 \\
ESO 506-G027	& XRT & 00035273003	& 2005-08-15 & 1938 & 20 \\
ESO 506-G027	& XRT & 00035273004	& 2005-08-28 & 10808 & 242 \\
WKK 1263	& XMM & 0312191901	& 2006-02-01 & 8902, 11482, 11495 & 24195, 10198, 10120 \\
WKK 1263	& XRT & 00035268001	& 2005-12-15 & 3253 & 564 \\
WKK 1263	& XRT & 00035268002	& 2005-12-29 & 8690 & 1346 \\
SWIFT J1303.8+5345& XMM & 0312192001	& 2006-06-23 & 8408, 11470, 11490 & 86597, 27942, 27492	\\
SWIFT J1303.8+5345& XRT	& 00035269001 & 2005-12-19 & 7468 & 7723 \\
SWIFT J1303.8+5345& XRT	& 00035269004 & 2006-07-02 & 4752 & 3986 \\
NGC 4992 & XMM & 0312192101 & 2006-06-27 & 12849, -, - & 1756, -, - \\
NGC 6860 	& XMM & 0312192201 	&2006-03-19 & -, 11815, 11823 & -, 5485, 4910 \\
NGC 6860 	& XRT & 00035275001 	& 2005-12-12 & 3536 & 632 \\
NGC 6921	& XMM	& 0312192301	& 2006-04-23 & 8789, 11235, 11249 & 2243, 554, 513 \\
MCG +04-48-002 & XMM & 0312192301	& 2006-04-23 & 8789, 11235, 11249 & 845, 180, 163 \\
MCG +04-48-002	& XRT	& 00035276001	& 2005-12-16 & 4071 & 41 \\
MCG +04-48-002	& XRT	& 00035276002	& 2006-03-23 & 4600 & 40 \\
MCG +04-48-002	& XRT	& 0003072201	& 2006-06-03 & 6885 & 65 \\
\enddata

\tablenotetext{1}{For the XMM observations, the exposure times and total counts are listed
for the EPIC PN, MOS1, and MOS2 respectively.}

\end{deluxetable}

\clearpage

\begin{deluxetable}{lllllll}
\tabletypesize{\scriptsize}
\tablecaption{SWIFT XRT Spectral Analysis: Absorbed Power Law Fits ({\tt tbabs}*{\tt tbabs}*{\tt pow})\label{tbl-3}}
\tablewidth{0pt}
\tablehead{
\colhead{Source} & \colhead{Observation ID}  & \colhead{n$_H$\tablenotemark{1}} & \colhead{$\Gamma$} &
\colhead{$\chi^2$/dof} & \colhead{Soft Flux\tablenotemark{2}} & \colhead{Hard Flux\tablenotemark{2}} 
}

\startdata
MRK 352 &00035243001 & $0.01^{+0.02}_{-0.01}$ & $1.75^{+0.07}_{-0.06}$ & 181.3/133 & 4.04 & 7.75 \\ 
\\
MRK 352 &00035243002 & $0.00^{+0.01}_{-0.00}$ & $1.65^{+0.04}_{-0.04}$ & 152.0/152 & 2.69 & 5.95 \\ 
\\
SWIFT J0216.3+5128 &00035247001 & $2.37^{+0.33}_{-0.29}$ & $2.15^{+0.18}_{-0.17}$ & 56.1/67 & 1.35 & 10.75 \\ 
\\
SWIFT J0216.3+5128 & 00035247002 & $1.15^{+0.25}_{-0.23}$ & $1.77^{+0.21}_{-0.20}$ & 46.0/44 & 1.30 & 11.10 \\
\\
SWIFT J0216.3+5128 & 00035247003 & $1.29^{+0.75}_{-0.65}$ & $1.88^{+0.59}_{-0.55}$ & 15.7/11 & 1.06 & 9.17 \\
\\
NGC 1142\tablenotemark{3} &00035248001 & 83.45 & 3.88 & 35.5/9 & 0.00 & 5.07 \\ 
\\
NGC 1142 &00035248002 & $28.19^{+20.59}_{-10.99}$ & $1.51^{+1.89}_{-1.18}$ & 9.3/7 & 0.00 & 7.36 \\
\\
NGC 1142 &00035248003 & $95.73^{+84.77}_{-36.88}$ & $3.11^{+4.44}_{-2.08}$ & 2.8/4 & 0.00 & 6.59 \\ 
\\
SWIFT J0318.7+6828 &00035249001 & $3.70^{+1.74}_{-1.83}$ & $1.73^{+0.48}_{-0.55}$ & 17.4/17 & 0.15 & 4.11 \\ 
\\
SWIFT J0318.7+6828 &00035249002 & $3.66^{+1.44}_{-1.41}$ & $1.44^{+0.40}_{-0.46}$ & 11.6/16 & 0.13 & 4.78 \\ 
\\
ESO 548-G081 &00035250001 & -- & $1.92^{+0.04}_{-0.05}$ & 131.7/136 & 13.91 & 19.62 \\ 
\\
ESO 548-G081 &00035250002 & -- & $2.06^{+0.03}_{-0.02}$ & 332.6/225 & 16.19 & 18.34 \\
\\
ESO 362-G018 &00035234001 & $0.00^{+0.01}_{-0.00}$ & $1.79^{+0.09}_{-0.09}$ & 36.6/33 & 7.66 & 12.69 \\
\\
ESO 362-G018 &00035234002 & -- & $1.45^{+0.05}_{-0.05}$ & 159.7/108 & 4.38 & 12.16 \\
\\
ESO 490-G026 &00035256001 & $0.27^{+0.04}_{-0.03}$ & $1.88^{+0.07}_{-0.07}$ & 143.8/177 & 7.07 & 18.91 \\ 
\\
ESO 490-G026 &00035256002 & $0.27^{+0.09}_{-0.09}$ & $1.70^{+0.14}_{-0.13}$ & 59.2/62 & 6.32 & 21.07 \\ 
\\
ESO 490-G026 &00035256003 & $0.49^{+0.15}_{-0.14}$ & $1.88^{+0.19}_{-0.18}$ & 27.2/42 & 4.28 & 14.49 \\ 
\\
SWIFT J0904.3+5538 &00035260001 & $0.00^{+0.01}_{-0.00}$ & $1.54^{+0.08}_{-0.08}$ & 39.9/40 & 2.04 & 5.02 \\ 
\\
SWIFT J0904.3+5538 &00035260002 & $0.00^{+0.02}_{-0.00}$ & $1.49^{+0.09}_{-0.09}$ & 34.2/32 & 1.51 & 4.01 \\ 
\\
MCG +04-22-042 &00035263001 & $0.01^{+0.00}_{-0.01}$ & $1.90^{+0.05}_{-0.05}$ & 173.9/185 & 7.99 & 11.69 \\ 
\\
MRK 417\tablenotemark{3}& 00035264003 & -- & -1.78 & 38.5/4 & 0.00 & 2.16 \\ 
\\
UGC 6728 &00035266001 & $0.01^{+0.02}_{-0.01}$ & $1.82^{+0.07}_{-0.07}$ & 106.6/117 & 6.88 & 11.78 \\ 
\\
UGC 6728 &00035266002 & $0.00^{+0.04}_{-0.00}$ & $1.76^{+0.15}_{-0.08}$ & 62.4/42 & 4.58 & 8.37 \\ 
\\
UGC 6728 &00035266003 & $0.04^{+0.08}_{-0.04}$ & $1.74^{+0.25}_{-0.20}$ & 19.6/19 & 4.80 & 9.67 \\ 
\\
SWIFT J1200.8+0650 &00035267001 & $11.23^{+3.50}_{-2.62}$ & $1.60^{+0.48}_{-0.41}$ & 39.5/28 & 0.01 & 4.99 \\ 
\\
SWIFT J1200.8+0650 &00035267002 & $4.48^{+13.37}_{-4.48}$ & $1.12^{+2.03}_{-1.27}$ & 2.7/3 & 0.00 & 5.68 \\ 
\\
ESO 506-G027\tablenotemark{3} &00035173004 & 36.38 & -0.22 & 33.7/9 & 0.00 & 5.02 \\ 
\\
WKK 1263 &00035268001 & $0.04^{+0.22}_{-0.04}$ & $1.63^{+0.27}_{-0.17}$ & 26.8/24 & 2.09 & 7.79 \\ 
\\
WKK 1263 &00035268002 & $0.12^{+0.11}_{-0.10}$ & $1.63^{+0.14}_{-0.13}$ & 55.1/59 & 1.73 & 6.98 \\ 
\\
MCG +09-21-096 &00035269001 & $0.00^{+0.01}_{-0.00}$ & $1.80^{+0.04}_{-0.03}$ & 250.2/240 & 13.67 & 22.13 \\ 
\\
MCG +09-21-096 &00035269004 & $0.00^{+0.02}_{-0.00}$ & $1.72^{+0.05}_{-0.04}$ & 143.9/151 & 11.41 & 21.09 \\ 
\\
NGC 6860 &00035275001 & $0.28^{+0.15}_{-0.11}$ & $1.17^{+0.20}_{-0.18}$ & 30.9/27 & 1.55 & 11.54 \\ 
\\
\enddata

\tablenotetext{1}{Cold hydrogen column density in units of $10^{22}$\,cm$^{-2}$ from the {\tt tbabs}
model.  This accounts for absorption beyond the Galactic values which are listed in Table~\ref{tbl-1}.
A dash in this column indicates that no extra absorption was necessary.  \\}
\tablenotetext{2}{The observed soft flux (0.3-2\,keV) and hard flux (2-10\,keV) are given in units of
$10^{-12}$\,erg\,cm$^{-2}$\,s$^{-1}$.\\}
\tablenotetext{3}{Errors at the 90\% confidence range can not be calculated for sources when
$\Delta\chi^2$/dof $> 2.0$.}
\end{deluxetable}

\begin{deluxetable}{lllllll}
\tabletypesize{\scriptsize}
\tablecaption{XMM-Newton Spectral Analysis: Absorbed Power Law Fits ({\tt tbabs}*{\tt tbabs}*{\tt pow})\label{tbl-4}}
\tablewidth{0pt}
\tablehead{
\colhead{Source} & \colhead{Observation ID}  & \colhead{n$_H$\tablenotemark{1}} & \colhead{$\Gamma$} &
\colhead{$\chi^2$/dof} & \colhead{Soft Flux\tablenotemark{2}} & \colhead{Hard Flux\tablenotemark{2}} 
}

\startdata
MRK 352 & 0312190101 & -- & $1.95^{+0.01}_{-0.01}$ & 3104.4/1569 & 6.87 &
 9.76 \\ 
\\
NGC 612 \tablenotemark{3} & 0312190201 & 79.90 & 0.31 & 274.2/83 & 0.00
 & 1.66 \\ 
\\
SWIFT J0216.3+5128 & 0312190301 & $1.74^{+0.06}_{-0.07}$ & $1.77^{+0.04}_{-0.04}$ & 1069.7/985 & 0.90 & 8.44 \\ 
\\
NGC 1142 \tablenotemark{3} & 0312190401 & 65.68 & 1.63 & 942.8/197 & 0.00 & 3.00 \\ 
\\
SWIFT J0318.7+6828 & 0312190501 & $3.20^{+0.31}_{-0.34}$ & $1.36^{+0.10}_{-0.11}$ & 812.3/772 & 0.23 & 7.43 \\ 
\\
ESO 548-G081 & 0312190601 & -- & $2.03^{+0.01}_{-0.01}$ & 2943.5/1642 & 11.34 & 13.48 \\ 
\\
ESO 362-G018 \tablenotemark{3} & 0312190701 & --& 1.47& 2534.6/549 & 0.69 & 1.83 \\ 
\\
ESO 490-G026 & 0312190801 & $0.24^{+0.01}_{-0.01}$ & $1.59^{+0.02}_{-0.02}$ & 1634.2/1757 & 5.30 & 19.79 \\ 
\\
SWIFT J0641.3+3257 & 0312190901 & $12.11^{+3.08}_{-1.79}$ & $0.98^{+0.30}_{-0.20}$ & 218.4/154 & 0.00 & 3.38 \\ 
\\
MRK 18 & 0312191001 & $13.06^{+7.20}_{-3.52}$ & $1.26^{+0.72}_{-0.38}$ & 413.4/327 & 0.00 & 1
.60 \\ 
\\
SWIFT J0904.3+5538 & 0312191101 & -- & $1.79^{+0.01}_{-0.02}$ & 1126.3/865 & 3.19 & 5.41 \\ 
\\
SWIFT J0911.2+4533 & 0312191201 & $27.63^{+27.50}_{-9.43}$ & $1.34^{+2.21}_{-0.43}$ & 19.9/17 & 0.00 & 1.55 \\ 
\\
MCG +04-22-042 & 0312191401 & -- & $2.00^{+0.01}_{-0.00}$ & 1835.8/1819 &
 12.45 & 15.52 \\ 
\\
MRK 417 \tablenotemark{3} & 0312191501 & 41.76 & 0.56 & 565.6/82 & 0.00
 & 1.52 \\ 
\\
UGC 6728 & 0312191601 & $0.01^{+0.00}_{-0.01}$ & $1.78^{+0.02}_{-0.02}$ & 1207.0/1160 & 3.59 
& 6.49 \\ 
\\
SWIFT J1200.8+0650 & 0312191701 & $9.31^{+0.72}_{-0.67}$ & $1.30^{+0.11}_{-0.10}$ & 484.0/357 & 0.02 & 5.17 \\ 
\\
ESO 506-G027 \tablenotemark{3} & 0312191801 & 75.91 & 1.14 & 543.4/172 
& 0.00 & 3.73 \\ 
\\
WKK 1263 & 0312191901 & $0.06^{+0.01}_{-0.02}$ & $1.53^{+0.03}_{-0.02}$ & 1257.3/1248 & 2.32 
& 9.99 \\ 
\\
MCG +09-21-096 & 0312192001 & -- & $1.76^{+0.01}_{-0.00}$ & 1291.2/1626 &
 9.27 & 15.85 \\ 
\\
NGC 4992 & 0312192101 & $76.20^{+16.36}_{-8.73}$ & $1.69^{+0.58}_{-0.30}$ & 139.6/82 & 0.00 &
 1.91 \\ 
\\
NGC 6860 & 0312192201 & $0.00^{+0.01}_{-0.00}$ & $0.28^{+0.04}_{-0.03}$ & 523.5/406 & 0.57 & 
9.84 \\ 
\\
NGC 6921 & 0312192301 & $98.33^{+14.57}_{-12.91}$ & $1.95^{+0.40}_{-0.37}$ & 287.3/168 & 0.00
 & 2.58 \\ 
\\
 MCG +04-48-002\tablenotemark{3} & 0312192301 & 0.0 & $1.98$ & 146.0/56 & 0.02
 & 0.06 \\ 
\\
\enddata

\tablenotetext{1}{Cold hydrogen column density in units of $10^{22}$\,cm$^{-2}$ from the {\tt tbabs}
model.  This accounts for absorption beyond the Galactic values which are listed in Table~\ref{tbl-1}.
A dash in this column indicates that no extra absorption was necessary.  \\}
\tablenotetext{2}{The observed soft flux (0.3-2\,keV) and hard flux (2-10\,keV) are given in units of
$10^{-12}$\,erg\,cm$^{-2}$\,s$^{-1}$.\\}
\tablenotetext{3}{Errors at the 90\% confidence range were not calculated for sources when
$\Delta\chi^2$/dof $> 2.0$, since these fits were unsatisfactory.}
\end{deluxetable}

\begin{deluxetable}{llllllll}
\tabletypesize{\scriptsize}
\tablecaption{XMM-Newton Detailed Fits: {\tt tbabs}*{\tt tbabs}*({\tt pow} + {\tt zgauss}) Model\label{tbl-5}}
\tablewidth{0pt}
\tablehead{
\colhead{Source} & \colhead{n$_H$\tablenotemark{1}}  & \colhead{$\Gamma$} & 
\colhead{Fe K eqw\tablenotemark{2}} & \colhead{Fe K norm.\tablenotemark{2}} &
\colhead{$\chi^2$/dof} & \colhead{Soft Flux\tablenotemark{3}} & \colhead{Hard Flux\tablenotemark{3}} 
}

\startdata
MCG +04-22-042  &  -- & $2.01^{+0.01}_{-0.01}$  & $130.38^{+33.95}_{-38.60}$ & $21.0^{+5.50}_{-6.20}$ & 1801.9/1818 & 12.44 & 15.65 \\
\\
UGC 6728  & $0.01_{-0.01}$ & $1.78^{+0.03}_{-0.02}$  & $83.33^{+56.58}_{-45.66}$ & $5.73^{+3.89}_{-3.14}$ & 1198.8/1159 & 3.59 & 6.51 \\
\\
WKK 1263  & $0.06^{+0.01}_{-0.02}$ & $1.54^{+0.02}_{-0.02}$  & $31.23^{+32.35}_{-23.67}$ & $3.69^{+3.83}_{-2.80}$ & 1253.0/1247 & 2.32 & 9.99 \\
\\
MCG +09-21-096 &  -- & $1.77^{+0.01}_{-0.01}$   & $55.94^{+31.09}_{-29.09}$ & $9.83^{+5.47}_{-5.11}$ & 1281.6/1625 & 9.26 & 15.89 \\
\\

\enddata
\tablenotetext{1}{Cold hydrogen column density in units of $10^{22}$\,cm$^{-2}$ from the {\tt tbabs}
model.  This accounts for absorption beyond the Galactic values which are listed in Table~\ref{tbl-1}.
A dash in this column indicates that no extra absorption was necessary.  \\}
\tablenotetext{2}{Equivalent width (in eV) and flux normalization for an inserted Gaussian line at
6.4\,keV (redshifted) with a set FWHM of 0.01\,keV.  The flux normalization is in units of
$10^{-6} \times$ total photons\,cm$^{-2}$\,s$^{-1}$ in the line.\\}
\tablenotetext{3}{The observed soft flux (0.3-2\,keV) and hard flux (2-10\,keV) are given in units of
$10^{-12}$\,erg\,cm$^{-2}$\,s$^{-1}$.}

\end{deluxetable}

\begin{deluxetable}{lllllllll}
\tabletypesize{\scriptsize}
\tablecaption{XMM-Newton Detailed Fits: {\tt tbabs}*{\tt tbabs}*({\tt bbody} + {\tt pow} + {\tt zgauss}) Model\label{tbl-6}}
\tablewidth{0pt}
\tablehead{
\colhead{Source} & \colhead{n$_H$\tablenotemark{1}}  & \colhead{kT\tablenotemark{2}} 
& \colhead{$\Gamma$} & 
\colhead{Fe K eqw\tablenotemark{3}} & \colhead{Fe K norm.\tablenotemark{3}} &
\colhead{$\chi^2$/dof} & \colhead{Soft Flux\tablenotemark{4}} & \colhead{Hard Flux\tablenotemark{4}} 
}

\startdata
MRK 352 &  --   & $0.096^{+0.003}_{-0.003}$ & $1.70^{+0.02}_{-0.01}$ & $60.29^{+28.09}_{-27.89}$ & $7.78^{+3.62}_{-3.60}$ & 1695.2/1566 & 6.50 & 11.89 \\
\\
ESO 548-G081  & --  & $0.087^{+0.003}_{-0.003}$ & $1.85^{+0.01}_{-0.01}$  & $114.35^{+38.41}_{-22.56}$ & $18.7^{+4.60}_{-4.70}$ & 1760.8/1639 & 10.84 & 15.83 \\
\\
ESO 490-G026 & $0.33^{+0.04}_{-0.02}$ & $0.074^{+0.005}_{-0.006}$ & $1.67^{+0.03}_{-0.02}$ & $59.37^{+23.36}_{-23.28}$ & $13.3^{+5.20}_{-5.23}$ & 1542.9/1754 & 5.31 & 19.49 \\ 
 \\ 
SWIFT J0904.3+5538 & $0.06^{+0.03}_{-0.02}$ & $0.070^{+0.004}_{-0.006}$ & $1.71^{+0.06}_{-0.03}$  & $40.71^{+71.76}_{-40.71}$ & $2.79^{+4.92}_{-2.79}$ & 903.2/862 & 3.05 & 5.96 \\ 
\\

\enddata
\tablenotetext{1}{Cold hydrogen column density in units of $10^{22}$\,cm$^{-2}$ from the {\tt tbabs}
model.  This accounts for absorption beyond the Galactic values which are listed in Table~\ref{tbl-1}.
A dash in this column indicates that no extra absorption was necessary.  \\}
\tablenotetext{2}{Temperature of the blackbody component (kT) in keV units.\\}
\tablenotetext{3}{Equivalent width and flux normalization for an inserted Gaussian line at
6.4\,keV (redshifted) with a set FWHM (in eV) of 0.01\,keV.  The flux normalization is in units of
$10^{-6} \times$ total photons\,cm$^{-2}$\,s$^{-1}$ in the line.\\}
\tablenotetext{4}{The observed soft flux (0.3-2\,keV) and hard flux (2-10\,keV) are given in units of
$10^{-12}$\,erg\,cm$^{-2}$\,s$^{-1}$.}
\end{deluxetable}

\begin{deluxetable}{llll}
\tablecaption{XMM-Newton Warm Absorber Model\label{tbl-11}}
\tablewidth{0pt}
\tablehead{
\colhead{Source} & \colhead{$\tau$ \ion{O}{7} \tablenotemark{a}}  & \colhead{$\tau$ \ion{O}{8}\tablenotemark{a}} & \colhead{$\Delta \chi^2$\tablenotemark{b}} 
}
\startdata
MRK 352 & 0.017 & 0.005 & -12.3 \\
\\
ESO 548-G081  & 0.032 & 0.005  & -0.06 \\
\\
ESO 490-G026 & $0.233^{+0.048}_{-0.083}$ & $0.095^{+0.047}_{-0.043}$ & 25 \\ 
 \\ 
SWIFT J0904.3+5538 & $0.186^{+0.098}_{-0.091}$ & 0.039 & 9.9 \\ 
\\
 MCG +04-22-042 & $0.063^{+0.022}_{-0.029}$ & $0.035^{+0.026}_{-0.025}$ & 41.5 \\
\\
UGC 6728 & 0.072 & 0.023 & 0.8 \\
\\
WKK 1263 & $0.046^{+0.087}_{-0.046}$ & 0.020 & 1.0 \\
\\
MCG +09-21-096 & $0.036^{+0.023}_{-0.025}$& 0.008 & 5.7 \\
\\
\enddata
\tablenotetext{a}{Optical depth with errors or upper limits for the additions of edge models ({\tt zedge}) at 0.74\,keV and 0.87\,keV.\\}
\tablenotetext{b}{The $\chi^2$ from the best fit model in Table~\ref{tbl-5} or~\ref{tbl-6} minus $\chi^2$ from the warm absorber model.}
\end{deluxetable}

\begin{deluxetable}{lllllllll}
\tabletypesize{\scriptsize}
\tablecaption{XMM-Newton Detailed Fits: {\tt tbabs}*{\tt pcfabs}*({\tt pow} + {\tt zgauss}) Model\label{tbl-7}}
\tablewidth{0pt}
\tablehead{
\colhead{Source} & \colhead{n$_H$\tablenotemark{1}}  & \colhead{fraction\tablenotemark{1}} 
& \colhead{$\Gamma$} & 
\colhead{Fe K eqw\tablenotemark{2}} & \colhead{Fe K norm.\tablenotemark{2}} &
\colhead{$\chi^2$/dof} & \colhead{Soft Flux\tablenotemark{3}} & \colhead{Hard Flux\tablenotemark{3}} 
}
\startdata
NGC 612 &$129.70^{+12.90}_{-8.30}$ & $0.999^{+0.001}_{-0.002}$ & $2.12^{+0.06}_{-0.33
}$ &  $108.10^{+63.74}_{-62.83}$ & $17.9^{+10.6}_{-10.4}$ & 116.7/81 & 0.02 & 1.50 \\ 
\\
NGC 1142 &$79.75^{+5.81}_{-3.05}$ & $0.996^{+0.001}_{-0.001}$ & $2.27^{+0.09}_{-0.17}
$ &  $219.21^{+57.30}_{-40.87}$ & $39.0^{+10.2}_{-7.27}$ & 260.5/195 & 0.08 & 3.01 \\ 
\\
SWIFT J0318.7+6828 &$4.10^{+0.48}_{-0.41}$ & $0.967^{+0.009}_{-0.009}$ & $1.52^{+0.12}_{-0.11}$ & $44.15^{+41.58}_{-39.43}$ & $5.35^{+5.04}_{-4.78}$ & 775.9/770 & 0.24 & 7.31 \\
\\
ESO 362-G018 &$26.64^{+2.72}_{-2.49}$ & $0.913^{+0.005}_{-0.011}$ & $2.13^{+0.04}_{-0
.04}$  & $421.82^{+35.15}_{-105.96}$ & $28.6^{+2.38}_{-7.18}$ & 824.1/547 & 0.63 & 3.34 \\ 
\\
SWIFT J0641.3+3257 &$16.01^{+2.68}_{-2.28}$ & $0.982^{+0.007}_{-0.009}$ & $1.24^{+0.26}_{-0
.23}$ &  $7.76^{+46.01}_{-7.76}$ & $0.51^{+3.00}_{-0.51}$ & 176.9/152 & 0.02 & 3.31 \\ 
\\
MRK 18 &$18.25^{+3.64}_{-2.71}$ & $0.97^{+0.02}_{-0.02}$ & $1.62^{+0.31}_{-0.22}$ 
&  $178.09^{+109.77}_{-107.72}$ & $5.47^{+3.37}_{-3.31}$
 & 322.4/325 & 0.03 & 1.57 \\ 
 \\
SWIFT J0911.2+4533 &$33.02^{+11.01}_{-12.76}$ & $0.994^{+0.004}_{-0.009}$ & $2.47^{+0.98}
_{-1.20}$ &  $582.76^{+425.24}_{-317.37}$ & $23.7^{+17.3}_{-12.9}$ &
 9.3/15 & 0.01 & 1.45 \\ 
 \\
MRK 417 &$85.69^{+12.73}_{-6.96}$ & $0.995^{+0.002}_{-0.002}$ & $2.25^{+0.15}_{-0.17}
$ &  $114.77^{+75.78}_{-80.97}$ & $10.8^{+7.11}_{-7.60}$ & 88.1/80
 & 0.06 & 1.44 \\ 
 \\
SWIFT J1200.8+0650 &$10.80^{+0.68}_{-0.82}$ & $0.991^{+0.002}_{-0.003}$ & $1.47^{+0.10}_{-0
.11}$ &  $84.18^{+47.13}_{-27.15}$ & $7.69^{+4.31}_{-2.48}$ & 340.3/355 & 0.04 & 5.08 \\ 
\\
ESO 506-G027 &$76.82^{+7.37}_{-6.79}$ & $0.986^{+0.006}_{-0.003}$ & $0.91^{+0.23}_{-0
.18}$ &  $428.89^{+50.01}_{-85.01}$ & $72.0^{+8.40}_{-14.3}$ & 228.9/170 & 0.03 & 3.87 \\ 
\\
NGC 4992 &$69.05^{+7.39}_{-2.38}$ & $0.9974^{+0.0015}_{-0.0016}$ & $1.61^{+0.13}_{-0.41}
$ & $320.22^{+80.93}_{-72.56}$ & $29.8^{+5.55}_{-5.39}$ & 120.8/132
 & 0.01 & 1.93 \\ 
 \\
 NGC 6860 &$4.53^{+1.33}_{-1.30}$ & $0.60^{+0.07}_{-0.10}$ & $0.79^{+0.11}_{-0.15}$ & $75.64^{+62.90}_{-64.48}$ & $9.45^{+7.86}_{-8.06}$ & 483.5/404 & 0.54 & 8.99 \\
 \\
NGC 6921 &$97.42^{+19.08}_{-3.78}$ & $0.9984^{+0.0007}_{-0.0012}$ & $2.34^{+0.11}_{-0.2
1}$ &  $54.56^{+47.68}_{-53.02}$ & $11.9^{+9.8}_{-10.3}$ & 190.7/154 & 0.03 & 2.53 \\
\\
MCG +04-48-002 & $96.00^{+51.97}_{-27.77}$ & $0.9963^{+0.0023}_{-0.0133}$ & $2.47^{+ 0.32}_{-0.44}$ &  $811^{+1219}_{-329}$ & $42.3^{+63.7}_{-17.2}$ & 62.8/52 & 0.025 & 0.31 \\
\\

\enddata
\tablenotetext{1}{Cold hydrogen column density in units of $10^{22}$\,cm$^{-2}$ from the {\tt pcfabs}
model.  This accounts for absorption beyond the Galactic values which are listed in Table~\ref{tbl-1}.
The fraction column is the partial covering fraction.  \\}
\tablenotetext{2}{Equivalent width (in eV) and flux normalization for an inserted Gaussian line at
6.4\,keV (redshifted) with a set FWHM of 0.01\,keV.  The flux normalization is in units of
$10^{-6} \times$ total photons\,cm$^{-2}$\,s$^{-1}$ in the line.\\}
\tablenotetext{3}{The observed soft flux (0.3-2\,keV) and hard flux (2-10\,keV) are given in units of
$10^{-12}$\,erg\,cm$^{-2}$\,s$^{-1}$.}
\end{deluxetable}

\begin{deluxetable}{lllllllllll}
\tabletypesize{\scriptsize}
\tablecaption{XMM-Newton Detailed Fits: {\tt tbabs}*({\tt tbabs}*{\tt pow} + {\tt tbabs}*({\tt pow} + {\tt zgauss})) Model\label{tbl-8}}
\tablewidth{0pt}
\tablehead{
\colhead{Source} & \colhead{n$_H$\tablenotemark{1}}  & \colhead{$\Gamma_1$} & \colhead{n$_H$} &
 \colhead{$\Gamma_2$} & \colhead{N$_{\Gamma_1}$/N$_{\Gamma_2}$\tablenotemark{2}} &
\colhead{Fe K eqw\tablenotemark{3}} & \colhead{Fe K norm.\tablenotemark{3}} &
\colhead{$\chi^2$/dof} & \colhead{Flux\tablenotemark{4}} 
}
\startdata
NGC 612 &$0.32^{+0.09}_{-0.18}$ & $4.27^{+1.78}_{-1.15}$ & $79.79^{+19.47}_{-12.30}$ & $0.28^{+0.21}_{-0.19}$ & 0.215 & $177.00^{+86.96}_{-74.36}$ & $12.8^{+6.25}_{-5.37}$ & 82.0/79 & 1.69 \\ 
 \\ 
NGC 1142 &$0.15^{+0.07}_{-0.06}$ & $3.10^{+0.31}_{-0.71}$ & $62.94^{+10.87}_{-5.72}$ & $1.49^{+0.31}_{-0.17}$ & 0.028 & $268.10^{+62.48}_{-52.50}$ & $35.4^{+8.31}_{-6.90}$ & 245.8/193 &  3.19 \\ 
 \\ 
SWIFT J0318.7+6828 &$1.11^{+1.42}_{-1.11}$ & $1.51^{+2.63}_{-0.72}$ & $6.10^{+8.15}_{-6.10}$ & $1.74^{+1.10}_{-0.24}$ & 0.139 & $45.29^{+62.89}_{-37.03}$ & $5.63^{+5.36}_{-4.79}$ & 772.8/768 & 7.41 \\ 
 \\ 
ESO 362-G018 &$0.00^{+0.01}_{-0.00}$ & $2.30^{+0.07}_{-0.06}$ & $6.30^{+2.32}_{-2.66}$ & $0.67^{+0.18}_{-0.23}$ & 1.753 & $432.23^{+110.50}_{-32.59}$ & $22.1^{+3.31}_{-3.40}$ & 736.0/545 &  4.26 \\ 
 \\ 
SWIFT J0641.3+3257 &$0.00^{+0.04}_{-0.00}$ & $0.82^{+1.38}_{-0.76}$ & $16.96^{+9.75}_{-3.30}$ & $1.45^{+0.78}_{-0.42}$ & 0.012 & $1.58^{+55.02}_{-1.58}$ & $0.20^{+3.37}_{-0.20}$ & 176.5/150 &  3.30 \\ 
 \\ 
MRK 18 &$0.36^{+0.57}_{-0.29}$ & $3.74^{+2.79}_{-1.67}$ & $18.28^{+7.22}_{-5.71}$ & $1.78^{+0.65}_{-0.56}$ & 0.048 & $188.57^{+113.34}_{-109.56}$ & $5.91^{+3.33}_{-3.49}$ & 315.4/323 & 1.53 \\ 
 \\ 
SWIFT J0911.2+4533 &$3.84^{+9.73}_{-3.84}$ & $2.56^{+7.44}_{-5.52}$ & $48.03^{+28.32}_{-24.61}$ & $3.35^{+6.59}_{-2.04}$ & 0.002 & $695.56^{+1213.04}_{-437.52}$ & $32.0^{+55.8}_{-20.1}$ & 8.2/13  & 1.36 \\ 
 \\ 
MRK 417 &$0.00^{+0.04}_{-0.00}$ & $2.36^{+0.16}_{-0.16}$ & $54.15^{+25.04}_{-11.30}$ & $0.88^{+0.97}_{-0.42}$ & 0.101 & $179.20^{+87.71}_{-86.64}$ & $9.65^{+4.72}_{-4.67}$ & 78.8/78  & 1.62 \\ 
 \\ 
SWIFT J1200.8+0650 &$0.00^{+0.06}_{-0.00}$ & $1.75^{+0.93}_{-0.43}$ & $10.57^{+0.79}_{-0.98}$ & $1.43^{+0.10}_{-0.08}$ & 0.009 & $94.46^{+35.70}_{-35.96}$ & $8.58^{+3.24}_{-3.27}$ & 338.9/353 & 5.14 \\ 
 \\ 
ESO 506-G027 &$0.00^{+0.08}_{-0.00}$ & $0.81^{+0.28}_{-0.31}$ & $80.37^{+7.77}_{-8.68}$ & $0.99^{+0.48}_{-0.24}$ & 0.011 & $418.22^{+69.39}_{-59.66}$ & $73.6^{+12.2}_{-10.5}$ & 228.1/168  & 3.89 \\ 
 \\ 
NGC 4992 &$0.32^{+0.73}_{-0.31}$ & $2.68^{+3.73}_{-1.69}$ & $68.47^{+14.94}_{-9.62}$ & $1.41^{+0.53}_{-0.14}$ & 0.006 & $318.31^{+90.94}_{-74.91}$ & $27.9^{+7.96}_{-6.56}$ & 70.4/78  & 1.97 \\ 
 \\ 
NGC 6860 &$0.00^{+0.01}_{-0.00}$ & $0.47^{+0.10}_{-0.06}$ & $0.60^{+9.45}_{-0.13}$ & $0.48^{+1.56}_{-0.08}$ & 0.745 & $65.36^{+60.57}_{-62.15}$ & $8.17^{+7.57}_{-7.77}$ & 459.7/402 & 9.97 \\ 
 \\ 
NGC 6921 &$0.57^{+0.20}_{-0.21}$ & $5.08^{+2.53}_{-1.25}$ & $90.25^{+10.75}_{-10.63}$ & $1.71^{+0.53}_{-0.25}$ & 0.025 & $89.01^{+54.93}_{-58.37}$ & $14.7^{+9.11}_{-9.67}$ & 149.3/164  & 2.64 \\ 
 \\ 
 MCG +04-48-002 &$0.13^{+0.36}_{-0.13}$ & $3.06^{+1.07}_{-0.78}$ & $223.58^{+48.06}_{-88.16}$ & $10.00_{-6.63}$ & 1.6e-9 & $496^{+954}_{-279}$ & $34.2^{+66.6}_{-19.1}$ & 58.4/50  & 0.26 \\ 
 \\ 
\enddata
\tablenotetext{1}{Cold hydrogen column density in units of $10^{22}$\,cm$^{-2}$ from the {\tt tbabs}
model.  This accounts for absorption beyond the Galactic values which are listed in Table~\ref{tbl-1}.
In this model, there is a separate column density component for each of the two power laws.\\}
\tablenotetext{2}{Ratio of the flux normalization value for power law one versus power law two.  For
nearly all of these sources, the low absorption power law is much weaker.\\}
\tablenotetext{3}{Equivalent width (in eV) and flux normalization for an inserted Gaussian line at
6.4\,keV (redshifted) with a set FWHM of 0.01\,keV.  The flux normalization is in units of
$10^{-6} \times$ total photons\,cm$^{-2}$\,s$^{-1}$ in the line.\\}
\tablenotetext{4}{The observed total flux (0.5 -- 10\,keV) in units of
$10^{-12}$\,erg\,cm$^{-2}$\,s$^{-1}$.}
\end{deluxetable}

\begin{deluxetable}{llllll}
\tabletypesize{\scriptsize}
\tablecaption{Compton-thick Reflection Model: {\tt tbabs}*({\tt tbabs}*{\tt pow} + {\tt tbabs}*({\tt pexrav} + {\tt zgauss}))\label{tbl-10}}
\tablewidth{0pt}
\tablehead{
\colhead{Source} & \colhead{n$_H$\tablenotemark{1}}  & \colhead{$\Gamma$} & \colhead{cutoff E\tablenotemark{2}} &
 \colhead{BAT norm\tablenotemark{3}} & \colhead{$\chi^2$/dof} 
}
\startdata
NGC 612\tablenotemark{4} & $62.55^{+11.90}_{-23.92}$ & $0.95^{+0.36}_{-0.80}$ & $48.58^{+83.42}_{-24.18}$ & $0.08^{+0.09}_{-0.05}$ & 81.7/85\\
\\
ESO 362-G018 & $0.62^{+0.90}_{-0.26}$ & $1.99^{+0.15}_{-0.27}$ & 100 & $0.782^{+0.27}_{-0.22}$ & 622.8/547 \\
\\
MRK 417\tablenotemark{4} & $18.19^{+9.00}_{-13.57}$ & $0.73^{+0.34}_{-0.53}$ & $106.7^{+112.10}_{-54.70}$ & $0.19^{+0.12}_{-0.09}$  & 114.6/83 \\
\\
ESO 506-G027 &  $65.49^{+11.93}_{-10.40}$ & $1.88^{+0.22}_{-0.26}$ & $100$ & $0.94^{+0.24}_{-0.10}$  & 226.7/174 \\
\\
NGC 6860 & $0.00^{+1.04}$ & $2.31^{+0.14}_{-0.17}$ & 100 &  $1.21^{+0.32}_{-0.19}$ &  444.5/408\\
\\
\enddata
\tablenotetext{1}{Cold hydrogen column density in units of $10^{22}$\,cm$^{-2}$ from the {\tt tbabs}
model.  This accounts for absorption beyond the Galactic values which are listed in Table~\ref{tbl-1}.
In this model, there is a separate column density component for each of the two power laws.
The column densities listed are for the more heavily absorbed source.\\}
\tablenotetext{2}{Cutoff energy for the {\tt pexrav}/reflection model.  For this model, we assumed
that this component was a pure reflection component and allowed the cutoff energy to vary.  Where the cutoff energy could not be constrained (the model parameter increases to very large, unphysical values), we fixed this parameter at the default value of 100\,keV.\\}
\tablenotetext{3}{Using the constant model, we allowed the BAT normalization to vary by a multiplicative factor.  The recorded
value is the factor variation of the BAT flux to the PN flux (normalized at 1), except for NGC 6860 which has no PN data (normalized to MOS1).}
\tablenotetext{4}{For NGC 612 and MRK 417, the best fit to the data, in terms of reduced $\chi^2$, was a model with low BAT normalization.  Both of these sources showed curvature in the BAT spectrum, which was not well fit by the {\tt pexrav} model.  We include, in the text, a discussion of this as well as a description of the fits with BAT normalization set to 1 (similar to the best fits for the remaining 3 Compton-thick candidates).}
\end{deluxetable}

\begin{deluxetable}{llll}
\tabletypesize{\scriptsize}
\tablecaption{Variability in individual \xmm Observations\label{tbl-13}}
\tablewidth{0pt}
\tablehead{
\colhead{Source} & \colhead{$<$Ct Rate$>$\tablenotemark{a}}  & \colhead{$\sigma^{2}_{rms}$\tablenotemark{b}} & \colhead{$\chi^2$/dof\tablenotemark{c}} 
}
\startdata
MRK 352  & 7.023 & $7.010\pm0.0014$ & 576.71/99 \\ 
NGC 612  & 0.119 & $8.583\pm5.640$ & 123.93/107 \\
SWIFT J0216.3+5128 & 1.470 & $1.043\pm0.0048$ & 82.57/100 \\  
NGC 1142  & 0.283 & $6.755\pm0.8434$ & 82.51/100 \\  
SWIFT J0318.7+6828 & 1.166 & $49.05\pm0.5085$ & 727.06/71 \\ 
ESO 548-G081  & 11.730 & $4.497\pm3\times10^{-4}$ & 620.33/100 \\
ESO 362-G018  & 0.949 & $1.067\pm0.01185$ & 107.24/100 \\
ESO 490-G026  & 5.128 & $0.00523\pm2\times10^{-6}$ & 105.16/103 \\  
SWIFT J0641.3+3257  & 0.243 & $62.76\pm8.863$ & 273.35/120 \\  
MRK 18  & 0.507 & $94.36\pm3.251$ & 995.16/113 \\     
SWIFT J0904.3+5538 & 0.960 & $0.7704\pm7\times10^{-4}$ & 133.70/121 \\  
MCG +04-22-042 & 13.151 & $0.08061\pm6\times10^{-5}$ & 88.10/80 \\
MRK 417  & 0.145 & $19.83\pm12.28$ & 103.53/77 \\  
UGC 6728  & 3.982 & $29.45\pm0.02293$ & 1069.19/81 \\  
SWIFT J1200.8+0650  & 0.430 & $1.860\pm0.09154$ & 100.27/110 \\  
ESO 506-G027  & 0.268 & $4.805\pm0.6714$ & 117.02/100 \\
WKK 1263  & 2.446 & $0.1615\pm3\times10^{-4}$ & 94.24/89 \\ 
MCG +09-21-096 & 9.348 & $0.02412\pm3\times10^{-6}$ & 96.63/99 \\ 
NGC 4992  & 0.136 & $9.399\pm3.531$ & 163.66/143 \\
NGC 6860  & 0.478 & $49.60\pm1.823$ & 377.40/119 \\  
NGC 6921  & 0.268 & $105.8\pm12.07$ & 500.70/122 \\   
\enddata
\tablenotetext{a}{Average count rate for the \xmm observation in the 0.3 -- 10\,keV band.\\}
\tablenotetext{b}{Corresponding excess variability measurements, as defined in \citet{nan97} $\times 10^{-3}$.\\}
\tablenotetext{c}{$\chi^2$ value divided by the number of bins for variability.}
\end{deluxetable}

\begin{deluxetable}{llllllll}
\tabletypesize{\scriptsize}
\tablecaption{Variability between \xmm and XRT Observations\label{tbl-9}}
\tablewidth{0pt}
\tablehead{
\colhead{Source} & \colhead{Soft var.\tablenotemark{1}}  & \colhead{$\Delta$ t$_{soft}$\tablenotemark{2}} & 
\colhead{Hard var.\tablenotemark{1}} &  \colhead{$\Delta$ t$_{hard}$\tablenotemark{2}} & \colhead{n$_H$ var.?} &
 \colhead{$\Gamma$ var.?} & \colhead{Flux var.?} 
}
\startdata
MRK 352 & 0.92 & 125.7 &0.49  & 125.7 & yes & no & yes \\
SWIFT J0216.3+5128 & 0.39 & 16.5 & 0.27 & 14.7 & no & no & yes \\
NGC 1142 & 0.00 & 151.9 & 0.79 &  159.4 & no & yes & yes \\
SWIFT J0318.7+6828 & 0.59 & 65.5 & 0.61 & 59.4 & no & no & yes \\
ESO 548-G081 & 0.35 & 49.7 & 0.36 & 49.7 & no &yes &yes  \\
ESO 362-G018 & 1.64 & 91.8 & 1.22 & 63.5 & yes & no & yes \\
ESO 490-G026 & 0.49 & 101.7 & 0.35 & 4.8 & yes &yes &yes  \\
SWIFT J0904.3+5538 & 0.75 & 84.1 & 0.29 & 84.1 & yes &yes &yes  \\
MCG +04-22-042 & 0.44 & 129.1 & 0.28 & 129.1 & yes &no &yes  \\
MRK 417 & 0.00 & 8.3 & 0.35 & 8.3 & no & no & no \\
UGC 6728 & 0.66 & 119.8 & 0.58 & 119.8 & no &no &yes  \\
SWIFT J1200.8+0650 & 2.00 & 188.4 & 0.13 & 9.7 & no & no & no \\
ESO 506-G027 & 0.00 & 149.7 & 0.29 & 149.7 &no & no & no \\
WKK 1263 & 0.29 & 35.0 & 0.36 & 35.0 & no &no &yes  \\
MCG +09-21-096 & 0.38 & 186.2 & 0.32 & 186.2 & no &yes &yes  \\
NGC 6860 & 0.92 & 97.7 & 0.16 & 97.7 & yes &yes &yes  \\
\enddata
\tablenotetext{1}{Comparing soft and hard flux from the \xmm and XRT observations listed
in Tables~\ref{tbl-3} and~\ref{tbl-4}. As noted in the Variability section, we used the statistic
$(F_{max} - F_{min})/F_{avg}$ to compare the individual \xmm and XRT fluxes in the
soft (0.5 -- 2.0\,keV) and hard (2.0 -- 10.0\,keV) bands. For a few of the high column density
sources (n$_H > 10^{23}$\,cm$^{-2}$), the soft band flux was not able to be measured accurately
due to a lack of counts.  Therefore, the values in the Soft var. column are unreliable for these
sources (NGC 1142, MRK 417, SWIFT J1200.8+0650, and ESO 506-G027).}
\tablenotetext{2}{The corresponding value t$_{max} - $t$_{min}$ in days for the $(F_{max} - F_{min})/F_{avg}$ values in each band.}
\end{deluxetable}

\end{document}